\documentclass[12pt,preprint, twosided]{article}
\usepackage[utf8]{inputenc}
\usepackage[T1]{fontenc}
\usepackage{amsmath}
\usepackage{mathrsfs}
\usepackage{graphicx}
\usepackage{bm}
\usepackage{latexsym,epsfig,dcolumn} 
\usepackage{amsfonts}
\usepackage{amssymb}
\usepackage{amsfonts}
\usepackage{amsthm}
\usepackage{mathtools}
\usepackage{url}
\usepackage{tabularx}
\usepackage{latexsym}
\usepackage{amsmath,amssymb,dsfont}
\usepackage{ifpdf}
\usepackage{hyperref}
\usepackage{nicefrac}
\usepackage[onehalfspacing]{setspace}
\usepackage{here}
\usepackage[round]{natbib}
\usepackage{float}
\usepackage{paralist}
\usepackage{lscape}
\usepackage{arydshln}
\usepackage{authblk}

\usepackage{geometry}
\usepackage[usenames,dvipsnames,svgnames,table]{xcolor}
\title{Fast symmetric additive covariance smoothing}
\author{Jona Cederbaum, Fabian Scheipl, Sonja Greven\footnote{\textbf{Address for correspondence:} Sonja Greven, 
             Department of Statistics, 
             Faculty of Mathematics, Computer Science and Statistics,
             LMU Munich,
             Ludwigstr.~33, 80539 Munich,
             Germany.\\
			 \textbf{E-mail:} Sonja.Greven@stat.uni-muenchen.de}}
\date{}
\affil{Department of Statistics,
	  Faculty of Mathematics, Computer Science and Statistics, \\
	  LMU Munich,
	  Germany}

\newcommand{\pkg}[1]{\textsf{#1}}
\newcommand{\bvec}{\left[\begin{array}{c}}
\newcommand{\evec}{\end{array}\right]}
\newcommand{\bmat}[1]{\left[\begin{array}{*{#1}{c}}}
\newcommand{\emat}{\end{array}\right]}

\newcommand{\tocite}[1]{\emph{\textcolor{red}{(CITE)}}}

\newcommand{\bi}{\begin{itemize}}
\newcommand{\ei}{\end{itemize}}
\newcommand{\ben}{\begin{enumerate}}
\newcommand{\een}{\end{enumerate}}
\newcommand{\beq}{\begin{equation}}
\newcommand{\eeq}{\end{equation}}
\newcommand{\bea}{\begin{eqnarray}}
\newcommand{\eea}{\end{eqnarray}}
\newcommand{\bc}{\begin{center}}
\newcommand{\ec}{\end{center}}
\newcommand{\beastern}{\begin{eqnarray*}}
\newcommand{\eeastern}{\end{eqnarray*}}

\newcommand{\pr}{\prime}

\newcommand{\EV}{\mathds{E}}
\newcommand{\Cov}{\operatorname{Cov}}

\newcommand{\diag} {\operatorname{diag}}

\newcommand{\rrMSE}{\operatornamewithlimits{rrMSE}}

\newcommand{\dint}{\operatornamewithlimits{d}}

\newcommand{\Ya}{\tilde{Y}_i(t_{ij})}
\newcommand{\Yb}{\tilde{Y}_{i^\prime}(t_{i^\prime j^\prime})}
\newcommand{\Yc}{\tilde{Y}_m(t_{mo})}
\newcommand{\Yd}{\tilde{Y}_{m^\prime}(t_{m^\prime o^\prime})}

\newcommand{\mB}{\bm{B}}

\newcommand{\mC}{\bm{C}}

\newcommand{\mI}{\bm{I}}
\newcommand{\mK}{\bm{K}}

\newcommand{\mM}{\bm{M}}

\newcommand{\mQ}{\bm{Q}}

\newcommand{\mS}{\bm{S}}

\newcommand{\mU}{\bm{U}}

\newcommand{\mx}{\bm{x}}
\newcommand{\mW}{\bm{W}}

\newcommand{\mz}{\bm{z}}
\newcommand{\mZ}{\bm{Z}}

\newcommand{\mphi}{\bm{\phi}}

\newcommand{\mtheta}{\bm{\theta}}
\newcommand{\mTheta}{\bm{\Theta}}

\newcommand{\mdelta}{\bm{\delta}}
\newcommand{\malpha}{\bm{\alpha}}
\newcommand{\mSigma}{\bm{\Sigma}}

\numberwithin{equation}{section}

\definecolor{LmuGreen}{RGB}{0,148,64}

\begin{document}

\maketitle

\begin{abstract}
We propose a fast bivariate smoothing approach for symmetric surfaces that has a wide range of applications. We show how it can be applied to estimate the covariance function in longitudinal data as well as multiple additive covariances in functional data with complex correlation structures. Our symmetric smoother can handle (possibly noisy) data sampled on a common, dense grid as well as irregularly or sparsely sampled data. Estimation is based on bivariate penalized spline smoothing using a mixed model representation and the symmetry is used to reduce computation time compared to the usual non-symmetric smoothers. We outline the application of our approach in functional principal component analysis and demonstrate its practical value in two applications. The approach is evaluated in extensive simulations. We provide documented open source software implementing our fast symmetric bivariate smoother building on established algorithms for additive models.
\end{abstract}

\section{Introduction}
Covariance functions play a central role in many areas of statistics. They summarize the dependency between stochastic observations and encode smoothness assumptions about (observed or latent) random processes. We propose a fast bivariate smoothing approach for symmetric surfaces which can estimate covariance functions in a wide range of data situations. Our approach can handle dependent processes based on an additive decomposition of the covariance function and is also applicable to processes that are observed on irregular or sparse grids.

In functional data analysis \citep[FDA; see, e.g.][]{Ramsay.2005}, covariance functions are at the heart of functional principal component analysis (FPCA), a key tool for dimension reduction based on an eigen analysis of the covariance operator of a random process. FPCA is commonly used to estimate the model parameters in functional predictor and functional response regression models \citep[see][for an overview]{Morris.2015}. Other examples that are based on covariance functions include functional discriminant analysis \citep{James.2001} and functional canonical correlation analysis \citep{Leurgans.1993}. In longitudinal data analysis (LDA), where measurements are frequently recorded at irregularly spaced time points, the correct specification of the covariance benefits the estimation efficiency of the fixed effects and improves the individual predictions \citep[cf.][]{Fan.2007}. The covariance is also a crucial ingredient in time series analysis, e.g.~in risk models and portfolio allocation \citep[cf.][]{Tai.2009}. The interest commonly lies in a single time series in contrast to FDA (and LDA) where multiple curves are observed, e.g.~over time. In principle, our symmetric smoothing approach is also applicable to time series which is, however, not the focus in this paper. 

Covariance functions are commonly assumed to be smooth. Thus, when the observed curves are not sufficiently smooth (i.e.~observed with error) or not measured on a common dense grid, smoothing becomes necessary at some point during covariance estimation. Directly smoothing observed curves \citep[see, e.g.,][]{Besse.1986}, however, is very difficult or impossible for sparsely observed data which are frequently recorded both in FDA and LDA \citep{Yao.2005}. Moreover, pre-smoothing observed curves removes the measurement error, which is not accounted for in subsequent estimation steps. We pursue an  alternative approach and apply bivariate smoothing to the sample covariance of the observed data points.

Most existing work on non-parametric covariance estimation is either restricted to independent functional (or longitudinal) observations and/or only applies to data sampled on a common grid. Furthermore, most bivariate smoothing approaches are not specifically designed for covariances. They do not exploit the symmetry of the estimated surface and thus use redundant information in the available data. To the best of our knowledge, previous approaches have never addressed these issues simultaneously. They can be divided according to three main criteria: 1) the generality of the assumed correlation structure in the data, 2) the generality of possible sampling grids, and 3) the estimation procedure including the selection of the degree of smoothing. 

A number of approaches address covariance smoothing in LDA. They are restricted to independent curves but allow for general sampling grids. Smoothing is either accomplished by bivariate kernel smoothing \citep[e.g.][]{Staniswalis.1998, Yao.2003, Yao.2005} or by bivariate (penalized) spline smoothing \citep[e.g.][]{Kauermann.2011}. The degree of smoothing is either chosen by visual inspection \citep{Staniswalis.1998}, different leave-one-curve-out cross validation algorithms \citep[e.g.][]{Yao.2003, Yao.2005} or based on a mixed model representation \citep[e.g.][]{Kauermann.2011}. These approaches do not account for the symmetry of the estimated surface. \cite{James.2000} directly estimate the smooth eigenfunctions of the covariance function. They estimate a reduced rank mixed effects model via the EM algorithm and use B-spline basis functions to represent the eigenfunctions of the covariance operator. \cite{Peng.2009} estimate the same reduced rank model based on a more efficient Newton-Raphson procedure on the Stiefel manifold. The extension of these reduced rank methods to complex correlation structures is not straightforward. \cite{Xiao.2016b} recently proposed a bivariate smoother designed for covariance smoothing which can be used for sparsely observed, independent functions. They use bivariate penalized B-splines and enforce a symmetry constraint on the spline coefficients which we take up in our extension to correlated curves. Estimation is done by a three-step procedure which accounts for the covariance of the sample covariance. Their leave-one-curve-out cross validation procedure for selecting the smoothing parameter is not applicable for correlated functional data, however.

Other covariance smoothing approaches can be applied to correlated functions but are restricted to functions sampled on a common grid and considerably simpler correlation structures than ours. \cite{Di.2009} and \cite{Greven.2010} use bivariate penalized splines and 
select the smoothing parameter using restricted maximum likelihood \citep[REML;][]{Patterson.1971} estimation. \cite{Shou.2014} apply a method of moments approach based on symmetric sums represented in a sandwich form. For smoothing, they propose to use an extension of the fast covariance estimation algorithm of \cite{Xiao.2016} to correlated functions. \cite{Di.2014} extend the functional random intercept model of \cite{Di.2009} to sparsely sampled functional data, but the correlation structure remains less general than ours and an extension is not straightforward. More general correlation structures are allowed in the approach of \cite{Cederbaum.2016} that is also suitable for sparsely and irregularly sampled functional data. Their focus lies, however, on a model with crossed functional random effects and estimation is only discussed for this special case. Apart from considering less general correlation structures, all these approaches neither avoid the use of redundant information nor account for the symmetry of the smoothed surface. 

We propose a fast symmetric bivariate smoothing approach that applies to data with a broad range of possible correlation structures, much broader than existing methods. Furthermore, our approach is well-suited for (possibly noisy) data sampled on a common, dense grid as well as for irregularly or sparsely sampled data. Strength is borrowed by pooling information across different curves, which is particularly important for curves observed on sparse, unequal grids. The smoothing approach we present is widely applicable: In this paper, we demonstrate how it can be applied to longitudinal data as a special case of independent functional data as well as to correlated functional data with very general and complex correlation structures. For the latter, we extend our bivariate smoothing approach to smoothing additive covariance functions. To the best of our knowledge, all previous proposals in this field have been restricted to estimating much less general dependency structures.

We estimate the covariance functions using a smooth method of moments approach represented as a bivariate additive varying coefficient model. The estimation is based on bivariate penalized splines. We choose the smoothing parameters using REML, which allows the direct extension to additive bivariate smoothing of a superposition of multiple covariance functions. This allows our method to be used for a broad range of complex real-word data settings. It also frees us from having to pre-specify a discrete grid of candidate values for the smoothing parameters that is required for cross-validation based approaches like \cite{Xiao.2016b}.  
Smoothing the sample covariance quickly becomes a high-dimensional problem as the number of elements in the sample covariance increases quadratically with the number of grid points. We take advantage of the symmetry of the sample covariance and only estimate the upper triangle of the surface including the diagonal. The estimates are then reflected across the diagonal to obtain the entire estimated covariance, which is continuous but not necessarily smooth across the diagonal. To avoid boundary effects on the diagonal, we enforce smoothness by imposing a symmetry constraint on the spline coefficients, which for the simplest case of independent curves reduces to that of \cite{Xiao.2016b}. We show how the symmetry constraint can be applied separately to additive covariances and can even be used for any bivariate symmetric smoothing problem beyond covariance functions. Our approach modifies the covariance smoothing approach proposed in \cite{Cederbaum.2016} and extends it to more general models. It reduces both the data entering the estimation and the number of spline coefficients that have to be estimated, which leads to considerably faster estimation requiring less memory.

We provide software implementing our approach based on a novel constructor function for \pkg{R}-package \pkg{mgcv}, which provides a general framework for additive models allowing for a very flexible model specification \citep{R, Wood.2011}. 

We outline the application of our approach to FPCA and demonstrate its practical relevance by an application to sparse longitudinal observations of CD4 cell count trajectories and to densely but irregularly observed acoustic signals from a speech production study. This study requires crossed functional random effects due to repeated measurements for both speakers and target words and thus corresponds to a case of dependent functional data with additive covariance structure.

We organize our paper as follows: Section \ref{sec: FSCS} first develops our fast symmetric covariance smoothing approach for a simple special case  with only one smooth covariance function and additional measurement error. In Section \ref{sec: extensions}, the smoother is extended to complex dependency structures involving the smoothing of multiple additive covariance functions. Section \ref{sec: FPCA} outlines the application of our covariance smoother in FPCA. In Section \ref{sec: implementation}, details on the implementation are given. In Section \ref{sec: applications} and \ref{sec: simulations}, we evaluate our approach in an application to speech production data and in simulations, respectively. Section \ref{sec: discussion} closes with a discussion and outlook. Theoretical results and supplementary material are available in the appendix. R-code implementing our approach can be provided upon request.

\section{Fast symmetric covariance smoothing} \label{sec: FSCS}
For simplicity, we first explain our covariance smoothing approach for a simple special case with only one smooth auto-covariance and additional measurement error. This will be extended to additive covariance smoothing in Section \ref{sec: extensions}.
\subsection{Model with independent curves}\label{sec: indep model}
Consider the following model 
\bea \label{eq: indep model}
Y_i(t_{ij}) = \mu(t_{ij},\mx_i) + E_{i}(t_{ij}) + \varepsilon_{i}(t_{ij}),\ j=1,\ldots,D_i, \ i=1,\ldots,n,
\eea
where $Y_i(t_{ij})$ is the observed value of response curve $i$ observed at the $j$th observation point $t_{ij} \in \mathcal{T}$, a closed interval in $\mathbb{R}$. $\mu(t_{ij},\mx_i)$ is a global mean function depending on a vector of known covariates $\mx_i$. $E_i(t_{ij})$ is a smooth curve-specific deviation from the global mean and $\varepsilon_i(t_{ij})$ is additional independent and identically distributed white noise measurement error with constant variance $\sigma^2$ that accounts for random uncorrelated variation within curve $i$. The model can be seen as a function-on-scalar regression model \citep[e.g.][]{Faraway.1997, Ramsay.2005, Reiss.2010} where all $n$ curves are assumed to be independent and it is a special case of the general functional linear mixed model \citep[FLMM; see][for a discussion and further references]{Morris.2015} in Section \ref{sec: extensions} with one curve-specific functional random intercept. Model \eqref{eq: indep model} is often applied to longitudinal data with $\mathcal{T}$ denoting a time interval. Note that all curves may either be observed on a common, fine grid or on curve-specific, possibly sparse, $D_i$ evaluation points $t_{ij}$, $j=1,\ldots,D_i$, $i=1,\ldots,n$.

In the following, we assume that $E_i(\cdot)$ and $\varepsilon_i(\cdot)$, $i=1,\ldots,n$, are zero-mean, mutually uncorrelated random processes and that $E_i(\cdot)$ is square-integrable. We denote the auto-covariance function of $E_i(\cdot)$ by $ K^E(t,t^\prime)=\Cov\left[E_i(t),E_i(t^\prime)\right]$, $t,t^\prime\in \mathcal{T}$. We further assume that the mean and the auto-covariance are smooth in $t$ and in arguments $t$, $t^\prime$, respectively.

\subsection{Estimation in the independent case} \label{sec: indep estimation}
We apply the following smooth method of moments approach to estimate the auto-covariance $K^E(t,t^\prime)$. It modifies the approach presented in \cite{Cederbaum.2016} by accounting for the symmetry of covariances, which leads to a considerable reduction of computation times. In this simple case of independent functional responses, our smoother is closely related to that of \cite{Xiao.2016b}, who approach the problem from a slightly different perspective. Their approach, however, is not directly extendable to correlated data as will be discussed in Section \ref{sec: extensions}.

We focus in the following on the centered functional responses $\tilde{Y}_i(t_{ij}) = Y_i(t_{ij}) - \mu(t_{ij},\mx_i)$ with expectation zero and denote their realizations by $\tilde{y}_i(t_{ij})$. We exploit the fact that the expectation of the centered cross products corresponds to the covariance of the functional response which is given as
\bea \label{eq: indep covariance model}
\EV\left[\tilde{Y}_i(t_{ij})\tilde{Y}_i(t_{ij^\prime})\right]= \Cov\left[\tilde{Y}_i(t_{ij}),\tilde{Y}_i(t_{ij^\prime})\right]  = K^E(t_{ij},t_{ij^\prime}) + \sigma^2\delta_{jj^\prime}, \\ \nonumber j,j^\prime=1,\ldots,D_i,\ i=1,\ldots,n,
\eea
with $\delta_{jj^\prime}$ equal to one if $j=j^\prime$ and zero otherwise.
Equation \eqref{eq: indep covariance model} can be seen as a special case of a bivariate additive varying coefficient model for the empirical covariances $\tilde{y}_{it_{ij}}\tilde{y}_{it_{ij^\prime}}$, in which the auto-covariance and the error variance are the unknown components. We estimate the smooth auto-covariance function $K^E(t,t^\prime)$ and the error variance $\sigma^2$ simultaneously using a bivariate spline representation for $K^E(t,t^\prime)$ under working assumptions of independence and homoscedasticity.

For this, let $\mC$ denote the $\mathcal{C}\times 1$ stacked vector of all cross products, with $\mathcal{C} = \sum_{i=1}^n D_i^2$. Then, Model \eqref{eq: indep covariance model} can be represented as
\bea \label{eq: indep additive model}
\EV\left(\mC\right)
&=& \left[\mM^E| \mdelta^\varepsilon \right] \left({\mtheta^{E}}^\top,\sigma^2\right)^\top \eqqcolon \mM \malpha, 
\eea
where $\mM^E$ denotes the $\mathcal{C}\times \left(F^E\right)^2$ bivariate spline design matrix, containing the evaluations of any bivariate spline basis with $\left(F^E\right)^2$ basis functions that are symmetric across the diagonal, i.e.~across $t_{ij}=t_{ij^\prime}$. We use bivariate tensor product B-splines, but other bases are possible. See Section \ref{sec: implementation} and Appendix \ref{appendix: supplementary details on estimation} for details. $\mdelta^\varepsilon$ is an indicator vector of length $\mathcal{C}$ whose elements take values $\delta_{jj^\prime}$. $\mtheta^E$ is a spline coefficient vector of length $\left(F^E\right)^2$. 
To avoid over-fitting, we use an isotropic quadratic smoothness penalty of the form 
\bea  \label{eq: indep penalty}
\operatorname{pen}(\lambda) =  \lambda {\mtheta^{E}}^\top \mS^E \mtheta^E,
\eea
where $\lambda$ denotes the smoothing parameter that controls the bias-variance tradeoff and $\mS^E$ is a suitable penalty matrix, see Section \ref{sec: implementation} for a discussion.

The development above uses all $\mathcal{C}$ cross products for the estimation of the covariance as is commonly done in published smooth method of moments approaches for covariance estimation \citetext{\citealp{Yao.2003}; \citealp{Yao.2005}; for correlated curves, e.g.,~\citealp{Staniswalis.1998}; \citealp{Di.2009}; \citealp{Greven.2010}}. As it is quadratic in the number of function evaluations, $\mathcal{C}$ quickly becomes extremely large in practice and often poses significant computational challenges. Since the cross products $\tilde{y}_{it_{ij}}\tilde{y}_{it_{ij^\prime}}$ and $\tilde{y}_{it_{ij^\prime}}\tilde{y}_{it_{ij}}$ are identical, we, like \cite{Xiao.2016b}, avoid the use of this redundant information and only estimate the upper triangle of the auto-covariance surface including the diagonal. Moreover, using this redundant data twice violates the implicit working assumption of independent observations in our additive model. A detailed discussion of working assumptions and possible strategies for handling violations is given in Section \ref{sec: covariance of cross products}. 

We assume in the following that $\mC$ is sorted such that it can be partitioned as $\mC = \left({\mC_{t<t^\prime}}^{\top},{\mC_{t=t^\prime}}^{\top},{\mC_{t>t^\prime}}^{\top}\right)^\top$, where $\mC_{t<t^\prime}$, $\mC_{t=t^\prime}$, $\mC_{t>t^\prime}$ comprise all cross products $\tilde{y}_{it_{ij}}\tilde{y}_{it_{ij^\prime}}$, $i=1,\ldots,n$, with $t_{ij}<t_{ij^\prime}$, $t_{ij}=t_{ij^\prime}$, and $t_{ij}>t_{ij^\prime}$, respectively. Then, with suitable sorting within the three partitions of $\mC$, the symmetry of the cross products implies that $\mC_{t<t^\prime}= \mC_{t>t^\prime}$. In order to speed up estimation, we only use $\mC^{\Delta}\coloneqq \left({\mC_{t<t^\prime}}^{\top},{\mC_{t=t^\prime}}^{\top}\right)^\top$ for the estimation in the bivariate model \eqref{eq: indep additive model}. The total number of cross products thus amounts to $\mathcal{C}^{\Delta} := \sum_{i=1}^n D_i(D_i+1)/2$. The design matrix of the bivariate additive model can accordingly be partitioned as $\mM = \left({\mM_{t<t^\prime}}^\top,{\mM_{t=t^\prime}}^\top,{\mM_{t>t^\prime}}^\top\right)^\top$. Let $\mM^{E\Delta}$ and $\mdelta^{\epsilon \Delta}$ denote the submatrix and subvector corresponding to $\mC^{\Delta}$. Then, the bivariate additive model \eqref{eq: indep additive model} reduces to $\EV\left(\mC^{\Delta}\right) = \left[\mM^{E \Delta}|\mdelta^{\varepsilon \Delta}\right] \left({\mtheta^{E}}^\top,\sigma^2\right)^\top$. Reflecting the upper triangle across the diagonal ensures that the obtained surface estimate is symmetric and continuous in both directions $t$ and $t^\prime$, but it does not guarantee smoothness across the diagonal without additional constraint. 
Boundary effects can occur because the coefficients can vary more freely, the smaller the overlap of the support of their associated basis functions with the area where $t\leq t^\prime$, since less data are available for their estimation. This may lead to wiggly estimates in the area of the diagonal. An expected consequence is that the separation of the smooth auto-covariance surface and the error variance on the diagonal becomes more challenging, which we indeed observe in our application in Section \ref{sec: applications} and in our simulations in Sections \ref{sec: simulations}. 

We enforce smoothness across the diagonal in order to avoid such boundary effects. As a symmetric surface implies a symmetric spline coefficient matrix $\mTheta^E = \left[\theta^E_{bb^\prime}\right]_{b,b^\prime = 1,\ldots,F^E} = {\mTheta^{E}}^\top$, 
where $\mtheta^E = \left({\mtheta_{b<b^\prime}^{E}}^\top, {\mtheta_{b=b^\prime}^{E}}^\top,{\mtheta_{b>b^\prime}^{E}}^\top\right)^\top$ contains first the entries of $\mTheta^E$ below the diagonal ($\theta^E_{bb^\prime}$, $b<b^\prime$), then the diagonal entries ($\theta^E_{bb^\prime}$, $b=b^\prime$) and lastly the entries above the diagonal ($\theta^E_{bb^\prime}$, $b>b^\prime$), we impose a symmetry constraint on $\mTheta^E$. Thus, our approach differs in two crucial points from the naive covariance estimation in \eqref{eq: indep additive model} and from most previous covariance smoothing approaches. First, we reduce the number of cross products that enter the estimation and second, imposing the symmetry constraint almost halves the number of spline coefficients that have to be estimated. Both aspects greatly speed up the computation as we show in Sections \ref{sec: applications} and \ref{sec: simulations}.

With suitable sorting within the partitions $\mtheta^E_{b<b^\prime}$ and $\mtheta^E_{b>b^\prime}$, the above symmetry constraint on the coefficient matrix corresponds to the following symmetry constraint on the coefficient vector
\bea\label{eq: symmetry constraint}
\theta_{bb^\prime}^E &=& \theta_{b^\prime b}^E,\ b,b^\prime=1,\ldots,F^E\ \Leftrightarrow \mtheta_{b<b^\prime}^E = \mtheta_{b>b^\prime}^E,
\eea
which corresponds to the constraint used in \cite{Xiao.2016b}.
This allows us to consider the reduced coefficient vector $\mtheta^{Er} = \left({\mtheta_{b<b^\prime}^{E}}^\top, {\mtheta_{b=b^\prime}^{E}}^\top\right)^\top$ of length $F^E\left(F^E+1\right)/2$.
For the implementation of the above symmetry constraint, we use that under the constraint we have $\mM^{E \Delta}\mtheta^E = \left(\mM^{E \Delta}_{b<b^\prime} +\mM^{E \Delta}_{b>b^\prime}\right) \mtheta_{b<b^\prime}^E + \mM^{E \Delta}_{b=b^\prime}\mtheta_{b=b^\prime}^E$, with $\mM^{E \Delta}_{b<b^\prime}$, $\mM^{E \Delta}_{b=b^\prime}$, and $\mM^{E \Delta}_{b>b^\prime}$ containing the respective columns of $\mM^{E \Delta}$. Thus, the constraint is equivalent to adding up columns $\mM^{E \Delta}_{b<b^\prime}$ and $\mM^{E \Delta}_{b>b^\prime}$ of the design matrix. This can be achieved by right multiplication of $\mM^{E \Delta}$ with the $\left(F^E\right)^2 \times F^E\left(F^E+1\right)/2$ constraint matrix 
\bea \nonumber \label{eq: constraint matrix}
\mW^E = \left[\begin{array}{*{2}{l}}
\mI_{\frac{F^E(F^E-1)}{2}} & \bm{0}_{\frac{F^E(F^E-1)}{2}\times F^E} \\
\bm{0}_{F^E\times \frac{F^E(F^E-1)}{2}} & \mI_{F^E} \\
\mI_{\frac{F^E(F^E-1)}{2}} & \bm{0}_{\frac{F^E(F^E-1)}{2} \times F^E}
\end{array}\right], 
\eea 
where $\mI_{x}$ is an identity matrix of dimension $x$ and $\bm{0}_{x\times y}$ is a null matrix of dimension $x\times y$.
We denote the reduced design matrix by $\mM^{E \Delta r}:= \mM^{E \Delta}\mW^E$. Under the symmetry constraint, $\mM^{E \Delta r} \mtheta^{Er} =\mM^{E \Delta} \mtheta^E$. The penalty matrix in \eqref{eq: indep penalty} also needs to be adjusted to the reduced coefficient vector $\mtheta^{Er}$ as $\mS^{Er} := {\mW^{E}}^\top \mS^E \mW^E$, corresponding to ${\mtheta^{Er}}^\top\mS^{Er}\mtheta^{Er} = {\mtheta^{E}}^\top\mS^{E}\mtheta^{E}$.

Altogether, the bivariate additive model for the reduced response vector with symmetry constraint \eqref{eq: symmetry constraint} is given by
\bea \label{eq: indep additive model constraint}
\EV\left(\mC^{\Delta}\right) &=& \left[\mM^{E \Delta r}| \mdelta^{\varepsilon \Delta} \right] \left({\mtheta^{E r}}^\top,\sigma^2\right)^\top \eqqcolon \mM^{\Delta r} \malpha^r.
\eea
Note that in Model \eqref{eq: indep additive model constraint}, each product $\tilde{y}_{it_{ij}}\tilde{y}_{it_{ij^\prime}}$, $t_{ij} \leq t_{ij^\prime}$, enters the estimation with the same weight. This is not the case when all products $\tilde{y}_{it_{ij}}\tilde{y}_{it_{ij^\prime}}$, $j,j^\prime=1,\ldots,D_i$, are used as in Model \eqref{eq: indep additive model}, where all products appear twice except for those on the diagonal ($t_{ij}=t_{ij^\prime}$). Our implementation allows to estimate Model \eqref{eq: indep additive model constraint} with the same weights as in Model \eqref{eq: indep additive model} by putting a weight of 0.5 on the products on the diagonal. There is room for debate on whether it is desirable to down-weigh the data on the diagonal compared to the rest. One would expect that this leads to wigglier estimates but our simulations in Section \ref{sec: simulations} show that the difference is not very large. 

In contrast to \cite{Xiao.2016b}, who derive a leave-one-curve-out generalized cross-validation (GCV) algorithm to choose the smoothing parameter for independent curves, we choose the smoothing parameter as variance component ratio using REML. REML has been shown to be more stable than GCV \citep{Wood.2011} and to be more robust to error correlation misspecification than prediction error methods \citep{Krivobokova.2007}. Even more importantly, it allows us to directly extend our symmetric smoothing approach to additive smoothing needed for functional data with complex dependency structures as will be shown in Section \ref{sec: extensions}. In more general designs, where the responses cannot be decomposed into independent subvectors, it is not clear how to perform smoothing parameter selection based on GCV and optimizing multiple smoothing parameters would require a computationally costly multi-dimensional grid search.
 
Details on the implementation are given in Section \ref{sec: implementation} and in Appendix \ref{appendix: supplementary details on estimation}.

\section{Fast symmetric additive covariance smoothing} \label{sec: extensions}
Simultaneous REML estimation of multiple smoothing parameters allows direct extension of our approach to more general models with complex correlation structures, for which we derive appropriate symmetry constraint matrices.

\subsection{General functional linear mixed model}
The general FLMM \citep[see, e.g.,][]{Morris.2015} can be seen as the functional analogue to the linear mixed model \citep[LMM; see, e.g.,][]{Pinheiro.2000}, which is often applied to scalar correlated data. The random effects in the linear mixed model are replaced by functional random effects in order to account for the functional nature of the response. A functional random intercept (fRI) for a subject, for example, is a subject-specific deviation from the mean in form of a function. The FLMM is given by
\bea \label{eq: general model}
Y_i(t_{ij}) = \mu(t_{ij},\mx_i) + \mz_i^\top\mU(t_{ij}) + E_i(t_{ij}) + \varepsilon_i(t_{ij}), \ j=1,\ldots,D_i, \ i=1,\ldots,n,
\eea
where $Y_i(t_{ij})$ denotes the response of curve $i$ at observation point $t_{ij}$, which can be additively decomposed as in Model \eqref{eq: indep model}. Model \eqref{eq: general model}, however, additionally accounts for correlation between (groups of) curves by the vector-valued random process $\mU(t_{ij})$ which is multiplied by $\mz_i$, a known covariate vector of length $q$. Examples for $\mz_i^\top\mU(t_{ij})$ yielding FLMMs with e.g.~crossed and hierarchical functional random effects are given in Section \ref{sec: crossed model} and in Appendix \ref{appendix: supplementary details on estimation}. 

We assume that $\mU(\cdot)$, $E_i(\cdot)$, and $\varepsilon_i(\cdot)$ are zero mean, mutually uncorrelated random processes and that $\mU(\cdot)$ and $E_i(\cdot)$ are square-integrable. As for Model \eqref{eq: indep model}, we denote the auto-covariance of $E_i(\cdot)$ by $K^E(t,t^\prime) = \Cov\left[E_i(t),E_i(t^\prime)\right]$, $t,t^\prime \in \mathcal{T}$. The $q\times q$ matrix-valued auto-covariance of $\mU(\cdot)$ is denoted by $\mK^U(t,t^\prime) = \Cov\left[\mU(t),\mU(t^\prime)\right]$. The covariances are assumed to be smooth (for each component in the case of $\bm{U}(t)$).

Let $G$ denote the number of grouping variables. Then, $\mU(t_{ij})$ can be divided into $G$ independent blocks $\mU_g(t_{ij})$, $g=1,\ldots,G$, which again contain blocks of $L^{U_g}$ independent copies $\mU_{gl}(t_{ij})$, $l=1,\ldots,L^{U_g}$, where $L^{U_g}$ is the number of levels of the $g$th grouping variable. $\mU_{gl}(t_{ij}) = \left(U_{gl1}(t_{ij}),\ldots,U_{gl\rho^{U_g}}(t_{ij})\right)^\top$, in turn, is a vector-valued random process of $\rho^{U_g}$ components for each level of this grouping variable, for example $\rho^{U_g}=2$ if the $g$th grouping variable is associated with a fRI and a functional random slope. The total number of entries in $\mU(t_{ij})$ is given by $q = \sum_{g=1}^G L^{U_g} \rho^{U_g}$. The $\rho^{U_g} \times \rho^{U_g}$ matrix-valued covariance of $\mU_{gl}(\cdot)$, $\mK^{U_g}(t,t^\prime) = \left[K^{U_g}_{ss^\prime}(t,t^\prime)\right]_{s,s^\prime=1,\ldots,\rho^{U_g}} = \Cov\left[\mU_{gl}(t),\mU_{gl}(t^\prime)\right]$, with $K^{U_g}_{ss^\prime}(t,t^\prime) = K^{U_g}_{s^\prime s}(t^\prime,t)$, is the same for all levels, $l=1,\ldots,L^{U_g}$, of the $g$th grouping variable. We can thus write the block-diagonal auto-covariance of $\mU(\cdot)$ as
\bea \nonumber 
\mK^U(t,t^\prime) = \diag\left(\underbrace{\mK^{U_1}(t,t^\prime),\ldots,\mK^{U_1}(t,t^\prime)}_{L^{U_1} \footnotesize{\mbox{ times}}},\ldots,\underbrace{\mK^{U_G}(t,t^\prime),\ldots,\mK^{U_G}(t,t^\prime)}_{L^{U_G} \footnotesize{\mbox{ times}}}\right).
\eea

\subsection{Estimation in the general functional linear mixed model}
Our fast symmetric covariance smoothing approach can be extended to the general model \eqref{eq: general model} by generalizing it to a matrix of covariances as described above and applying it to each additive component separately.

In analogy to Model \eqref{eq: indep model}, we base the covariance smoothing on the following decomposition of the expectation of the cross products of the centered functional responses
\bea \label{eq: general covariance model}
\EV\left[\tilde{Y}_i(t_{ij})\tilde{Y}_{i^\prime}(t_{i^\prime j^\prime})\right] &=& \Cov\left[\tilde{Y}_i(t_{ij}),\tilde{Y}_{i^\prime}(t_{i^\prime j^\prime})\right]\\ \nonumber &=& \mz_i^\top \mK^U(t_{ij},t_{i^\prime j^\prime}) \mz_{i^\prime} + \left[K^E(t_{ij},t_{i^\prime j^\prime}) + \sigma^2\delta_{jj^\prime}\right] \delta_{ii^\prime},
\eea
where, in contrast to Model \eqref{eq: indep model}, products are now also computed across different curves $i$, $i^\prime$.

Let $U_g(i)$ denote the level of $U_g$ for observation $i$. Similar to $\mU(\cdot)$, the covariate vector $\mz_i$ can be divided into $G$ blocks $\mz_i^\top = \left({\mz^{U_1}_{i}}^\top,\ldots,{\mz^{U_G}_{i}}^\top \right)$, where the blocks $\mz^{U_g}_{i}$, $g=1,\ldots,G$, can again be written as ${\mz^{U_g}_{i}}^\top = \left({\mz^{U_g}_{i1}}^\top,\ldots,{\mz^{U_g}_{iL^{U_g}}}^\top\right)$ with ${\mz^{U_g}_{il}}^\top = \left(z^{U_g}_{il1},\ldots,z^{U_g}_{il\rho^{U_g}}\right)$, $l=1,\ldots,L^{U_g}$. The scalars $z^{U_g}_{ils}$ take the value of the respective covariate $\omega^{U_g}_{is}$ times an indicator $\delta_{U_g(i)l}$, specifying whether observation $i$ belongs to level $l$ of grouping variable $g$. Based on this partition, the expectation in \eqref{eq: general covariance model} can be rewritten as
\bea \label{eq: general covariance model rewritten} 
\EV\left[\tilde{Y}_i(t_{ij})\tilde{Y}_{i^\prime}(t_{i^\prime j^\prime})\right]
 &=& \sum_{g=1}^G\sum_{l=1}^{L^{U_g}} \sum_{s=1}^{\rho^{U_g}}\sum_{s^\prime=1}^{\rho^{U_g}} z^{U_g}_{ils}z^{U_g}_{i^\prime ls^\prime} K^{U_g}_{ss^\prime}(t_{ij},t_{i^\prime j^\prime})\\ \nonumber &+&  \left[K^E(t_{ij},t_{i^\prime j^\prime}) + \sigma^2 \delta_{jj^\prime}\right] \delta_{ii^\prime}.
\eea
For example in the case of an FLMM with only one fRI ($G=1$, $\rho^{U_1}=1$), the products $z^{U_g}_{ils}z^{U_g}_{i^\prime ls^\prime}$ are indicators for whether the two observations in a cross products belong to the same level of the grouping variable.

We exploit the symmetry of covariances $\mK^{U_g}(t_{ij},t_{i^\prime j^\prime}) = {\mK^{U_g}(t_{i^\prime j^\prime}, t_{ij})}^\top$, $g=1,\ldots,G$, and of $K^E(t_{ij}, t_{i^\prime j^\prime}) = K^E(t_{i^\prime j^\prime}, t_{ij})$ and only use the products $\tilde{y}_{i t_{ij}}\tilde{y}_{i^\prime t_{i^\prime j^\prime}}$ with $t_{ij}\leq t_{i^\prime j^\prime}$, suitably sorted in the long vector $\mC^\Delta$. Let $\cdot$ denote the Hadamard (pointwise) product. As in the case with independent curves, Model \eqref{eq: general covariance model rewritten} can be represented as bivariate additive varying coefficient model, here of the form
\bea  \label{eq: general additive model}
\EV\left(\mC^{\Delta}\right) &=&\left[\mM^{U_1\Delta}|\ldots|\mM^{U_G\Delta}|\mM^{E\Delta}|\mdelta^{\varepsilon \Delta}\right] \left({\mtheta^{U_1}}^\top,\ldots,{\mtheta^{U_G}}^\top,{\mtheta^{E}}^\top,\sigma^2 \right)^\top,
\eea 
where $\mM^{U_g\Delta}$, $g=1,\ldots,G$, contain the column-wise concatenated submatrices $\mM^{U_g \Delta}_{ss^\prime}$, corresponding to the covariances $K^{U_g}_{ss^\prime}(t,t^\prime)$, $s,s^\prime=1,\ldots,\rho^{U_g}$. The submatrices $\mM^{U_g \Delta}_{ss^\prime}$ are given by $\mQ^{U_g \Delta}_{ss^\prime} \cdot \mB^{U_g \Delta}_{ss^\prime}$, where $\mQ^{U_g \Delta}_{ss^\prime}$ contain suitably sorted and repeated entries $\delta_{U_g(i)U_g(i^\prime)}\cdot\omega^{U_g}_{is}\omega^{U_g}_{i^\prime s^\prime}$ and $\mB^{U_g\Delta}_{ss^\prime}$ denote the bivariate spline design matrices. $\mM^{E\Delta}$ is analogously given by $\mQ^{E\Delta} \cdot \mB^{E\Delta}$, with bivariate spline design matrix $\mB^{E\Delta}$ and $\mQ^{E\Delta}$ a new indicator matrix, which reduces to an all-ones matrix in the model with independent curves, for which thus $\mM^{E \Delta} = \mB^{E \Delta}$. The concrete form of $\mQ^{U_g \Delta}_{ss^\prime}$ and $\mQ^{E\Delta}$, as well as the bivariate spline design matrices for tensor product B-splines are provided in Appendix \ref{appendix: supplementary details on estimation}.

We assume that for each, $g=1,\ldots,G$, the coefficient vector $\mtheta^{U_g}$ is sorted correspondingly to the columns of $\mM^{U_g \Delta} = \left[\mM^{U_g\Delta}_{11}|\ldots|\mM^{U_g\Delta}_{1\rho^{U_g}}|\ldots|\mM^{U_g\Delta}_{\rho^{U_g}1}|\ldots|\mM^{U_g\Delta}_{\rho^{U_g}\rho^{U_g}}\right]$. Moreover, with suitable sorting, each submatrix $\mM^{U_g\Delta}_{ss^\prime}$ can be partitioned as in the case of independent curves, $\mM^{U_g\Delta}_{ss^\prime} = \left[\mM^{U_g\Delta}_{ss^\prime,b<b^\prime}|\mM^{U_g\Delta}_{ss^\prime,b=b^\prime}|\mM^{U_g\Delta}_{ss^\prime,b>b^\prime}\right]$. Let $\mTheta^{U_g}_{ss^\prime}$ denote the correspondingly sorted coefficient matrices, where $\mtheta^{U_g}_{ss^\prime} = \left({\mtheta^{U_g}_{ss^\prime,b<b^\prime}}^\top,{\mtheta^{U_g}_{ss^\prime,b=b^\prime}}^\top,{\mtheta^{U_g}_{ss^\prime,b>b^\prime}}^\top\right)^\top$ contains first the entries of $\mTheta^E$ below the diagonal, then the diagonal entries and lastly the entries above the diagonal. Assume further that within the three blocks of $\mtheta^{U_g}_{ss^\prime}$, the entries $\theta_{ss^\prime,bb^\prime}$ are sorted correspondingly for all $s,s^\prime=1,\ldots,\rho^{U_g}$.

As a modular component, the symmetry constraint 
\bea \label{eq: general symmetry constraint}
\mTheta^{U_g}_{ss^\prime} = {\mTheta^{U_g}_{s^\prime s}}^\top,\ s,s^\prime=1,\ldots,\rho^{U_g},
\eea
can be applied to each, $g=1,\ldots,G$, due to the symmetry of covariances $\mK^{U_g}(t,t^\prime)$, yielding the reduced coefficient vectors $\mtheta^{U_g r}_{ss^\prime}$, $s \leq s^\prime$, and thus the reduced long coefficient vector $\mtheta^{U_g r}$. As in the case of independent curves, the constraint \eqref{eq: general symmetry constraint} is equivalent to adding up the respective columns of the large design matrix $\mM^{U_g \Delta}$. This can be achieved by right-multiplication of $\mM^{U_g \Delta}$ with a suitable constraint matrix $\mW^{U_g}$, yielding the reduced design matrix $\mM^{U_g\Delta r}=\left[\mM^{U_g\Delta r}_{11}|\ldots|\mM^{U_g\Delta r}_{1\rho^{U_g}}|\mM^{U_g\Delta r}_{22}|\ldots|\mM^{U_g\Delta r}_{\rho^{U_g}-1 \rho^{U_g}-1}|\mM^{U_g\Delta r}_{\rho^{U_g}-1 \rho^{U_g}}|\mM^{U_g\Delta r}_{\rho^{U_g}\rho^{U_g}}\right]$. 
Each $\mM^{U_g\Delta r}_{ss^\prime}$, $s\leq s^\prime$, consists of column-wise concatenated matrices 
\bea \nonumber
\mM^{U_g\Delta r}_{ss^\prime} = \left[\mM^{U_g\Delta}_{ss^\prime,b<b^\prime} + \mM^{U_g\Delta}_{s^\prime s,b>b^\prime}| \mM^{U_g\Delta}_{ss^\prime,b=b^\prime} + \mM^{U_g\Delta}_{s^\prime s,b=b^\prime}\delta_{s<s^\prime}\right],\ s\leq s^\prime = 1,\ldots,\rho^{U_g}.
\eea
The constraint matrix $\mW^{U_g}$ consists of $\left(\rho^{U_g}\right)^2 \times \frac{\left(\rho^{U_g}\right)^2+1}{2}$ blocks, most of which are zero. The block-rows of $\mW^{U_g}$ correspond to the constraint on the spline coefficients of the covariances $K^{U_g}_{ss^\prime}(t,t^\prime)$, $s,s^\prime=1,\ldots,\rho^{U_g}$, sorted as in $\mM^{U_g\Delta}$. The columns are sorted correspondingly to the reduced matrix $\mM^{U_g\Delta r}$. For the auto-covariances ($s=s^\prime$), the blocks are of the same form as the constraint matrix $\mW^E$ for independent curves. For the cross-covariances ($s< s^\prime$), the blocks either correspond to diagonal block matrices or to anti-diagonal block matrices, depending on whether the respective rows correspond to $s<s^\prime$ or $s>s^\prime$, respectively. The specific form of $\mW^{U_g}$ and examples for $\rho^{U_g}=2,3$ are given in Appendix \ref{appendix: supplementary details on estimation}.

A quadratic smoothness penalty associated with each smooth term controls the bias-variance tradeoff. Each penalty matrix $\mS^{U_g}$, consisting of blocks for each $K^{U_g}_{ss^\prime}(t,t^\prime)$, is accordingly reduced by left-and right-multiplication with the constraint matrix $\mW^{U_g}$. Smoothing the components in the upper triangle of $\mK^{U_g}(t,t^\prime)$ separately, allows to define different penalties for the auto-covariances and the cross-covariances, respectively. In particular, it is possible  to apply anisotropic penalties for the cross-covariances.

Reflecting the estimated triangular covariance surfaces across the diagonal yields estimates for the whole covariance surfaces $\mK^{U_g}(t,t^\prime)$ and $K^E(t,t^\prime)$, with smoothness assured also across the diagonal. Note that smoothing multiple covariances using our approach reduces computation time compared to estimating all spline coefficients even more than in the independent case of Section \ref{sec: FSCS}.

\subsection{Functional linear mixed model with crossed random intercepts} \label{sec: crossed model}
Motivated by our application to the phonetics data in Section \ref{sec: phonetics}, we now illustrate the specification of the FLMM for the special case of an FLMM with crossed fRIs accounting for the repeated measurements on two grouping variables (e.g.~speakers and target words) in a crossed design. In this model, we have $G=2$ grouping variables with $\rho^{U_1}=\rho^{U_2}=1$ associated random effects for each grouping variable, i.e.~one fRI each. $L^{U_1}$, $L^{U_2}$ are the numbers of levels of the first (e.g.~speakers) and second (e.g.~target words) grouping variable, respectively. The covariate vector $\mz_i$ only consists of indicators taking value one or zero to code group membership for the two grouping variables. The explicit specification of the covariate vector for crossed and hierarchical functional random effects is given in Appendix \ref{appendix: supplementary details on estimation}. For better readability, we rename in the following the components of the vector-valued random process as $B:=U_1$ and $C:=U_2$. The total number of components in $\mU(t_{ij})$ is $q=L^B+L^C$. Note that this model corresponds to the model in \cite{Cederbaum.2016}.

For this model, equation \eqref{eq: general covariance model rewritten} can be simplified as
\bea  \label{eq: crossed covariance model}
\EV\left[\tilde{Y}_i(t_{ij})\tilde{Y}_{i^\prime}(t_{i^\prime j^\prime})\right]
 &=& K^B(t_{ij},t_{i^\prime j^\prime}) \delta_{B(i)B(i^\prime)} + K^C(t_{ij},t_{i^\prime j^\prime})\delta_{C(i)C(i^\prime)} \\ \nonumber &+& \left[K^E(t_{ij},t_{i^\prime j^\prime}) + \sigma^2 \delta_{jj^\prime} \right] \delta_{ii^\prime},
\eea
where $\delta_{B(i)B(i^\prime)}$ and $\delta_{C(i)C(i^\prime)}$ take value one when the two curves $i$ and $i^\prime$ belong to the same level of the respective grouping variable and zero otherwise. As can be seen in Equation \eqref{eq: crossed covariance model}, the products for which neither $\delta_{B(i)B(i^\prime)}$ nor $\delta_{C(i)C(i^\prime)}$ equals one do not have to be considered due to expectation zero. Equation \eqref{eq: general additive model} then reduces to
\bea \label{eq: crossed additive model}
\EV\left(\mC^{\Delta}\right) &=&\left[\mM^{B\Delta}|\mM^{C \Delta}|\mM^{E\Delta}|\mdelta^{\varepsilon \Delta}\right] \left({\mtheta^{B}}^\top,{\mtheta^{C}}^\top,{\mtheta^{E}}^\top,\sigma^2 \right)^\top
\eea 
and the symmetry constraint can be applied to each $\mtheta^B$, $\mtheta^C$, and $\mtheta^E$.

\subsection{Covariance of cross products} \label{sec: covariance of cross products}
We estimate the auto-covariances as unknown, smooth functions in a bivariate additive varying coefficient model using a quadratic loss function. Since this is equivalent to a penalized likelihood criterion for Gaussian data, we implicitly assume independence of the cross products with homoscedastic Gaussian measurement error. As already mentioned in Section \ref{sec: indep estimation} and shortly discussed in \cite{Cederbaum.2016}, these are working assumptions which do not hold as the products frequently involve two points on the same curve or on correlated curves. Nevertheless, these implicit assumptions are made by many existing works \citep[e.g.][]{Yao.2005, Di.2009, Greven.2010}. We now briefly discuss how the covariance of the cross products can be accounted for.

For the model with independent curves \eqref{eq: indep model}, \cite{Xiao.2016b} derive an expression for the covariance of the cross products in terms of $K^E(t,t^\prime)$ and the error variance $\sigma^2$ under the assumption of Gaussian responses. They apply a three-step algorithm in which they first estimate $K^E(t,t^\prime)$ and $\sigma^2$ under the working assumptions. Second, they estimate the covariance of the cross products by plugging in the estimates for $K^E(t,t^\prime)$ and $\sigma^2$. In the third step, they re-estimate $K^E(t,t^\prime)$ and $\sigma^2$ using the estimated covariance of the cross products as a working covariance. 

We derive an expression for the covariance of the cross products for the general model \eqref{eq: general model} based on results from \cite{Isserlis.1918} on fourth moment rules for multivariate Gaussian random variables. The covariance of the cross products can be written as
\bea \label{eq: covariance of cross products}
&&\Cov\left[\tilde{Y}_i(t_{ij})\tilde{Y}_{i^\prime}(t_{i^\prime j^\prime}),\tilde{Y}_{m}(t_{mo})\tilde{Y}_{m^\prime}(t_{m^\prime o^\prime})\right]  \\ \nonumber
&=&\left\{\mz_i^\top \mK^U(t_{ij},t_{mo})\mz_m + \left[ K^E(t_{ij},t_{mo}) + \sigma^2\delta_{jo} \right] \delta_{im} \right\}\\\nonumber
&&\cdot  \left\{\mz_{i^\prime}^\top \mK^U(t_{i^\prime j^\prime },t_{m^\prime o^\prime})\mz_{m^\prime} + \left[ K^E(t_{i^\prime j^\prime},t_{m^\prime o^\prime}) + \sigma^2\delta_{j^\prime o^\prime} \right] \delta_{i^\prime m^\prime} \right\}\\\nonumber
&+& \left\{\mz_{i}^\top \mK^U(t_{ij},t_{m^\prime o^\prime})\mz_{m^\prime } + \left[ K^E(t_{ij},t_{m^\prime o^\prime}) + \sigma^2\delta_{j o^\prime} \right] \delta_{im^\prime} \right\}\\ \nonumber
&&\cdot  \left\{\mz_{i^\prime}^\top \mK^U(t_{i^\prime j^\prime },t_{mo})\mz_{m} + \left[ K^E(t_{i^\prime j^\prime},t_{mo}) + \sigma^2\delta_{j^\prime o} \right] \delta_{i^\prime m} \right\}.
\eea
The derivation and simplifications for the model with crossed fRIs are given in Appendix \ref{appendix: proofs}. The covariance \eqref{eq: covariance of cross products} is a function of the unknown covariances $\mK^U(t,t^\prime)$, $K^E(t,t^\prime)$ and $\sigma^2$, giving rise to the need for an iterative procedure as proposed in \cite{Xiao.2016b} for the simpler model \eqref{eq: indep model}. 
Our implementation with \pkg{R}-package \pkg{mgcv} allows to directly include the variance of the cross products by a specification of the \pkg{weights} argument in function \pkg{bam}. Our simulations showed, however, that accounting for only the heterogeneous variance does not lead to a substantial improvement in estimation accuracy. The dependencies could also be accounted for by pre-multiplication with the inverse square root of the covariance of the cross products, for which its construction and inversion would become necessary. 
For our application to the phonetics data, the covariance matrix of the cross products would be a $52,346,570 \times 52,346,570$ dense and unstructured matrix whose construction is not feasible with current technology ($\approx$ 17,500 Terabytes storage space would be required) and we thus do not focus on this extension in the following. However, \eqref{eq: covariance of cross products} allows the inclusion of the covariance of the cross products in less complex settings and for (much) smaller data sets and can then result in more efficient estimates. Despite the violations of working assumptions, we achieve good results with our approach in simulations (cp.~Section \ref{sec: simulations}).
Note that a relevant improvement of the covariance estimation can only be obtained if the working covariance is reasonably well estimated and thus more than one iteration of the algorithm might be necessary. Especially for multiple iterations, our fast smoothing algorithm considerably speeds up the estimation compared to smoothing the entire covariances.

\section{Application in functional principal component analysis} \label{sec: FPCA}
An important application of our fast symmetric additive smoothing approach is  functional principal component analysis. FPCA is a key tool for dimension reduction in FDA that extracts the dominant modes of variation in the data and provides an explicit variance decomposition. In the following, we briefly outline the four main steps of FPCA for the general FLMM \eqref{eq: general model} using our newly proposed covariance estimation approach. Applying FPCA to model 
\eqref{eq: general model} yields parsimonious representations of each random process $\mU_g(t)$, $g=1,\ldots,G$, and $E_i(t)$, in bases of eigenfunctions of the respective, previously estimated smooth auto-covariances.
In addition, we briefly describe how our covariance smoothing approach can be combined with the general framework of functional additive mixed models \citep[FAMM;][]{Scheipl.2015} allowing for approximate statistical inference for the mean conditional on the FPCA. For a more detailed description, see \cite{Cederbaum.2016}.

In the first step, the smooth mean function is commonly estimated based on a working independence assumption. We use penalized splines implemented in \pkg{R}-package \pkg{mgcv}, which can be used to estimate a large variety of covariate and interaction effects in the mean function. For the subsequent steps, the curves are then centered by subtracting the estimated mean from the functional observations. 

For the second step, we propose to simultaneously estimate the upper triangles of the covariances $\mK^{U_g}(t,t^\prime)$, $g=1,\ldots,G$, and $K^E(t,t^\prime)$ and the error variance using our novel covariance smoothing approach. The triangular covariance surfaces are then reflected across the diagonal, yielding estimates for the complete covariance surfaces $\mK^{U_g}(t,t^\prime) = \left[K^{U_g}_{ss^\prime}(t,t^\prime)\right]_{s,s^\prime = 1,\ldots,\rho^{U_g}}$, $g=1,\ldots,G$, and $K^E(t,t^\prime)$. Negative estimated values of $\sigma^2$ are set to zero.

In the third step, we use spectral decompositions of the estimated covariance surfaces based on Mercer's Theorem \citep{Mercer.1909}
\bea \nonumber 
\mK^{U_g}(t,t^\prime) = \sum_{k=1}^{\infty} \nu^{U_g}_{k}\mphi^{U_g}_{k}(t){\mphi^{U_g}_{k}(t^\prime)}^\top, \ \mK^E(t,t^\prime) = \sum_{k=1}^{\infty} \nu^E_k \phi^E_k(t) \phi^{E}_k(t^\prime),
\eea
with eigenvalues $\nu^{U_g}_k$, $\nu^E_k$ and (vector-valued) eigenfunctions $\mphi^{U_g}_k(t)=\left[\phi^{U_g}_{ks}(t)\right]_{s=1,\ldots,\rho^{U_g}}$, $\phi^E_k(t)$, respectively. In practice, the covariance surfaces are evaluated on a dense grid $T=\lbrace t_1,\ldots,t_D \rbrace \in \mathcal{T}$ of pre-specified length $D$. We obtain estimated eigenvalues $\hat{\nu}^{U_g}_{k}$, $k=1,\ldots,D\rho^{U_g}$, and $\hat{\nu}^E_k$, $k=1,\ldots,D$, as well as orthonormal eigenfunctions evaluated on $T$, $\hat{\mphi}_{k}^{U_g}=\left[\hat{\phi}^{U_g}_{ks}(t)\right]_{s=1,\ldots,\rho^{U_g},t\in T}$ $\in \mathbb{R}^{D\rho^{U_g}}$ and $\hat{\mphi}_k^E=\left[\hat{\phi}_k^E(t)\right]_{t\in T}$ $\in \mathbb{R}^D$, of each corresponding covariance operator. The multivariate eigenvectors $\hat{\mphi}^{U_g}_k$ consist of blocks for the respective eigenfunction components. The eigenvectors $\hat{\mphi}^{U_g}_{k}$, $\hat{\mphi}^E_k$ (and accordingly the eigenvalues) are rescaled to ensure orthonormality with respect to the additive scalar product $\langle(f_1,\ldots,f_{\rho^{U_g}}),(g_1,\ldots,g_{\rho^{U_g}})\rangle = \sum_{s=1}^{\rho^{U_g}}\int_{\mathcal{T}} f_s(t)g_s(t) \dint t$, and with respect to the $L^2$-scalar product $\langle f,g\rangle = \int_{\mathcal{T}} f(t)g(t) \dint t$, respectively. For more details on multivariate FPCA, see e.g., \cite{Ramsay.2005}. To guarantee positive semi-definiteness of the covariances, which is not ensured by our smoothing approach, negative eigenvalues can be set to zero, which has been shown to improve accuracy in terms of the $L^2$-norm \citep{Hall.2008} and to work well in practice \citep{Yao.2003}.
Dimension reduction is achieved by truncating the number of eigenfunctions. We choose the truncation levels based on the proportion of variance explained \citep[see][for an overview]{Greven.2010} and denote them by $N^{U_g}$, $g=1,\ldots,G$, and $N^E$, respectively. The truncated Karhunen-Loève (KL) expansion \citep{Loeve.1945,Karhunen.1947} allows parsimonious representations of the random processes in truncated bases of the corresponding eigenfunctions
\bea \label{eq: KL expansions}
\mU_{gl}(t) \approx \sum_{k=1}^{N^{U_g}} \xi^{U_g}_{lk}\mphi^{U_g}_k(t),\ E_i(t) \approx \sum_{k=1}^{N^E} \xi^{E}_{ik}\phi^{E}_k(t),
\eea
 with uncorrelated zero-mean random basis weights $\xi^{U_g}_{lk}$, $l=1,\ldots,L^{U_g}$, $k=1,\ldots,N^{U_g}$, and $\xi^E_{ik}$, $i=1,\ldots,n$, $k=1,\ldots,N^E$, with variance $\nu^{U_g}_{k}$ and $\nu^E_k$, respectively. 
 
In the fourth step, we predict the random basis weights, which give insight into the individual structure of each grouping level. Replacing the random processes in Model \eqref{eq: general model} by their truncated KL-expansions in \eqref{eq: KL expansions} allows to approximate the model by a scalar linear mixed model with random effects corresponding to the random basis weights $\xi^{U_g}_{lk}$, $\xi^E_{ik}$ \citep{Di.2009}. The basis weights can then be predicted as empirical best linear unbiased predictors by simply plugging-in the estimated eigenfunctions, eigenvalues, and the estimated error variance \citep{Di.2009,Greven.2010, Cederbaum.2016}.

Alternatively, we can represent our model as a FAMM using our estimated eigenfunctions and -values in basis expansions of the random processes as proposed by \cite{Scheipl.2015}. The random basis weights are predicted together with a re-estimation of the mean function in a mixed model framework. This allows for more efficient mean estimation due to taking the covariance structure into account and  for approximate statistical inference conditional on the FPCA, such as pointwise confidence bands for the mean and for covariate effects. For more details on the combination with the FAMM approach and an extensive comparison of the two ways to predict the basis weights, see \cite{Cederbaum.2016}.

\section{Implementation} \label{sec: implementation}
We base our implementation on \pkg{R}-package \pkg{mgcv} \citep{Wood.2011} which allows to add user-defined spline bases and penalties. In this framework, we define a novel class for bivariate smooths estimated subject to our symmetry constraint \eqref{eq: symmetry constraint} called '\pkg{symm}' by providing a new constructor method function \pkg{smooth.construct.symm.smooth.spec} and a corresponding predictor method function for the estimation of smooth surfaces in additive models. The class can be applied to any bivariate smooth term in a \pkg{gam}-formula. It is not restricted to symmetric data although we here consider symmetric data in the form of cross products in the smoothing of the covariance. It can be applied to (possibly noisy) data sampled on a regular grid as well as to irregularly or sparsely sampled data. As a modular component, our constructor can be applied separately to the auto-covariances of independent functional random effects in an FLMM. The case of correlated functional random effects is currently not covered in the implementation. In our application, we show how the constructor can be applied to complex designs on the basis of the FLMM with crossed fRIs as in \eqref{eq: crossed additive model}. One main advantage of using standard software is that it allows for flexible extensions. The current implementation is based on tensor product B-splines with difference penalties \citep{Eilers.2003}, but extensions to other bivariate bases that are symmetric across the diagonal and other penalties are possible.

The spline degree and the number of basis functions can be chosen. The user currently has the choice between two different quadratic penalties. One can either use the Kronecker sum penalty $\operatorname{pen}(\lambda)$ $= \lambda \mtheta^\top \left[\left(\mS_t \otimes \mI_{F}\right) + \left(\mI_F \otimes \mS_{t^\prime}\right)\right] \mtheta$ or alternatively a Kronecker product penalty of the form $\operatorname{pen}(\lambda)$ = $\lambda \mtheta^\top \left(\mS_t \otimes \mS_{t^\prime}\right)\mtheta = \lambda \mtheta^\top \left[\left(\mS_t \otimes \mI_{F}\right) \cdot \left(\mI_F \otimes \mS_{t^\prime}\right)\right] \mtheta$, where $\mS_t=\mS_{t^\prime}$ denote the marginal penalty matrices and $\mtheta$ is the coefficient vector. Note that both penalties are isotropic, reflecting the symmetry of the surface. The main difference between the two penalty matrices is that the null space of the latter is larger, more likely leading to wigglier estimates. Other possible penalties could be added by the user. 
 
To speed up estimation, function \pkg{bam} can compute the computationally expensive steps in parallel on multiple cores. 

In addition to the \pkg{R}-code for our constructor, we provide code for the FPCA based on our (additive) covariance smoothing approach for three special cases of the FLMM (Model \eqref{eq: indep model}, a model with an fRI and a smooth error curve, and the model with two crossed fRIs and a smooth error curve as in Section \ref{sec: crossed model}).

\section{Applications}\label{sec: applications}
We demonstrate the practical relevance of our approach in two distinct applications. We consider a standard data set consisting of sparse longitudinal observations as well as functional data with a complex correlation structure and different grids between curves. 

\subsection{CD4 cell count data}
In order to compare our approach to that of \cite{Xiao.2016b} for the common special case of longitudinal data, we apply FPCA to analyze the CD4 cell count trajectories in HIV positive individuals which are available in \pkg{R}-package \pkg{refund} \citep{refund}. 
As our focus here is on the more complex case of correlated functional data, to which the approach of \cite{Xiao.2016b} does not apply, the application to the CD4 cell count data is given in Appendix \ref{appendix: supplementary application details and results}.

\subsection{Phonetics data}\label{sec: phonetics}
We apply FPCA based on our covariance smoother to acoustic signal data with a crossed correlation structure. We show the increase of the computational efficiency compared to the approach of \cite{Cederbaum.2016}, for a case where their approach is applicable. To the best of our knowledge, their smooth method of moments approach is the only competitor for covariance smoothing of irregularly observed correlated curves with a crossed design structure. No alternative approach is available for general models \eqref{eq: general model}, where our approach is the first available for additive covariance smoothing.

\begin{figure}[ht]
\begin{center}
\includegraphics[width=0.8\textwidth]{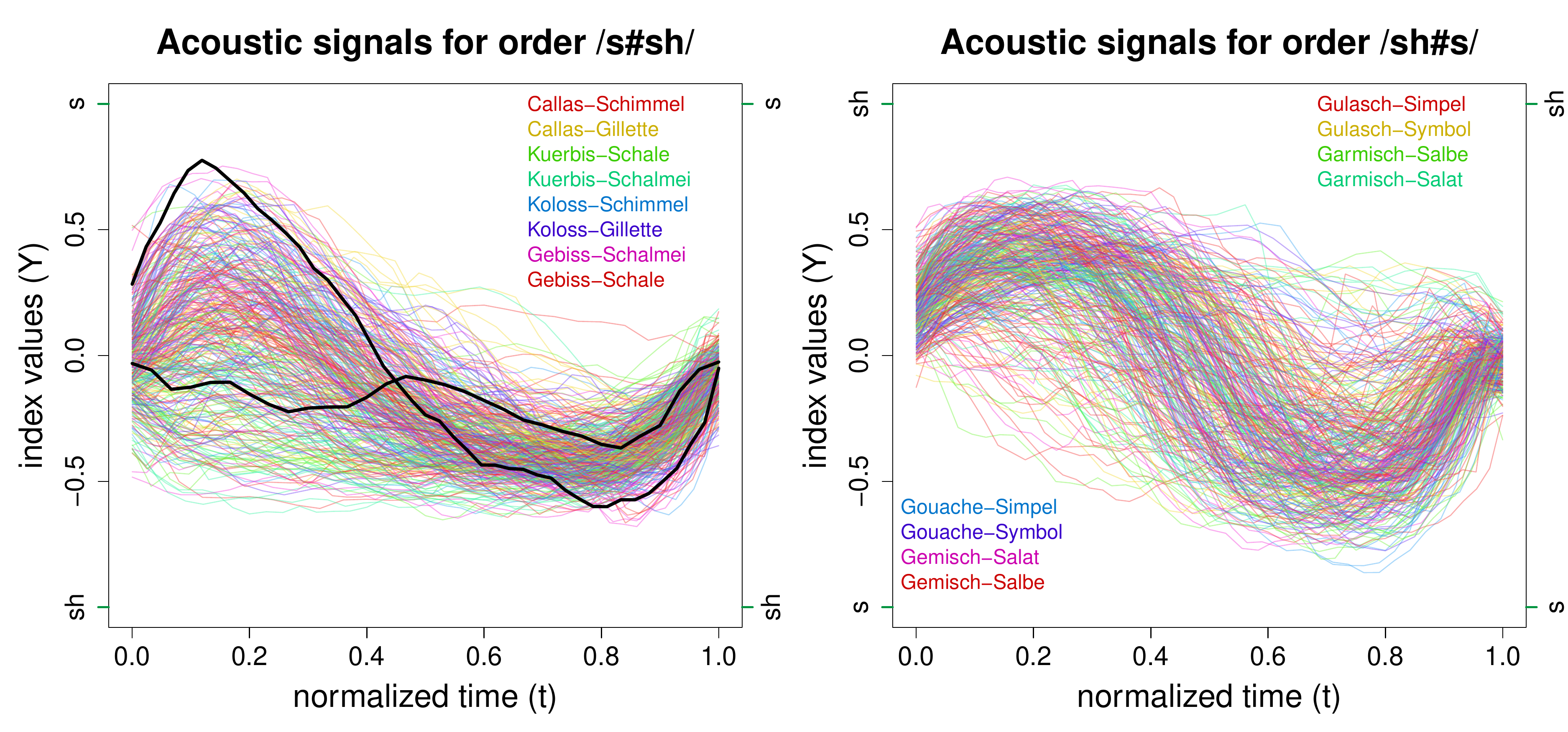}
\caption{Acoustic signal curves of the phonetics data over time \citep[cf.][]{Cederbaum.2016}. Left [right]: Signal curves of consonant order /s\#sh/ [sh\#s] colored by target word. Example curves with no/little assimilation and with strong assimilation highlighted (black).}
\label{fig: phonetics data}
\end{center}
\end{figure}

In phonetic research, the term consonant assimilation refers to the phenomenon that the articulation of two consonants becomes phonetically more alike when they appear subsequently in fluent speech. Consonant assimilation is accompanied by a complex interaction of language-specific, perceptual and articulatory factors which makes it an important topic in speech production research. 
The data we consider are part of a large study which was conducted by \cite{Pouplier.2016} in order to investigate among others the assimilation of the consonants /s/ and /sh/ as a function of their order (/s\#sh/ versus /sh\#s/, where \# denotes a word boundary), syllable stress and vowel context in the German language. The same data were previously analyzed by \cite{Cederbaum.2016}. \cite{Pouplier.2016} recorded the audio signals for nine native speakers which repeated the same sixteen target words each five times. The target words consisted of (semantically nonsensical) bisyllabic noun-noun compound words with abutting consonants /s/ and /sh/ in either order, e.g.~`Calla\textbf{s}-\textbf{Sch}immel' and `Gula\textbf{sch}-\textbf{S}impel', and with either stressed or unstressed syllables and varying vowel combinations. The time interval during the duration of the two consonants of interest was cut out manually by the phoneticians and standardized to a [0,1] interval in which the recorded acoustic signals were summarized in a functional index over time. The $n=707$ index curves (shown in Figure \ref{fig: phonetics data}) take values between $+1$ and $-1$, with positive [negative] values indicating proximity of the signal to a reference signal for the first [second] consonant of the target word, respectively. To illustrate the effect of consonant assimilation, two acoustic signals are highlighted in Figure \ref{fig: phonetics data}. The curve without or with little assimilation shows a clear transition from a strong positive to a strong negative value, whereas the curve with strong assimilation is quite flat and mostly takes negative values. The curves differ in the number and location of the observation points, ranging from 22--57 with a median of 34 points per curve. For a more detailed description of the data (including pre-processing steps), see \cite{Pouplier.2016, Cederbaum.2016}.

\begin{figure}[h!]
\begin{center}
\begin{tabular}{cccc}
TRI-CONSTR & TRI-CONSTR-W & WHOLE & TRI\\[-2ex]
\includegraphics[width=0.23\textwidth, page=1]{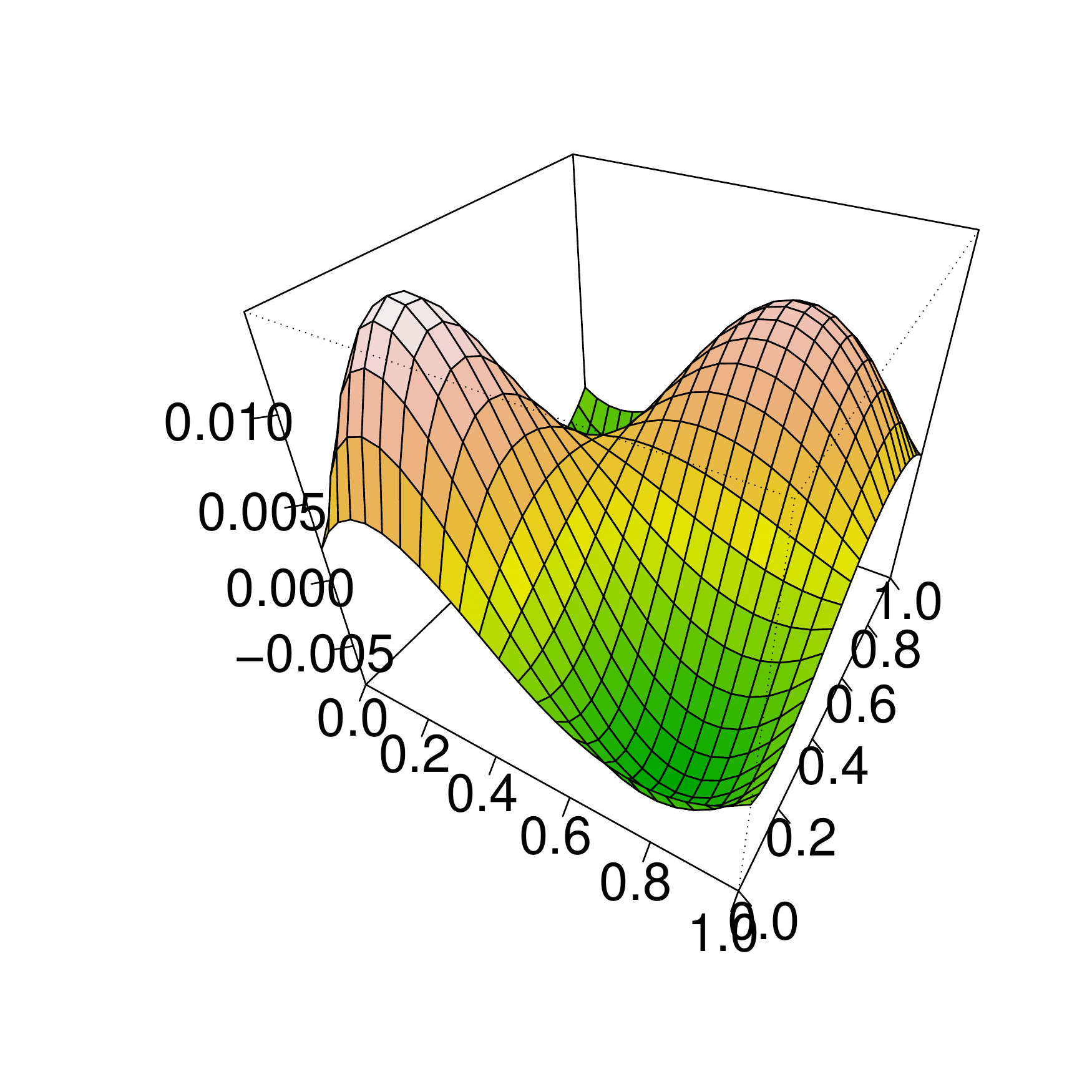} &
\includegraphics[width=0.23\textwidth, page=2]{covsB_estimated_persp_image_s_sh_acoustic_25_May_all_covs_cov_paper.pdf}&
\includegraphics[width=0.23\textwidth, page=3]{covsB_estimated_persp_image_s_sh_acoustic_25_May_all_covs_cov_paper.pdf} &
\includegraphics[width=0.23\textwidth, page=4]{covsB_estimated_persp_image_s_sh_acoustic_25_May_all_covs_cov_paper.pdf}\\[-4ex]
\includegraphics[width=0.23\textwidth, page=5]{covsB_estimated_persp_image_s_sh_acoustic_25_May_all_covs_cov_paper.pdf} &
\includegraphics[width=0.23\textwidth, page=6]{covsB_estimated_persp_image_s_sh_acoustic_25_May_all_covs_cov_paper.pdf}&
\includegraphics[width=0.23\textwidth, page=7]{covsB_estimated_persp_image_s_sh_acoustic_25_May_all_covs_cov_paper.pdf}&
\includegraphics[width=0.23\textwidth, page=8]{covsB_estimated_persp_image_s_sh_acoustic_25_May_all_covs_cov_paper.pdf}
\\[-2ex]
\includegraphics[width=0.23\textwidth]{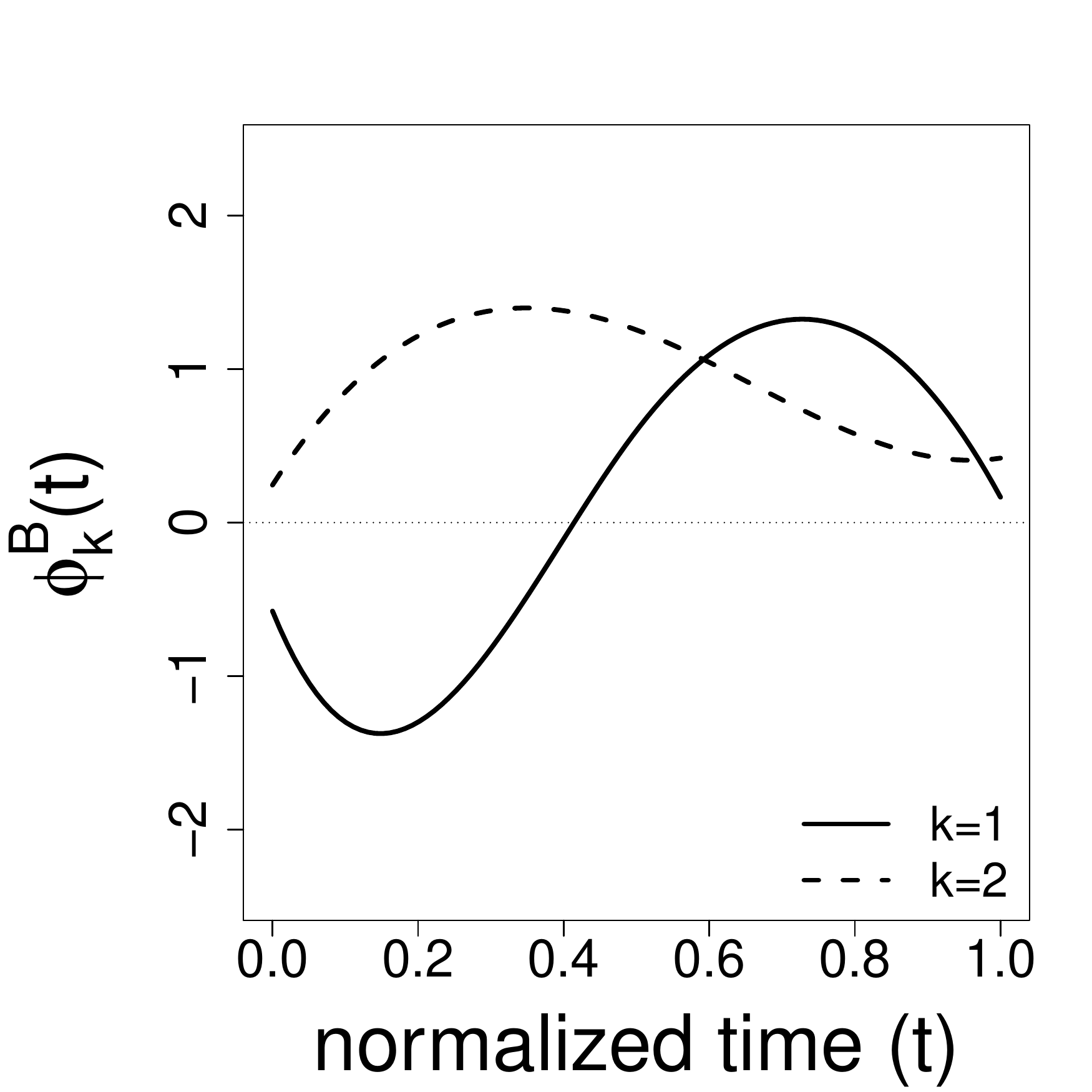}&
\includegraphics[width=0.23\textwidth]{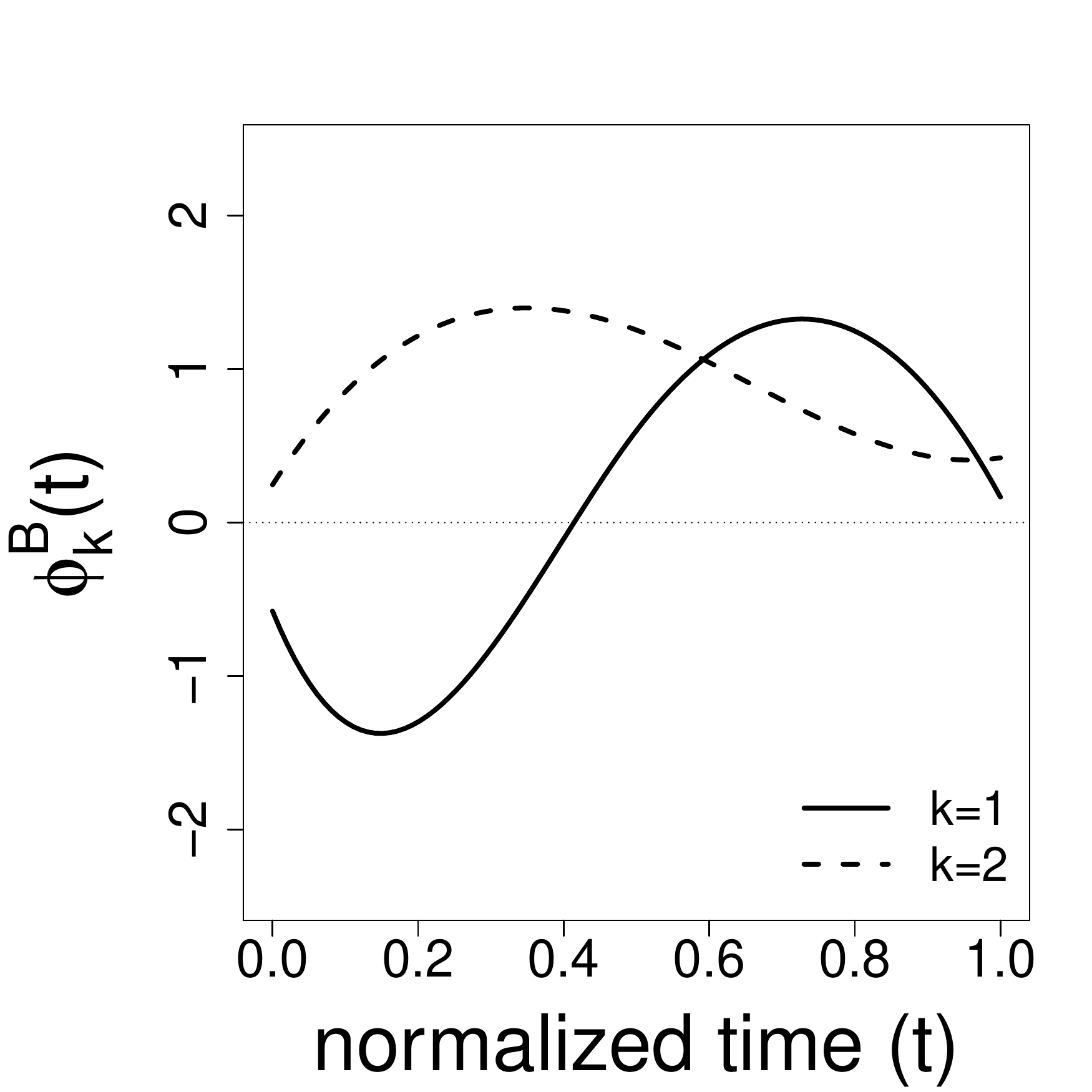}&
\includegraphics[width=0.23\textwidth]{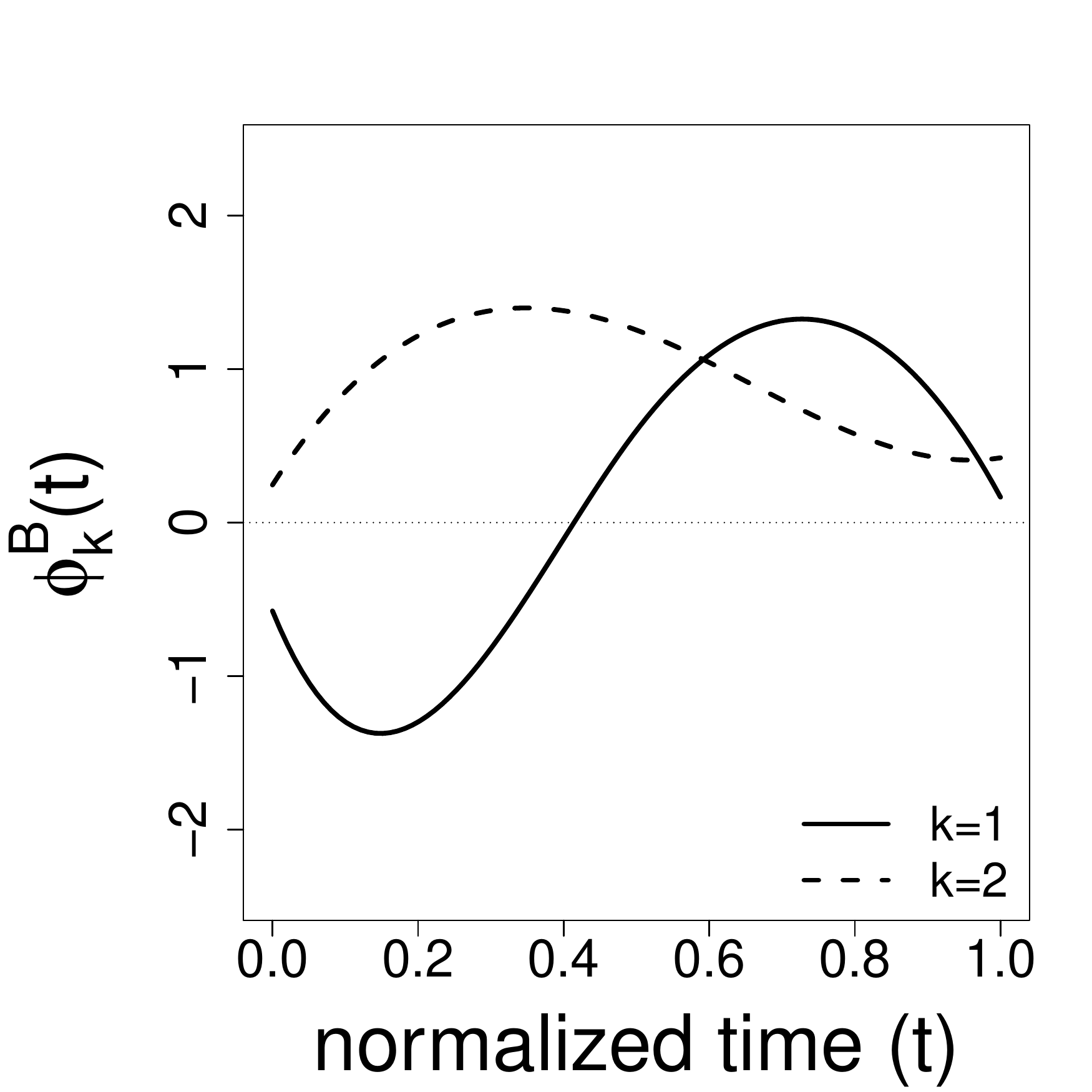}&
\includegraphics[width=0.23\textwidth]{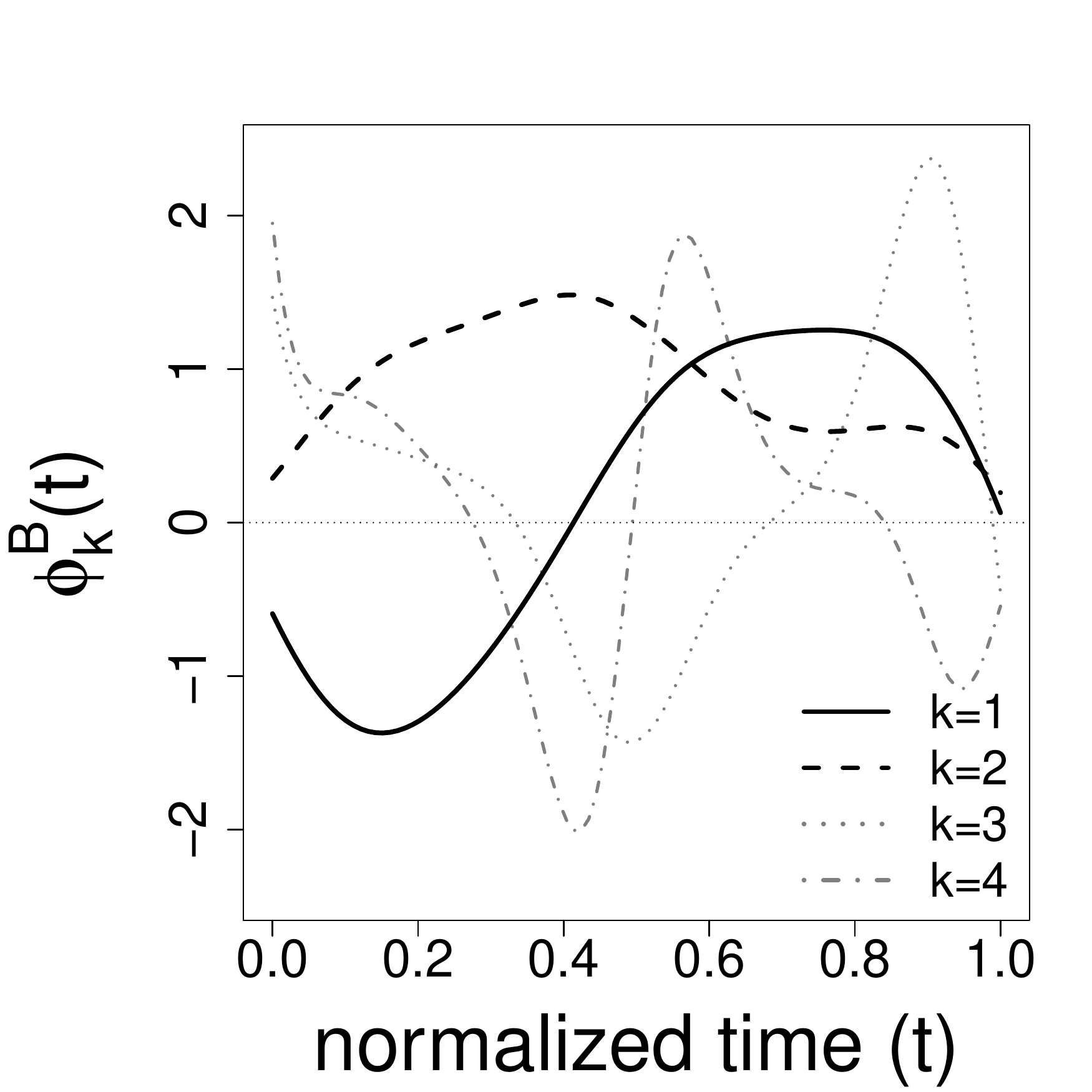}
\end{tabular}
\caption{Results for the fRI for speakers using the four smoothing methods. Top row: estimated covariance surfaces. Middle row: contours of the estimated covariance surfaces. Bottom row: estimated corresponding eigenfunctions $\phi^B_k(t)$.}
\label{fig: phonetics covariances}
\end{center}
\end{figure}

We fit an FLMM with crossed fRIs as described in Section \ref{sec: crossed model} and previously considered in \cite{Cederbaum.2016} to account for the repeated measurements of both speakers and target words. The mean function $\mu(t,\mx_i)$ includes effects and interaction effects of the dummy-coded covariates consonant order, syllable stress, and vowel context, smoothly varying over time. As the focus here is on the smooth auto-covariances of the functional random effects, $K^B(t,t^\prime)$, $K^C(t,t^\prime)$, and $K^E(t,t^\prime)$, we refer to \cite{Cederbaum.2016} for more details on and interpretations of the covariate effects. For each auto-covariance, we use cubic marginal B-spline bases with marginal third order difference penalty matrices and use the Kronecker sum penalty for bivariate smoothing (cp.~Section \ref{sec: implementation}). We compare the results from our novel symmetric bivariate smoother (denoted by TRI-CONSTR and by TRI-CONSTR-W with weights of 0.5 on the cross products on the diagonal, cp.~Section \ref{sec: indep estimation}) with the results we obtain using the smoothing approach of \cite{Cederbaum.2016} (denoted by WHOLE). The latter does not exploit the symmetry of the estimated surface and is equivalent to TRI-CONSTR-W except for the estimation of the smoothing parameter and numerical differences. To highlight the need for a symmetry constraint when only the upper triangle is considered, we further compare with the results obtained by estimating the upper triangle without a symmetry constraint (denoted by TRI) which does not guarantee smoothness across the diagonal (cp.~Section \ref{sec: indep estimation}). 
The estimated auto-covariances are evaluated on a pre-specified grid of length $D=100$. During the FPCA, we choose the truncation levels based on a pre-specified proportion of explained variance of $0.95$, yielding two [four] eigenfunctions for the auto-covariance of the fRI for speakers, $K^B(t,t^\prime)$, and three [twelve] eigenfunctions for the auto-covariance of the smooth error, $K^E(t,t^\prime)$, for our two approaches and WHOLE [for TRI]. For all four smoothing methods, no eigenfunction is chosen for the fRI for target words, which is due to the high number of covariates that describe the target words sufficiently \citep[cf.][]{Cederbaum.2016}.
Figure \ref{fig: phonetics covariances} shows the estimated surfaces and contours of the auto-covariance of the fRI for speakers, reconstructed after truncation from the estimated eigenvalues and eigenfunctions, which are shown in the bottom of the figure. 

We can see from Figure \ref{fig: phonetics covariances} that 
the estimated covariance $\hat{K}^B(t,t^\prime)$ based on our symmetric smoother (TRI-CONSTR-W) is very similar to the one obtained by using all cross products (WHOLE). Moreover, it shows that for this application, the weights on the diagonal cross products do not make a great difference. As expected, we observe wigglier estimates especially on the diagonal for TRI, for which the error variance is estimated to be zero. This is also reflected in the wiggliness and higher number of chosen eigenfunctions for TRI. Similar results can be found for $K^E(t,t^\prime)$. These are given in Appendix \ref{appendix: supplementary application details and results}, where additional estimation details and results including the complete variance decompositions are provided. For all four approaches, the first eigenfunction in Figure \ref{fig: phonetics covariances} (solid line) corresponds to the discrimination of the speaker between the first and the second consonant and the second eigenfunction (dashed line) mainly leads to a vertical shift of the signal curves. Accounting for the symmetry of the covariances leads to a considerable reduction of computation times. TRI-CONSTR and TRI-CONSTR-W have the shortest computation times for smoothing the three auto-covariances, using five kernels in parallel, amounting to 24.51 and 25.82 minutes, respectively. Smoothing the auto-covariances using WHOLE takes more than twice as long (55.76 minutes) and using TRI still amounts to 32.95 minutes, which partly results from the fact that more spline coefficients have to be estimated. In addition, WHOLE and TRI require the estimation of two (instead of one) smoothing parameters for each auto-covariance using the Kronecker sum penalty implemented in \pkg{R}-package \pkg{mgcv} of the form $\operatorname{pen}(\lambda) = \lambda_{t} \mtheta^\top \left(\mS_t \otimes \mI_{F}\right)\mtheta + \mtheta^\top\lambda_{t^\prime}\left(\mI_F \otimes \mS_{t^\prime}\right)\mtheta$.
\section{Simulations} \label{sec: simulations}
\subsection{Simulation designs}
To investigate the performance of our covariance smoothing approach, we conduct an extensive simulation study based on two different data generating processes. For the first scenario (Scenario 1), we simulate data consisting of $n=100$ independent curves from Model \eqref{eq: indep model} with $\mu(t) = \sin(t)+t$. The data for the second scenario (Scenario 2) are generated from an FLMM with crossed fRIs as in Section \ref{sec: crossed model}, such that they mimic the irregularly observed phonetics data. Note that no covariate mean effects are included such that we additionally obtain one eigenfunction for the fRI for target words and thus really have crossed fRIs. Additional generation details for this scenario are given in Appendix \ref{appendix: supplementary simulation details and results}. We compare the performance of our fast (additive) symmetric covariance smoother (TRI-CONSTR, TRI-CONSTR-W with weights of 0.5 on the diagonal cross products) as basis for FPCA with the performance of the smoothing approach proposed by \cite{Cederbaum.2016} (WHOLE). To evaluate the need of the symmetry constraint when only the upper triangle is estimated, we further compare with the results obtained without posing a symmetry constraint (TRI). For Scenario 1, we additionally provide a comparison with the smoothing approach of \cite{Xiao.2016b} (denoted by FACE), implemented in function \pkg{face.sparse} in \pkg{R}-package \pkg{face} \citep{FACE}. FACE does, however, not apply to additive covariance smoothing needed for correlated curves. Note that \cite{Xiao.2016b} compare their symmetric covariance smoother to a number of other approaches that are all restricted to independent curves. They show that their approach is superior or comparable in terms of median integrated squared errors and inter quartile ranges (IQR) to the approach implemented in function \pkg{fpca.sc} \citep{Di.2009} in \pkg{R}-package \pkg{refund} based on bivariate B-splines with a difference penalty and to a self-coded variant based on thin plate regression splines for covariance smoothing. Moreover, they demonstrate their supremacy over the geometric likelihood approach of \cite{Peng.2009} and the local polynomial approach proposed in \cite{Yao.2003}. We thus do not include these alternatives here.

Based on Scenario 1, we investigate the sensitivity of the estimates to varying model complexity in terms of the complexity of the underlying eigenfunctions, signal to noise levels as functions of error variances and eigenvalues, and degree of sparseness. We consider all possible combinations of
\begin{compactenum}
\item simple eigenfunctions: $\lbrace \phi_1(t)=1, \phi_2(t)=\sqrt{3}(2t-1)\rbrace$, complex eigenfunctions $\lbrace \phi_1(t)=\sin(2\pi t), \phi_2(t)=\cos(2\pi t)\rbrace$
\item error variance: $\sigma^2=0.5$, $\sigma^2=0.05$ 
\item eigenvalues: $\lbrace \nu_1=0.15, \nu_2=0.075 \rbrace$, $\lbrace \nu_1=2, \nu_2=1\rbrace$
\item number of observation points: drawn from uniform distributions $\mathcal{U}\left[ 40,60\right]$.
\end{compactenum}
For the simple eigenfunctions and error variance $\sigma^2 \in \lbrace 0.05,0.5\rbrace $, we additionally consider a sparse setting, in which the number of observation points is drawn from the uniform distribution $\mathcal{U}\left[3,10 \right]$ and for the complex eigenfunctions and eigenvalues $\nu_1=2, \nu_2=1$, we additionally reduce the value of the error variance to $\sigma^2 = 0.01$.
For all settings, we generate 200 data sets. The random basis weights are centered and decorrelated such that the weights empirically have zero mean and a correlation of zero \citep[see][for a discussion]{Cederbaum.2016}. 

For the estimation in Scenario 1, we use ten cubic B-splines each for the estimation of the mean function and as marginal bases for the auto-covariances. For the estimation in Scenario 2, we use eight and five cubic B-splines for the estimation of the mean function and as marginal bases for the auto-covariances, respectively. We use Kronecker sum penalties (cp.~Section \ref{sec: implementation}) of marginal third order difference penalties for bivariate smoothing. Estimation of the smoothing parameter is based on REML, except for FACE, which uses leave-one-curve-out cross validation. We use equidistant knots in function \pkg{face.sparse} instead of the default (quantile based knots) which would require an adapted penalty that is not implemented. The arguments that determine the smoothing parameter search in function \pkg{face.sparse} are left at their defaults. As function \pkg{face.sparse} does not allow to specify a fixed truncation level, we choose the number of eigenfunctions based on a pre-specified proportion of explained variance of $0.95$. Note that we use the proportion of explained variance in the functional observations, whereas \cite{FACE} use that in $E_i(t)$. In order to be able to differentiate the error incurred by the truncation of the covariance surface to a few leading FPCs from the pure covariance surface estimation error, we pre-specify the correct truncation lags for Scenario 2, for which no comparison with FACE is possible anyway.

\subsection{Simulation results}\label{sec: simulation results}
We present and discuss the results of both simulation scenarios. For Scenario 1, we focus our presentation of the results on the setting with complex eigenfunctions, an error variance of $\sigma^2=0.05$ and eigenvalues of size $\nu_1=2$, $\nu_2=1$ (denoted as Setting 1). As a measure of goodness of fit, we use root relative mean squared errors (rrMSEs) of the form $\sqrt{(\mbox{true}-\mbox{estimated})^2/\mbox{true}^2}$. The complete results for all settings and the specific forms of the rrMSE for all model components, are given in Appendix \ref{appendix: supplementary simulation details and results}.

\begin{figure}[p]
\begin{center}
\includegraphics[width= 1\textwidth]{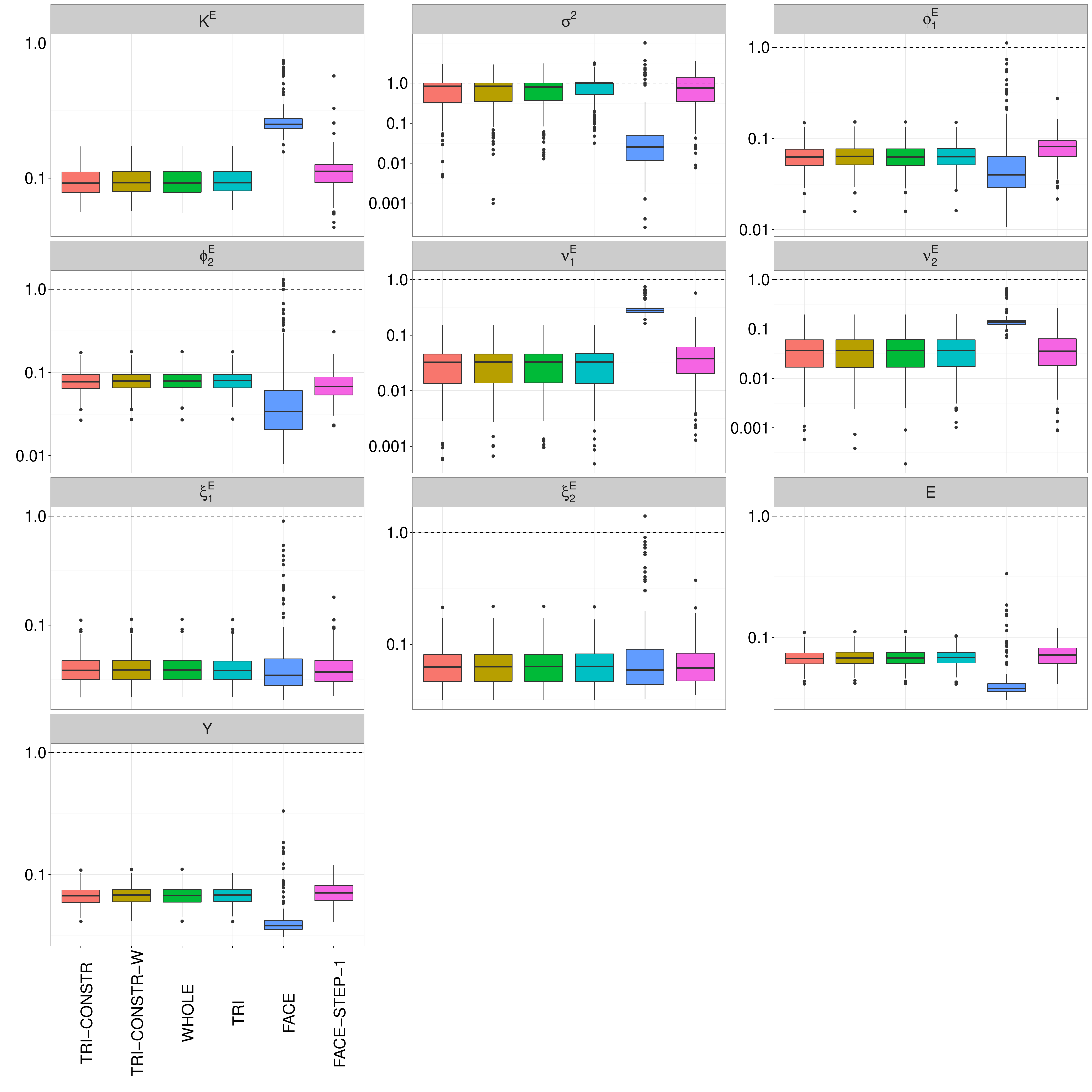}
\caption{Boxplots of the rrMSEs (log10 scale at y-axis) for the scenario with independent curves (Scenario 1) for the smoothing methods being compared. Top row: rrMSEs for auto-covariance $K^E(t,t^\prime)$, error variance $\sigma^2$, and the first eigenfunction $\phi^E_1(t)$. Second row: rrMSEs for the second eigenfunction $\phi^E_2(t)$ and eigenvalues $\nu^E_1$, $\nu^E_2$. Third row: rrMSEs for the random basis weights $\xi^E_1$, $\xi^E_2$ and process $E_i(t)$. Bottom row: rrMSEs for curves $Y_i(t)$.}
\label{fig: boxplots rrMSEs indep curves}
\end{center}
\end{figure}

Figure \ref{fig: boxplots rrMSEs indep curves} depicts boxplots of the rrMSEs for $200$ simulation runs for Setting 1. For each model component, it shows the boxplots for the compared smoothing methods and, in addition, for a modified version of FACE (denoted by FACE-STEP-1), in which the covariance of the cross products is not accounted for and thus only the first step of the three-step procedure is performed. FACE-STEP-1 is added to evaluate the effect of accounting for the covariance of the cross products (cf.~Section \ref{sec: covariance of cross products}). It shows that all components, except for the error variance, are estimated very well for our approach (TRI-CONSTR, TRI-CONSTR-W). The weights on the diagonal cross products do not have a great influence. Moreover, our approach yields similar rrMSEs as WHOLE. We also obtain similar results for TRI for most components, except for the error variance for which TRI has a higher median rrMSE. FACE yields a more than $2.7$ times higher median rrMSE for the auto-covariance $K^E(t,t^\prime)$ compared to our approach and consequently also worse results for the eigenvalues $\nu^E_k$, $k=1,2$. It yields smaller median rrMSEs for the eigenfunctions $\phi^E_k$, $k=1,2$, but the IQR is larger and more outliers occur than in all other approaches. The estimation of the error variance profits most from accounting for the covariance of the cross products, which is reflected in a much lower median rrMSE for FACE than for the other methods. This also leads to lower median rrMSEs for the random basis weights, $\xi^E_1$ and $\xi^E_2$, which depend on the error variance and consequently also to lower median rrMSEs for the reconstructed processes $E_i(t)$ and $Y_i(t)$. For all components, however, FACE yields a high number of outliers that range above the maximal rrMSEs obtained for the other methods. It is noticeable that the estimation of the auto-covariance is considerably better (in terms of rrMSE) for FACE when the covariance of the cross products is not accounted for. Moreover, we see that FACE-STEP-1 yields roughly similar results to our approach and WHOLE. For TRI-CONSTR, TRI-CONSTR-W and WHOLE, the truncation level is correctly estimated to be two in all $200$ simulation runs. A higher number of eigenfunctions (three to four) is chosen for TRI in eight simulation runs of this setting which corresponds to our results in the application to the phonetics data. FACE and FACE-STEP-1 choose more than two (three to five) eigenfunctions in 190 and 198 simulation runs, respectively.

To sum up the results for the other ten settings of Scenario 1, we can say that over all settings and components TRI-CONSTR, TRI-CONSTR-W and WHOLE yield pretty similar rrMSEs with a tendency to a supremacy of TRI-CONSTR, especially in the sparse settings. TRI yields similar to worse results compared to our method and WHOLE. Especially for the error variance it yields up to 82\% higher median rrMSEs (in one of the sparse settings). The estimation quality of all methods differs between the dense and the sparse settings. Our approach yields relatively similar results within the dense settings and higher rrMSEs in the sparse settings. Moreover, TRI-CONSTR and TRI-CONSTR-W tend to perform better in the settings with complex eigenfunctions and favor smaller error variances (except for the estimation quality of the error variance itself). In contrast, FACE tends to perform better in the settings with simple eigenfunctions and favors larger error variances. Our approach and WHOLE always select the correct truncation level, except in the sparse settings, where for some simulation runs more eigenfunctions are selected. TRI tends to select more eigenfunctions. For all settings, FACE and FACE-STEP-1 have simulation runs in which more than two eigenfunctions are selected.

\begin{figure}[h!]
\begin{center}
\includegraphics[width=1\textwidth]{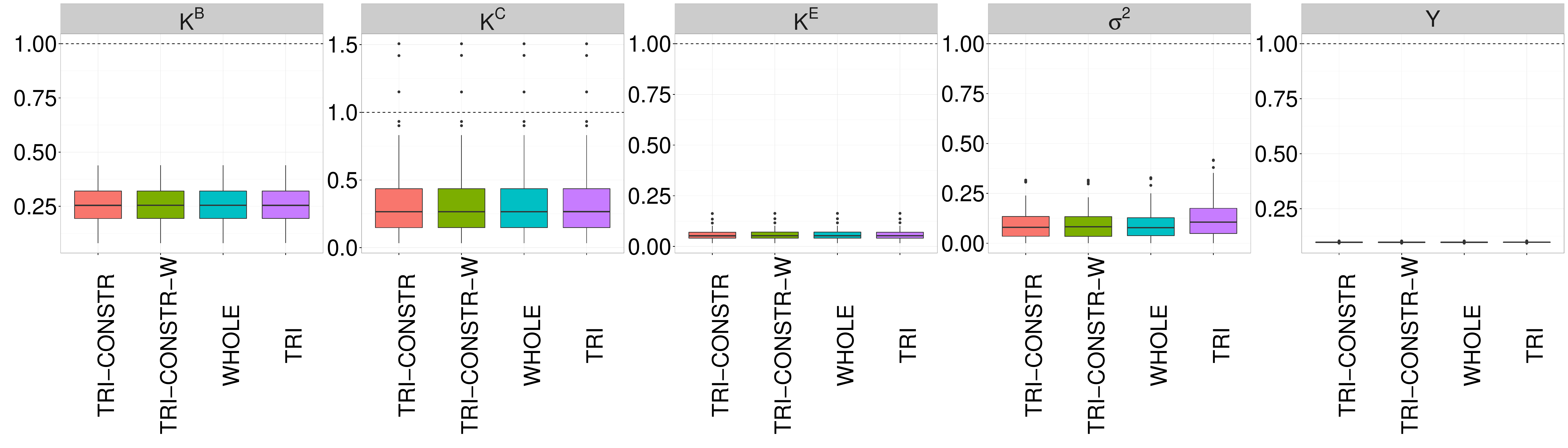}
\caption{Boxplots of the rrMSEs for the scenario with crossed fRIs (Scenario 2) for the smoothing methods being compared. Shown are boxplots of the rrMSEs for the three auto-covariances $K^B(t,t^\prime)$, $K^C(t,t^\prime)$, and $K^E(t,t^\prime)$ as well as for the error variance $\sigma^2$ and the curves $Y_i(t)$.}
\label{fig: boxplots rrMSEs crossed}
\end{center}
\end{figure}

In contrast to FACE and FACE-STEP-1, our approach is directly extendable to smoothing multiple auto-covariances simultaneously, which is required in Scenario 2. Figure \ref{fig: boxplots rrMSEs crossed} depicts the rrMSEs for the auto-covariances, the error variance and curves $Y_i(t)$ for Scenario 2. It shows that the three auto-covariances are estimated equally well for the four compared smoothing methods. For the error variance, however, TRI performs worse with a 32.5\% higher median rrMSE and a larger IQR which also results in slightly higher rrMSEs for the reconstructed curves. Over all, our methods perform very well considering the small number of levels for the fRIs $B$ (9 levels) and $C$ (16 levels). For all model components, TRI-CONSTR and TRI-CONSTR-W yield similar rrMSEs.
\pagebreak

\subsection{Computational efficiency}
\begin{figure}[h!]
\begin{center}
\includegraphics[width= 0.33\textwidth]{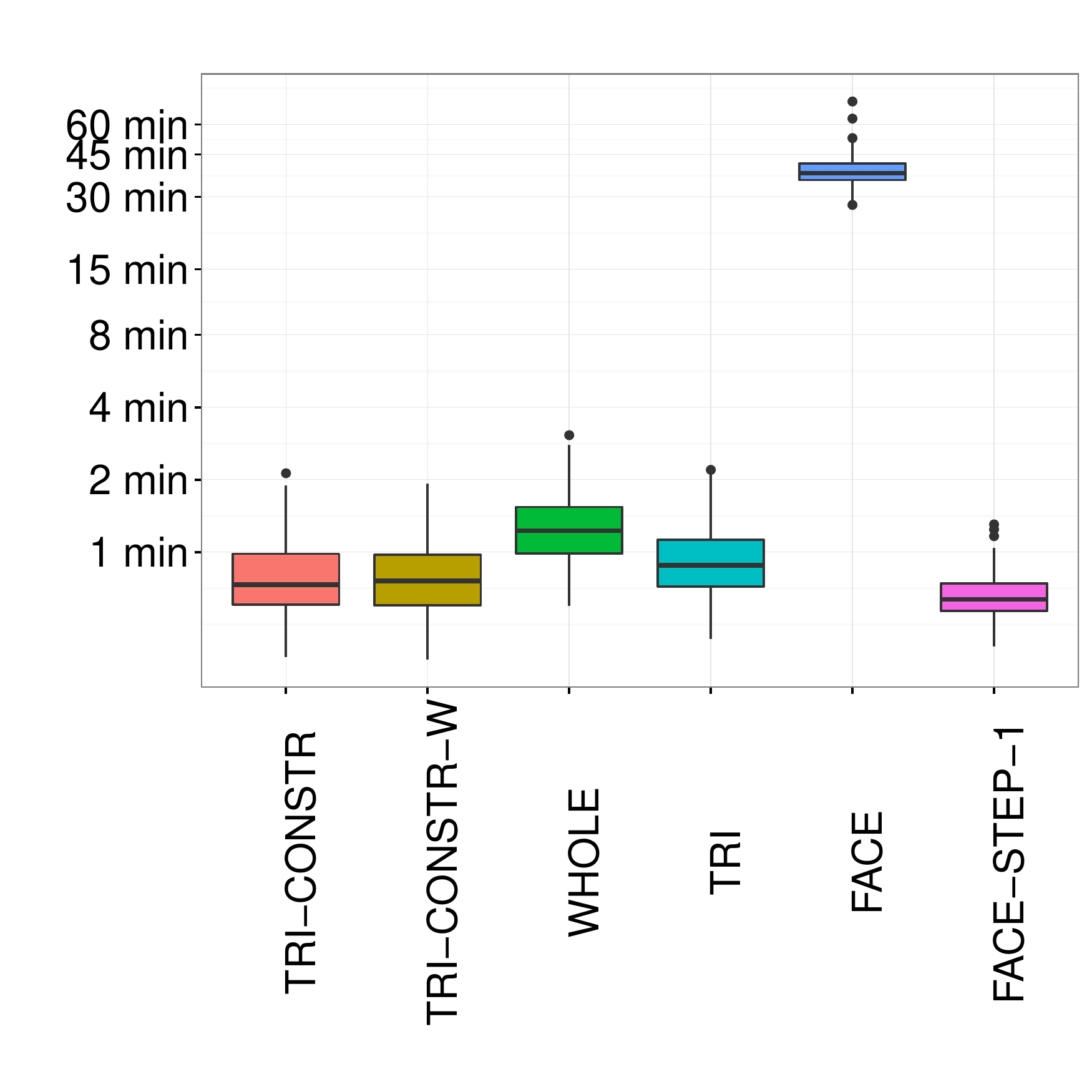}
\includegraphics[width= 0.33\textwidth]{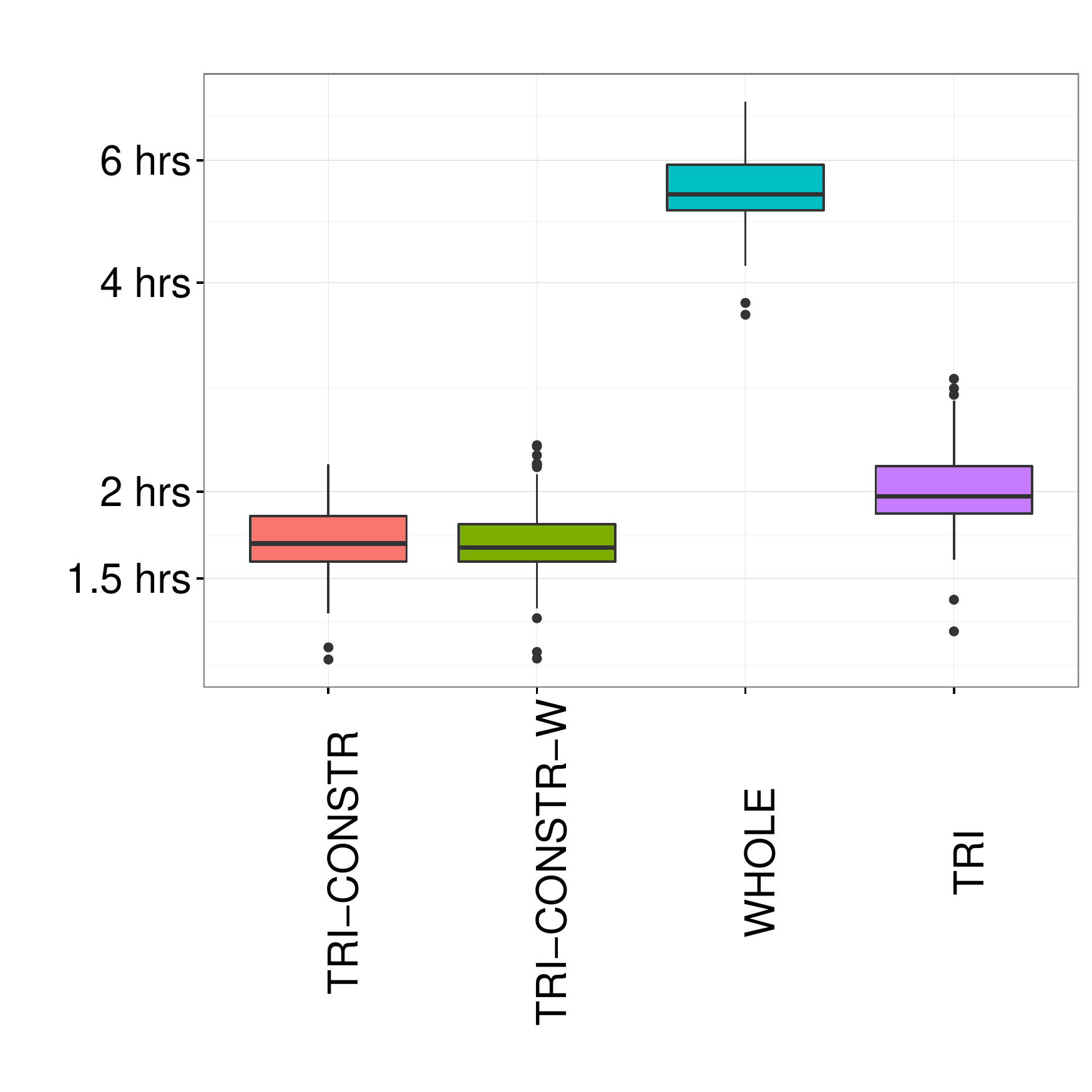}
\caption{Computation times (log10 scale at y-axis) for 200 simulation runs for all compared methods. Left: computation times for Setting 1 of the scenario with independent curves. Right: computation times for the scenario with crossed fRIs.}
\label{fig: boxplots times indep curves}
\end{center}
\end{figure}

Figure \ref{fig: boxplots times indep curves} shows computation times on a 64 Bit Linux platform with 660 Gb of RAM memory for the two settings discussed above. Our approach (with and without weights on the diagonal cross products) greatly speeds up the computation compared to WHOLE and also to TRI, especially in the scenario with crossed fRIs (right figure), where the computation times are longer and thus matter more. Note that in addition to accounting for the symmetry of covariances, we reduce the number of smoothing parameters to be estimated compared to WHOLE and TRI, for which two smoothing parameters are estimated for each auto-covariance using the Kronecker sum penalty implemented in \pkg{R}-package \pkg{mgcv}. In the scenario with independent curves (left figure), FACE is considerably slower than the other methods. This applies to all except the sparse settings, in which the computation times are extremely short anyway (max.~15 sec.). 
Note that variation for FACE and FACE-STEP-1 is considerably smaller as the smoothing parameter is chosen based on a fixed grid.

\section{Discussion and Outlook} \label{sec: discussion}
We have introduced a fast bivariate smoothing approach for symmetric surfaces which applies to a broad range of data situations. We focus on its application to estimate covariance functions in longitudinal data as well as multiple additive covariance functions in functional data with very general correlation structures. Our smoother can handle (possibly noisy) data sampled on a common, dense grid as well as irregularly or sparsely observed data, which are frequently encountered in practice. It extends the smooth methods of moments estimator of \cite{Cederbaum.2016} to more general correlation structures and additionally takes advantage of the symmetry of the sample covariances, which leads to considerably faster estimation requiring less memory. A symmetry constraint additionally ensures smoothness of the estimated covariance surfaces across the diagonal and further reduces computational costs. We show how our smoother can be applied as basis for FPCA, a key tool for dimension reduction in FDA, and demonstrate its practical relevance in a longitudinal data application and in an application to complexly correlated functional phonetics data. 

We provide software implementing our approach that builds on the established \pkg{R}-package \pkg{mgcv} \cite{Wood.2011} allowing for flexible extensions. Within this framework, we provide a novel constructor function for the estimation of smooth surfaces in additive models subject to the symmetry constraint. Our constructor can be applied as a modular component to general bivariate symmetric smoothing problems.

Simulation experiments (in Section \ref{sec: simulations}) show that the proposed method recovers the true functions very well and yields similar results as the estimation approach of \cite{Cederbaum.2016} while considerably speeding up the estimation and extending the range of possible model structures.

This work opens up a number of interesting directions for future research. A first direction concerns the working assumptions of the covariance estimation as additive varying coefficient model for the cross products (cp.~Section \ref{sec: covariance of cross products}). It would be interesting to investigate whether a suitable loss function for the cross products could be derived. Under the assumption of Gaussian responses, the cross products follow a product normal distribution, for whose PDF \cite{Nadarajah.2016} recently derived a closed-form expression based on the modified Bessel function. A second direction concerns the positive semi-definiteness of the covariance operator, which is not ensured in our approach and in most existing covariance smoothing approaches \citep[e.g.][]{Yao.2003, Hall.2008, Di.2009, Greven.2010}. Although \cite{Hall.2008} show that setting negative eigenvalues to zero improves the estimation quality and \cite{Yao.2003} demonstrate that this works well in practice, it could be desirable to ensure positive semi-definiteness in the estimation. \cite{Wu.2003} estimate a positive semi-definite covariance based on an auto-regressive model with regression coefficients corresponding to the components of a modified Cholesky decomposition of the covariance. Also the approach of \cite{Peng.2009} ensures positive semi-definiteness. It remains an open question, however, how these approaches could be extended to smoothing multiple additive covariances for functional data with complex correlation structures. 

\subsection*{Acknowledgements}
The authors thank Marianne Pouplier and Phil Hoole for supplying and explaining the phonetics data. All authors were supported by the Emmy Noether grant GR 3793/1-1 from the German Research Foundation.

\bibliographystyle{apalike-url}
\bibliography{references}

\appendix
\newpage
\section{Derivations} \label{appendix: proofs}
\renewcommand{\thesubsection}{\thesection.\arabic{subsection}}
\subsection{Derivation for the covariance of cross products in the general FLMM}
Let $X = \left(X_1,\ldots,X_p\right)^\top$ be a $p$-dimensional Gaussian random variable with zero mean and covariance $\mSigma$. Then, given $p\geq 4$, we can express the fourth moment of $X$ based on Isserlis' Theorem \citep{Isserlis.1918} as 
\bea  \label{eq: fourth moment}
\EV \left(X_i X_j X_k X_l\right) = \mSigma_{ij}\mSigma_{kl} + \mSigma_{ik}\mSigma_{jl} +\mSigma_{il}\mSigma_{jk},
\eea
where $\mSigma_{ij}$ is the covariance of $X_i$ and $X_j$.

Consider the covariance of the cross products of the centered functional responses in the general FLMM
\bea \label{eq: covariance of cross products general}
&&\Cov\left[\Ya \Yb,\Yc \Yd\right]  \\ \nonumber
 &=& \EV\left[\Ya \Yb \Yc \Yd \right]- \underbrace{\EV\left[\Ya \Yb\right]}_{\Cov\left[\Ya,\Yb\right]} \underbrace{\EV\left[
\Yc \Yd\right]}_{\Cov\left[\Yc,\Yd\right]}\\ \nonumber
&\overset{\eqref{eq: fourth moment}}{=}& \Cov\left[ \Ya \Yb\right] \Cov\left[\Yc,\Yd\right] \\ \nonumber
&+& \Cov\left[ \Ya \Yc\right] \Cov\left[\Yb,\Yd\right] \\ \nonumber
&+& \Cov\left[ \Ya \Yd\right] \Cov\left[\Yb,\Yc\right] \\ \nonumber
&-& \Cov\left[\Ya, \Yb\right] \Cov\left[\Yc,\Yd\right] \\ \nonumber
&=& \Cov\left[ \Ya \Yc\right] \Cov\left[\Yb,\Yd\right] \\ \nonumber
&+& \Cov\left[ \Ya \Yd\right] \Cov\left[\Yb,\Yc\right] \\ \nonumber
&=& \left\{\mz_i^\top \mK^U(t_{ij},t_{mo})\mz_m + \left[ K^E(t_{ij},t_{mo}) + \sigma^2\delta_{jo} \right] \delta_{im} \right\}\\\nonumber
&&\cdot  \left\{\mz_{i^\prime}^\top \mK^U(t_{i^\prime j^\prime },t_{m^\prime o^\prime})\mz_{m^\prime} + \left[ K^E(t_{i^\prime j^\prime},t_{m^\prime o^\prime}) + \sigma^2\delta_{j^\prime o^\prime} \right] \delta_{i^\prime m^\prime} \right\}\\\nonumber
&+& \left\{\mz_{i}^\top \mK^U(t_{ij},t_{m^\prime o^\prime})\mz_{m^\prime } + \left[ K^E(t_{ij},t_{m^\prime o^\prime}) + \sigma^2\delta_{j o^\prime} \right] \delta_{im^\prime} \right\}\\ \nonumber
&&\cdot  \left\{\mz_{i^\prime}^\top \mK^U(t_{i^\prime j^\prime },t_{mo})\mz_{m} + \left[ K^E(t_{i^\prime j^\prime},t_{mo}) + \sigma^2\delta_{j^\prime o} \right] \delta_{i^\prime m} \right\}.\qed
\eea

\subsection{Simplification of the covariance of cross products for crossed fRIs}
For the special case of an FLMM with two crossed fRIs as in Section \ref{sec: crossed model}, the covariance of the cross products in equation \eqref{eq: covariance of cross products general} simplifies to
\bea \nonumber
&&\Cov\left[\Ya \Yb,\Yc \Yd\right]  \\ \nonumber
&=& \lbrace K^B(t_{ij},t_{mo}) \delta_{B(i)B(m)} + K^C(t_{ij},t_{mo})\delta_{C(i)C(m)} + \left[K^E(t_{ij},t_{mo}) + \sigma^2 \delta_{jo} \right] \delta_{im} \rbrace \\ \nonumber
&\cdot & \lbrace K^B(t_{i^\prime j^\prime},t_{m^\prime o^\prime})  \delta_{B(i^\prime)B(m^\prime)} + K^C(t_{i^\prime j^\prime},t_{m^\prime o^\prime})\delta_{C(i^\prime)C(m^\prime)}+ \left[K^E(t_{i^\prime j^\prime},t_{m^\prime o^\prime}) + \sigma^2 \delta_{j^\prime o^\prime} \right] \delta_{i^\prime m^\prime} \rbrace \\ \nonumber
&+& \lbrace K^B(t_{ij},t_{m^\prime o^\prime})  \delta_{B(i)B(m^\prime)} + K^C(t_{ij},t_{m^\prime o^\prime}) \delta_{C(i)C(m^\prime)} + \left[K^E(t_{ij},t_{m^\prime o^\prime}) + \sigma^2 \delta_{jo^\prime} \right] \delta_{im^\prime} \rbrace \\ \nonumber
&\cdot & \lbrace K^B(t_{i^\prime j^\prime},t_{mo}) \delta_{B(i^\prime)B(m)} + K^C(t_{i^\prime j^\prime},t_{mo}) \delta_{C(i^\prime)C(m)}  + \left[K^E(t_{i^\prime j^\prime},t_{mo}) + \sigma^2 \delta_{j^\prime o} \right] \delta_{i^\prime m} \rbrace,
\eea
where $\delta_{B(i)B(m)}$ and $ \delta_{C(i)C(m)}$ take value one when the two curves $i$ and $m$ belong to the same level of the respective grouping variable and zero otherwise.

\newpage
\section{Supplementary details on the estimation and implementation} \label{appendix: supplementary details on estimation}

\subsection{(Additive) varying coefficient model using tensor product B-splines}
Let $\otimes$ and $\cdot$ denote the Kronecker product and the Hadamard (pointwise) product, respectively. 

\paragraph{Model with independent curves}
Using tensor product B-splines yields the following form of Model \eqref{eq: indep additive model}
\bea \nonumber
\EV\left[\mC\right] 
&=& \left[\left(\mB^E_t \otimes {\mathds{1}_{F^E}}^\top\right) \cdot \left({\mathds{1}_{F^E}}^\top\otimes \mB^E_{t^\prime}\right)| \mdelta^\varepsilon\right] \left({\mtheta^{E}}^\top,\sigma^2\right)^\top \\ \nonumber
&=& \left[\mM^E| \mdelta^\varepsilon \right] \left({\mtheta^{E}}^\top,\sigma^2\right)^\top = \mM \malpha,
\eea
where $\mB^E_t$, $\mB^E_{t^\prime}$ are the $\mathcal{C}\times F^E$ marginal spline design matrices that contain the evaluated spline basis functions for the directions $t$ and $t^\prime$, respectively. $\mB^E_t$ and $\mB^E_{t^\prime}$ contain identical but permuted rows. $\mathds{1}_{F^E} = \left(1,\ldots,1\right)^\top$ is of length $F^E$, the number of marginal basis functions in each direction. The bivariate additive model for the reduced response vector is obtained by replacing the marginal spline design matrices by the reduced $\mathcal{C}^\Delta \times F^{E}$ matrices $\mB^{E \Delta}_t$, $\mB^{E\Delta}_{t^\prime}$ and the index vector by the reduced index vector $\mdelta^{\varepsilon \Delta}$. Note that in the model with independent curves, the bivariate spline design matrix $\mB^{E \Delta}$ corresponds to the design matrix $\mM^{E \Delta}$ as products are only computed on the same curves and thus the indicator matrix $\mQ^{E \Delta}$ reduces to an all-ones matrix.

\paragraph{General FLMM} 
For each $g=1,\ldots,G$, and each $s,s^\prime=1,\ldots\rho^{U_g}$, let $\mB^{U_g\Delta}_{ss^\prime, t}$ and $\mB^{U_g\Delta}_{ss^\prime, t^\prime}$ denote the marginal spline design matrices of dimensions $\mathcal{C}^\Delta \times F^{U_g}_{ss^\prime,t}$ and $\mathcal{C}^\Delta \times F^{U_g}_{ss^\prime,t^\prime}$, respectively. Due to the symmetry of the covariances $\mK^{U_g}(t,t^\prime)$, we assume $F^{U_g}_{ss^\prime,t} = F^{U_g}_{s^\prime s,t^\prime}$ and $F^{U_g}_{ss^\prime,t^\prime} = F^{U_g}_{s^\prime s,t}$, respectively. Then, the bivariate spline design matrices $\mB^{U_g\Delta}_{ss^\prime}$ are given by 
\bea \nonumber
\mB^{U_g\Delta}_{ss^\prime} &=& \left(\mB^{U_g\Delta}_{ss^\prime,t} \otimes {\mathds{1}_{F^{U_g}_{ss^\prime, t^\prime}}}^\top\right)\cdot \left({\mathds{1}_{F^{U_g}_{ss^\prime,t}}}^\top \otimes \mB^{U_g \Delta}_{ss^\prime ,t^\prime}\right).
\eea

The submatrices $\mM^{U_g^\Delta}_{ss^\prime}$ corresponding to the covariances $K^{U_g}_{ss^\prime}(t,t^\prime)$ are given as $\mM^{U_g \Delta}_{ss^\prime} = \mQ^{U_g^\Delta}_{ss^\prime} \cdot \mB^{U_g^\Delta}_{ss^\prime}$, where $\mQ^{U_g \Delta}_{ss^\prime}$ are $\mathcal{C}^\Delta \times F^{U_g}_{ss^\prime,t}F^{U_g}_{ss^\prime ,t^\prime}$ matrices with entries $\delta_{U_g(i)U_g(i^\prime)} \cdot \omega^{U_g}_{is}\omega^{U_g}_{i^\prime s^\prime}$, $i\leq i^\prime =1,\ldots,n$. $\delta_{U_g(i)U_g(i^\prime)}$ takes value one if the two curves $i$ and $i^\prime$ are of the same level of grouping variable $g$ and zero otherwise. The columns of $\mQ^{U_g \Delta}_{ss^\prime}$ are all identical and contain the suitably sorted and repeated entries. Suitably sorted and repeated in this context means that the sorting corresponds to the sorting in $\mC^{\Delta}$ and that the entries $\delta_{U_g(i)U_g(i^\prime)} \cdot \omega^{U_g}_{is}\omega^{U_g}_{i^\prime s^\prime}$ are repeated for all considered combinations of observation points $t_{ij}\leq t_{i^\prime j^\prime}$.

In analogy, matrix $\mQ^{E \Delta}$ is a $\mathcal{C}^\Delta \times \left(F^E\right)^2$ matrix with identical columns consisting of suitably sorted and repeated indicators $\delta_{E(i)E(i^\prime)}$, which take value one if the two points in the cross products belong to the same curve and zero otherwise.

\subsection{Examples for the general FLMM}
In the following, we provide examples for the specification of $\mz_i^\top \mU(t_{ij})$ in Model \eqref{eq: general model} yielding an FLMM with hierarchical (e.g.~subjects in groups) and crossed (e.g.~speakers and target words as in our phonetics application in Section (\ref{sec: phonetics})) functional random effects, respectively. 

Consider for simplicity the case of two independent grouping variables ($G=2$), with an fRI for the first grouping variable and an fRI and a functional random slope (in variable $\omega$) for the second grouping variable ($\rho^{U_1}=1$, $\rho^{U_1}=2$). Further assume that there are $n_i$ observations (curves) for each level of the second grouping variable, $i=1,\ldots,L^{U_2}$. We assume in the following that the fRIs and the functional random slopes are dependent. For independent effects, we would split them up and denote $G=3$. Let $\omega_i$ denote the value of variable $\omega$ for curve $i$ and $\delta_{U_g(i)l}$ takes value one if curve $i$ belongs to the $l$th level of grouping variable $g$ and zero otherwise. Then, $\mz_i^\top \mU(t_{ij})$, $i=1,\ldots,n$, is given by
\bea \nonumber
\footnotesize{
\mz_{i}^\top \mU(t_{ij}) = \left(\delta_{i11}^U,\ldots,\delta_{i1L^{U_1}}^U, \delta_{i21}^U,\delta_{i21}^U \omega_i,\ldots, \delta_{i2L^{U_2}}^U,\delta_{i2L^{U_2}}^U \omega_i\right) 
\left( \begin{array}{*{8}{c}}
U_{111}(t_{ij})\\
\vdots\\
U_{1L^{U_1}1}(t_{ij})\\
\vdots\\
U_{211}(t_{ij})\\
U_{212}(t_{ij})\\
\vdots\\
U_{2L^{U_2}1}(t_{ij})\\
U_{2L^{U_2}2}(t_{ij})
\end{array} \right).}
\eea
The hierarchical and the crossed functional functional random effects differ in the form of $\mz_i$. Let for simplicity assume that there are two and four levels of the two grouping variables ($L^{U_1} =2$, $L^{U_2}=4$), respectively. Then, in total $q=\sum_{g=1}^G \rho^{U_g}L^{U_g} = 10$ functional random effects are specified (apart from the smooth error $E_i(t)$).

In the case of hierarchical functional random effects (e.g.~four subjects in two groups), the $n\times q$ matrix  $\mZ$ consisting of the $\mz_i$, $i=1,\ldots,n$, then has the form
\bea \nonumber
\footnotesize{
\mZ = \left[\begin{array}{*{1}{c}}
\mz_1^\top\\
\vdots\\
\mz_n^\top \end{array}\right]
 = \left[\begin{array}{*{11}{c}}
1 &  & 1 & && &\vline &w_1&&\\
\vline  &  & \vline&& & &\vline & \vdots&& \\
\vline & & 1 & && &\vline & w_{n_1} &&\\
\vline & &  & 1 && & \vline & &w_{n_1+1}&&\\
\vline & & & \vline&& &\vline & &\vdots &&\\
1 &  &  & 1 && &\vline & &w_{n_1+n_2}&&\\
 & 1 &  &  &1& &\vline & & & w_{n_1+n_2+1}\\
 & \vline &&&\vline  &  &\vline & & &\vdots\\
 & \vline & & &1& &\vline & && w_{n_1+n_2+n_3}\\
 & \vline &  & &&1 &\vline &&&& w_{n_1+n_2+n_3+1} \\
 & \vline & & && \vline &\vline &&&& \vdots\\
 & 1 &  &  &&1 &\vline &&&& w_{n}
\end{array}\right],}
\eea
where we implicitly assume for ease of presentation that there are two subjects in each group.

For crossed functional random effects (e.g.~two speakers and four words), we assume for simplicity that half of the $n_i$ curves belong to the first and the other to the second level of the first grouping variable. Then, matrix $\mZ$ is given by
\bea \nonumber
\footnotesize{
\mZ = \left[\begin{array}{*{1}{c}}
\mz_1^\top\\
\vdots\\
\mz_n^\top \end{array}\right]
 = \left[\begin{array}{*{11}{c}}
1 &  & 1 &  && &\vline & w_1\\
\vline  &  & \vline & &&&\vline& \vdots \\
\vline & & 1 &  && &\vline &  w_{\frac{n_1}{2}}\\
\vline & &  & 1 && &\vline &&  w_{\frac{n_1}{2}+1}\\
\vline & & & \vline &&&\vline & & \vdots\\
\vline & &  & 1 && &\vline &&  w_{\frac{n_1}{2}+\frac{n_2}{2}}\\
\vline & &  &  &1& &\vline && &w_{\frac{n_1}{2}+\frac{n_2}{2}+1}\\
\vline & & & &\vline &&\vline &&& \vdots\\
\vline & &  &  &1& &\vline &&& w_{\frac{n_1}{2}+\frac{n_2}{2}+\frac{n_3}{2}}\\
\vline &  &  &  &&1 &\vline &&&&  w_{\frac{n_1}{2}+\frac{n_2}{2}+\frac{n_3}{2}+1}\\
\vline & &  &  & &\vline &\vline &&&& \vdots\\
1 &  &  &  &&1  &\vline &&&& w_{\frac{n}{2}}\\
 & 1 & 1 &  && &\vline &  w_{\frac{n}{2}+1}\\
& \vline    & \vline & &&&\vline & \vdots\\
&\vline & 1 &  &&&\vline  & w_{\frac{n}{2}+\frac{n_1}{2}}\\
&\vline  &  & 1 && &\vline && w_{\frac{n}{2}+\frac{n_1}{2}+1}\\
&\vline  & & \vline &&&\vline && \vdots\\
&\vline  &  & 1 && &\vline && w_{\frac{n}{2}+\frac{n_1}{2} + \frac{n_2}{2}}\\
&\vline  &  &  &1&& \vline &&& w_{\frac{n}{2}+\frac{n_1}{2} + \frac{n_2}{2}+1}\\
&\vline  & & &\vline &&\vline &&& \vdots \\
&\vline  &  &  &1& &\vline &&& w_{\frac{n}{2}+\frac{n_1}{2} + \frac{n_2}{2} + \frac{n_3}{2}} \\
&\vline   &  &  &&1&\vline &&&& w_{\frac{n}{2}+\frac{n_1}{2} + \frac{n_2}{2}+ \frac{n_3}{2} +1} \\
&\vline  &  &  & &\vline &\vline &&&& \vdots \\
 &1  &  &  &&1 &\vline &&&& w_n\\
\end{array}\right].}
\eea

\subsection{Form of the constraint matrix}
For each $g=1,\ldots,G$, the constraint matrix $\mW^{U_g}$ is a block matrix consisting of $\left(\rho^{U_g}\right)^2 \times \frac{\left(\rho^{U_g}\right)^2+1}{2}$ blocks, most of which are zero. The rows and columns of $\mW^{U_g}$ are sorted as in matrix $\mM^{U_g \Delta}$ and in the reduced matrix $\mM^{U_g \Delta r}$, respectively. The non-zero blocks can be divided into two groups: blocks corresponding to the auto-covariances $K^{U_g}_{ss}(t,t^\prime)$, $s=1,\ldots,\rho^{U_g}$, and blocks corresponding to the cross-covariances $K^{U_g}_{ss^\prime}(t,t^\prime)$, with $s< s^\prime$. 

Let $\left(F^{U_g}_{ss}\right)^2$ denote the number of spline basis functions used for smoothing the auto-covariance $K^{U_g}_{ss}(t,t^\prime)$. The blocks for the auto-covariances are of the same form as $\mW^E$ and given by the $\left(F^{U_g}_{ss}\right)^2 \times F^{U_g}_{ss}(F^{U_g}_{ss} +1)/2$ matrices
\bea \nonumber
 \left[\begin{array}{*{2}{l}}
\mI_{\frac{F^{U_g}_{ss}(F^{U_g}_{ss}-1)}{2}} & \bm{0}_{\frac{F^{U_g}_{ss}(F^{U_g}_{ss}-1)}{2}\times F^{U_g}_{ss}} \\
\bm{0}_{F^{U_g}_{ss}\times \frac{F^{U_g}_{ss}(F^{U_g}_{ss}-1)}{2}} & \mI_{F^{U_g}_{ss}} \\
\mI_{\frac{F^{U_g}_{ss}(F^{U_g}_{ss}-1)}{2}} & \bm{0}_{\frac{F^{U_g}_{ss}(F^{U_g}_{ss}-1)}{2} \times F^{U_g}_{ss}}
\end{array}\right], 
\eea
where $\mI_{x}$ is an identity matrix of dimension $x$ and $\bm{0}_{x\times y}$ is a null matrix of dimension $x\times y$. 

Consider for simplicity the case of bivariate tensor product spline bases, where we can denote $F^{U_g}_{ss^\prime, t}$ and $F^{U_g}_{ss^\prime, t^\prime}$ the number of marginal spline basis functions for smoothing the cross-covariance $K^{U_g}_{ss^\prime}(t,t^\prime)$, $s<s^\prime$, in direction $t$ and $t^\prime$, respectively. Due to the symmetry, we have $F^{U_g}_{ss^\prime,t} = F^{U_g}_{s^\prime s,t^\prime}$ and $F^{U_g}_{ss^\prime,t^\prime} = F^{U_g}_{s^\prime s,t}$.
Let $F^{U_g}_{ss^\prime, b=b^\prime}$ denote the number of coefficients on the diagonal in $\mTheta^{U_g}_{ss^\prime}$, which corresponds to the minimum of $F^{U_g}_{ss^\prime, t}$ and $F^{U_g}_{ss^\prime, t^\prime}$ and denote the number of coefficients below and above the diagonal as $F^{U_g}_{ss^\prime, b<b^\prime} \coloneqq \sum_{i=1}^{F^{U_g}_{ss^\prime,b=b^\prime}}\left(F^{U_g}_{ss^\prime,t^\prime}-i\right)$ and $F^{U_g}_{ss^\prime, b>b^\prime} \coloneqq \sum_{i=1}^{F^{U_g}_{ss^\prime,b=b^\prime}}\left(F^{U_g}_{ss^\prime,t}-i\right)$, respectively. The blocks for the cross-covariances are then $F^{U_g}_{ss^\prime,t}F^{U_g}_{ss^\prime,t^\prime} \times F^{U_g}_{ss^\prime,t}F^{U_g}_{ss^\prime,t^\prime}$ diagonal block matrices of the form
\bea \nonumber
 \left[\begin{array}{*{3}{l}}
\mI_{F^{U_g}_{ss^\prime, b<b^\prime}} & \bm{0}_{F^{U_g}_{ss^\prime, b<b^\prime} \times F^{U_g}_{ss^\prime b=b^\prime}} & \bm{0}_{F^{U_g}_{ss^\prime, b<b^\prime} \times F^{U_g}_{ss^\prime, b>b^\prime}}\\
\bm{0}_{F^{U_g}_{ss^\prime, b=b^\prime} \times F^{U_g}_{ss^\prime, b<b^\prime}} & \mI_{F^{U_g}_{ss^\prime, b=b^\prime}} & \bm{0}_{F^{U_g}_{ss^\prime, b=b^\prime}\times F^{U_g}_{ss^\prime, b>b^\prime}}\\
\bm{0}_{F^{U_g}_{ss^\prime, b>b^\prime}\times F^{U_g}_{ss^\prime, b<b^\prime}} & \bm{0}_{F^{U_g}_{ss^\prime, b>b^\prime}\times F^{U_g}_{ss^\prime, b=b^\prime}} & \mI_{F^{U_g}_{ss^\prime, b>b^\prime}}
\end{array}\right], 
\eea
when the respective rows correspond to $s<s^\prime$ and $F^{U_g}_{ss^\prime,t}F^{U_g}_{ss^\prime,t^\prime} \times F^{U_g}_{ss^\prime,t}F^{U_g}_{ss^\prime,t^\prime}$ anti-diagonal block matrices of the form 
\bea \nonumber
 \left[\begin{array}{*{3}{l}}
\bm{0}_{F^{U_g}_{ss^\prime, b<b^\prime}\times F^{U_g}_{ss^\prime, b<b^\prime}}& \bm{0}_{F^{U_g}_{ss^\prime, b<b^\prime}\times F^{U_g}_{ss^\prime, b=b^\prime}} & \mI_{F^{U_g}_{ss^\prime, b>b^\prime}}\\
\bm{0}_{F^{U_g}_{ss^\prime, b=b^\prime}\times F^{U_g}_{ss^\prime, b<b^\prime}} & \mI_{F^{U_g}_{ss^\prime, b=b^\prime}} & \bm{0}_{F^{U_g}_{ss^\prime, b=b^\prime}\times F^{U_g}_{ss^\prime, b>b^\prime}}\\
\mI_{F^{U_g}_{ss^\prime, b<b^\prime}} & \bm{0}_{F^{U_g}_{ss^\prime, b<b^\prime}\times F^{U_g}_{ss^\prime, b=b^\prime}} & \bm{0}_{F^{U_g}_{ss^\prime, b<b^\prime}\times F^{U_g}_{ss^\prime, b>b^\prime}}
\end{array}\right], 
\eea
when the respective rows correspond to $s>s^\prime$. 

\paragraph{Example with two random effects}
Consider a grouping variable $g$ with $\rho^{U_g}=2$ components. Omitting the dimensions of the submatrices for better readability, the constraint matrix $\mW^{U_g}$ is given by
\begin{scriptsize}
\bea \nonumber
\mW^{U_g} =  
\begin{matrix}
(1,1)\\
\phantom{()}\\
\phantom{()}\\
(1,2)\\
\phantom{()}\\
\phantom{()}\\
(2,1)\\
\phantom{()}\\
\phantom{()}\\
(2,2)\\
\end{matrix}
\left[\begin{array}{*{9}{l}}
\mI & \bm{0} &\vline &&&& \vline\\
\bm{0} & \mI &\vline&&&& \vline\\
\mI & \bm{0} &\vline&&&& \vline\\
\hline 
&&\vline &  \mI & \bm{0} & \bm{0} &\vline\\
&&\vline &  \bm{0} & \mI & \bm{0}&\vline\\
&&\vline &  \bm{0} & \bm{0} & \mI&\vline\\
\hline
&&\vline & \bm{0} & \bm{0} & \mI &\vline\\
&&\vline   & \bm{0} & \mI & \bm{0}&\vline\\
&&\vline   & \mI & \bm{0}& \bm{0} &\vline\\
\hline
&&\vline   & & & & \vline &\mI & \bm{0}\\
&&\vline   & & & &  \vline &\bm{0} & \mI \\
&&\vline   & & & &\vline &\mI & \bm{0} \\
\end{array}\right],
\eea
\end{scriptsize}

yielding the reduced design matrix $\mM^{U_g\Delta r}$
\begin{scriptsize}
\bea \nonumber
\left[\underbrace{\mM^{U_g\Delta}_{11,b<b^\prime}  +\mM^{U_g\Delta}_{11,b>b^\prime} | \mM^{U_g\Delta}_{11,b=b^\prime}}_{s=s^\prime=1}|\underbrace{\mM^{U_g\Delta}_{12,b<b^\prime}  +\mM^{U_g\Delta}_{21,b>b^\prime} | \mM^{U_g\Delta}_{12,b=b^\prime} + \mM^{U_g\Delta}_{21,b=b^\prime}|
\mM^{U_g\Delta}_{12,b>b^\prime}  +\mM^{U_g\Delta}_{21,b<b^\prime}}_{s< s^\prime (s=1,s^\prime=2)} |\underbrace{\mM^{U_g\Delta}_{22,b<b^\prime}  +\mM^{U_g\Delta}_{22,b>b^\prime} | \mM^{U_g\Delta}_{22,b=b^\prime}}_{s=s^\prime = 2} \right].
\eea
\end{scriptsize}

\paragraph{Example with three random effects} 
Analogously, for a grouping variable $g$ with $\rho^{U_g}=3$ components the constraint matrix $\mW^{U_g}$ is given by
\begin{scriptsize}
\bea \nonumber
\mW^{U_g} =  \begin{matrix}
(1,1)\\
\phantom{()}\\
\phantom{()}\\
(1,2)\\
\phantom{()}\\
\phantom{()}\\
(1,3)\\
\phantom{()}\\
\phantom{()}\\
(2,1)\\
\phantom{()}\\
\phantom{()}\\
(2,2)\\
\phantom{()}\\
\phantom{()}\\
(2,3)\\
\phantom{()}\\
\phantom{()}\\
(3,1)\\
\phantom{()}\\
\phantom{()}\\
(3,2)\\
\phantom{()}\\
\phantom{()}\\
(3,3)\\
\end{matrix}
\left[\begin{array}{*{21}{l}}
\mI & \bm{0} &\vline &&&& \vline &&&&\vline &&&\vline&&&&\vline\\
\bm{0} & \mI &\vline&&&& \vline&&&&\vline&&&\vline &&&&\vline\\
\mI & \bm{0} &\vline&&&& \vline&&&&\vline&&&\vline &&&&\vline\\
\hline 
&&\vline &  \mI & \bm{0} & \bm{0} &\vline&&&&\vline&&&\vline&&&&\vline\\
&&\vline &  \bm{0} & \mI & \bm{0}&\vline&&&&\vline &&&\vline &&&&\vline\\
&&\vline &  \bm{0} & \bm{0} & \mI&\vline&&&&\vline &&&\vline&&&&\vline\\
\hline 
&&\vline&&&&\vline &  \mI & \bm{0} & \bm{0} &\vline&&&\vline&&&&\vline\\
&&\vline&&&&\vline &  \bm{0} & \mI & \bm{0}&\vline&&&\vline&&&&\vline\\
&&\vline&&&&\vline &  \bm{0} & \bm{0} & \mI&\vline&&&\vline&&&&\vline\\
\hline
&&\vline &  \bm{0} & \bm{0} & \mI &\vline&&&&\vline &&&\vline&&&&\vline\\
&&\vline &  \bm{0} & \mI & \bm{0}&\vline&&&&\vline&&&\vline&&&&\vline\\
&&\vline &  \mI & \bm{0} & \bm{0}&\vline&&&&\vline&&&\vline&&&&\vline\\
\hline 
&&\vline &&&&\vline &&&&\vline &\mI & \bm{0} &\vline &&&& \vline&&&\\
&&\vline &&&&\vline &&&&\vline &\bm{0} & \mI &\vline&&&& \vline&&&\\
&&\vline &&&&\vline &&&& \vline &\mI & \bm{0} &\vline&&&& \vline&&&\\
\hline
&&\vline &&&&\vline  &&&&\vline &&&\vline &\mI & \bm{0} & \bm{0} & \vline\\
&&\vline &&&&\vline&&&&\vline &&&\vline &\bm{0} & \mI & \bm{0}& \vline\\
&&\vline &&&&\vline  &&&& \vline &&&\vline & \bm{0} & \bm{0} & \mI& \vline\\
\hline
&&\vline&&&&\vline &  \bm{0}& \bm{0} & \mI &\vline&&&\vline&&&& \vline\\
&&\vline&&&&\vline &  \bm{0} & \mI & \bm{0}&\vline&&&\vline &&&&\vline\\
&&\vline&&&&\vline &  \mI & \bm{0} & \bm{0}&\vline&&&\vline &&&&\vline\\
\hline
&&\vline&&&&\vline &&&& \vline& &&\vline& \bm{0}& \bm{0} & \mI &\vline&&\\
&&\vline&&&&\vline &&&& \vline& &&\vline& \bm{0} & \mI & \bm{0}&\vline&&\\
&&\vline&&&&\vline & &&& \vline&&&\vline& \mI & \bm{0} & \bm{0}&\vline&&\\
\hline
&&\vline &&&&\vline &&&&\vline&&&\vline&&&&\vline &\mI & \bm{0}\\
&&\vline &&&&\vline &&&&\vline&&&\vline&&&&\vline &\bm{0} & \mI\\
&&\vline &&&&\vline &&&&\vline&&&\vline&&&& \vline &\mI & \bm{0}
\end{array}\right].
\eea
\end{scriptsize}
\pagebreak

\section{Supplementary application details and results} \label{appendix: supplementary application details and results}
\subsection{CD4 cell count data} \label{sec: cd4}
\begin{figure}[h]
\begin{center}
\includegraphics[width=0.4\textwidth]{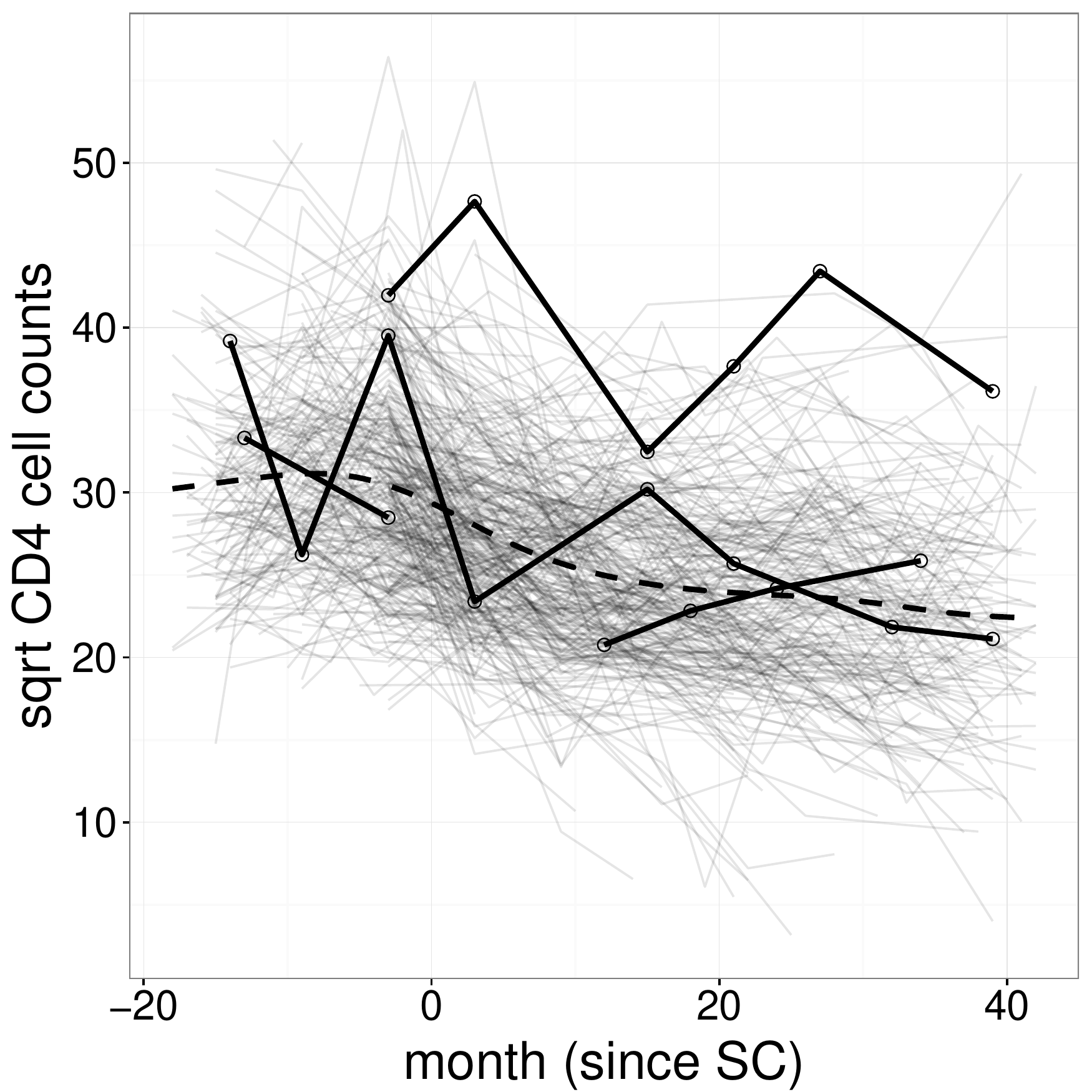}
\caption{Square root of observed CD4 cell count trajectories plotted against the months since SC. Shown are the trajectories of 366 HIV infected subjects. Some trajectories are highlighted for better display and an estimated smooth mean function (dashed) is shown.}
\label{fig: cd4}
\end{center}
\end{figure}

In AIDS research, the CD4 cell counts as a function of time since seroconversion (SC) -- the time at which HIV becomes detectable -- often serve as a longitudinally measured biomarker which provides insight into the progression of the disease. As the virus destroys the CD4 cells, a decreasing number of CD4 cells indicates a progress of the disease. The considered data set is part of the Multicenter AIDS Cohort Study \citep[MACS;][]{Kaslow.1987}. It contains the CD4 cell count trajectories of 366 HIV infected subjects collected from month -18 to month 42 since SC. Measurements were taken at roughly semi-annual visits yielding a total of 1888 CD4 cell counts per milliliter of blood, with between 1 to 11 counts per subject and a median of 5. 
To reduce skewness, we base our analysis on the square root of the CD4 cell counts, which are depicted in Figure \ref{fig: cd4}, with some trajectories highlighted for better display and an estimated overall mean function. We can see that on average, the CD4 cell counts are decreasing over time.
The data are available in \pkg{R}-package \pkg{refund} \citep{refund} and are further described in \cite{Goldsmith.2013}. Similar data from this study were previously analyzed in e.g.~\cite{Diggle.2002}, \cite{Yao.2005}, and \cite{Peng.2009}. 

We fit Model \eqref{eq: indep model} with only one fRI for each curve and an overall mean $\mu(t)$. In order to predict the continuous subject-specific trajectories with only few observations per subject available, we perform an FPCA based on our fast covariance smoothing approach (TRI-CONSTR and TRI-CONSTR-W with weights on the diagonal cross products). We demonstrate the similarity to the computationally less efficient approach proposed in \cite{Cederbaum.2016} (WHOLE), in which all cross products enter the estimation. Moreover, we show that boundary effects occur on the diagonal when only the triangular surface is estimated without a symmetry constraint (TRI). 
We compare our results to those obtained from applying the covariance smoothing approach proposed by \cite{Xiao.2016b} (FACE) using \pkg{R}-function \pkg{face.sparse} in package \pkg{face} \citep{FACE}. Moreover, we compare with FACE-STEP-1, a modification of FACE, in which the covariance of the cross products is not accounted for and thus only the first step of the three-step procedure is performed. 

We use 13 cubic B-spline basis functions for the estimation of the mean function and as marginal bases for the estimation of the auto-covariance surface using tensor products. To avoid over-fitting, we add a second order difference penalty. For our approach, we use the Kronecker sum penalty (cf.~Section \ref{sec: implementation}). We use equidistant knots in function \pkg{face.sparse} instead of the default (quantile based knots) which would require an adapted penalty that is not implemented.  
The equidistant grid, on which the mean and the auto-covariance are evaluated, is of length $D=100$, with values between -18 and 42. Note that so far function \pkg{face.sparse} assumes that the function argument takes values in the unit interval. We thus transformed the function argument and re-transformed the results to the original interval $[-18,42]$ after the estimation. Note that in order to ensure orthonormality with respect to the $L^2$-scalar product, we rescale the eigenfunctions and accordingly the eigenvalues after re-transforming the function argument. We truncate the number of eigenfunctions using a pre-specified proportion of explained variance of $L=0.99$. Note that we use the proportion of explained variance in the observed trajectories, whereas \cite{FACE} use that in $E_i(t)$.

\begin{figure}[h!]
\begin{center}
\begin{tabular}{cccc}
TRI-CONSTR & TRI-CONSTR-W & WHOLE & TRI \\[-2ex]
\includegraphics[width=0.23\textwidth, page = 1]{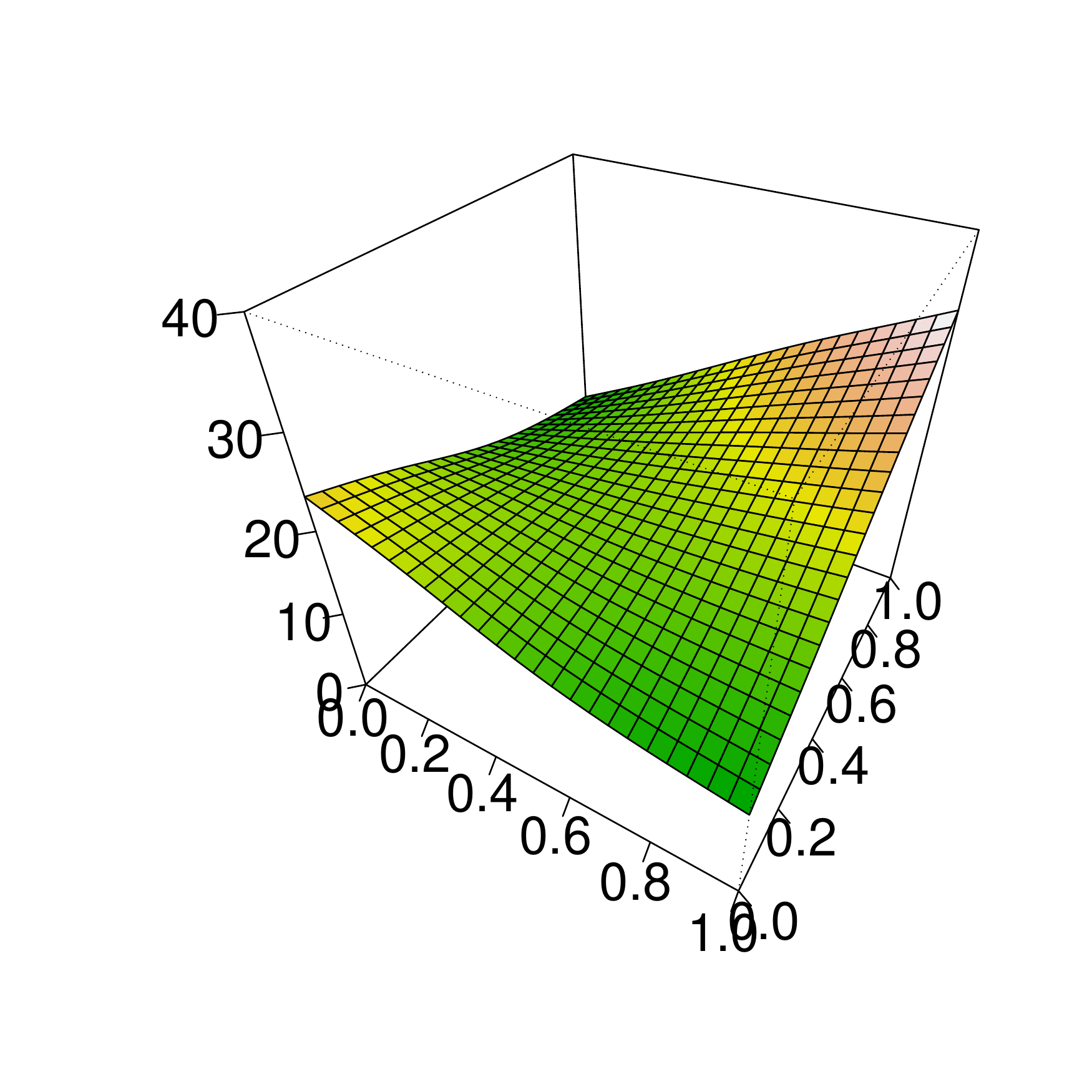} &
\includegraphics[width=0.23\textwidth, page = 2]{cov_estimated_persp_image_26_Aug_26_Aug_26_Aug_all_covs_cov_paper.pdf} &
\includegraphics[width=0.23\textwidth, page = 3]{cov_estimated_persp_image_26_Aug_26_Aug_26_Aug_all_covs_cov_paper.pdf}&
\includegraphics[width=0.23\textwidth, page = 4]{cov_estimated_persp_image_26_Aug_26_Aug_26_Aug_all_covs_cov_paper.pdf}
 \\[-4ex]
\includegraphics[width=0.23\textwidth, page = 7]{cov_estimated_persp_image_26_Aug_26_Aug_26_Aug_all_covs_cov_paper.pdf} &
\includegraphics[width=0.23\textwidth, page = 8]{cov_estimated_persp_image_26_Aug_26_Aug_26_Aug_all_covs_cov_paper.pdf} &
\includegraphics[width=0.23\textwidth, page = 9]{cov_estimated_persp_image_26_Aug_26_Aug_26_Aug_all_covs_cov_paper.pdf} &
\includegraphics[width=0.23\textwidth, page = 10]{cov_estimated_persp_image_26_Aug_26_Aug_26_Aug_all_covs_cov_paper.pdf} 
\\[-2ex]
\includegraphics[width=0.23\textwidth]{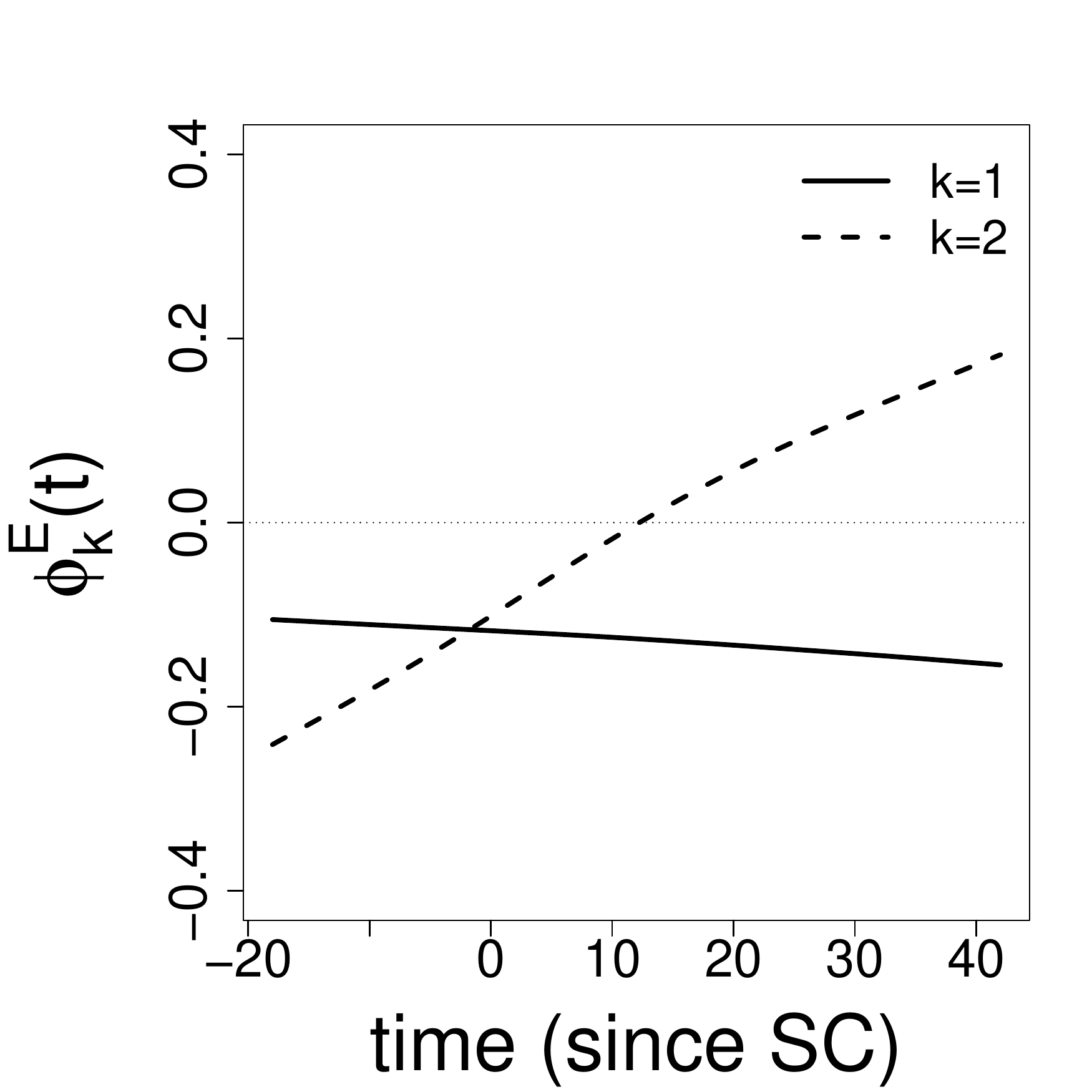} &
\includegraphics[width=0.23\textwidth]{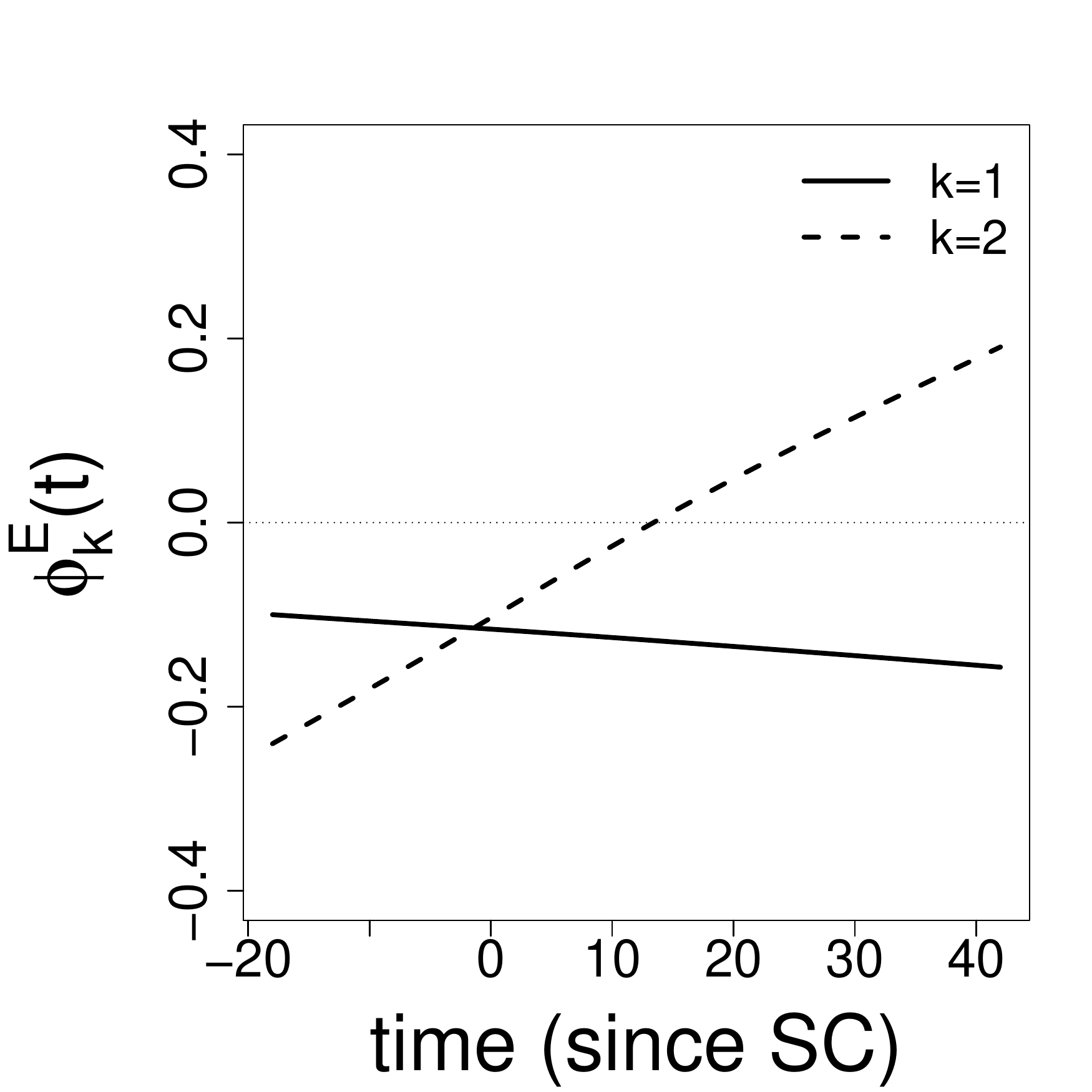} &
\includegraphics[width=0.23\textwidth]{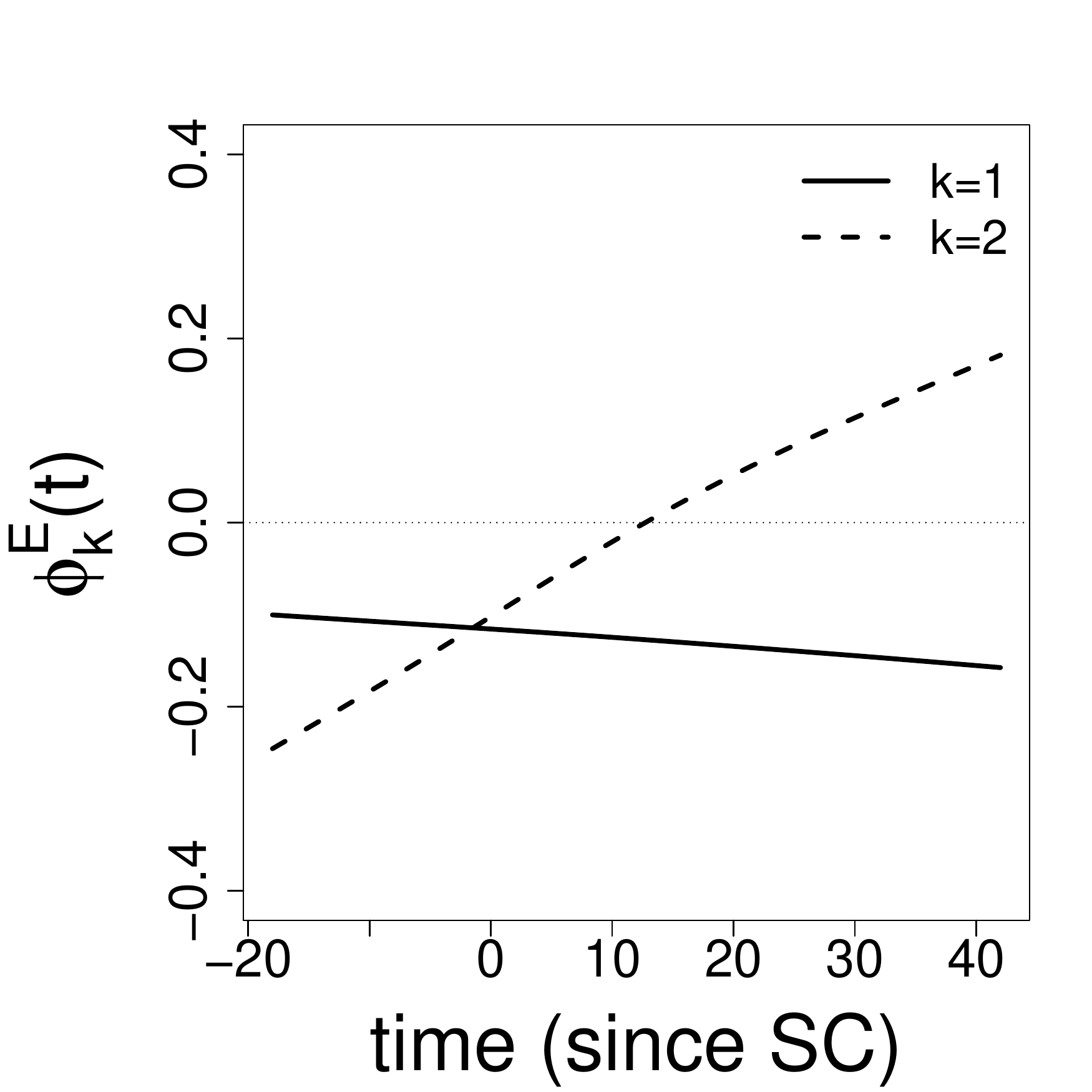}&
\includegraphics[width=0.23\textwidth]{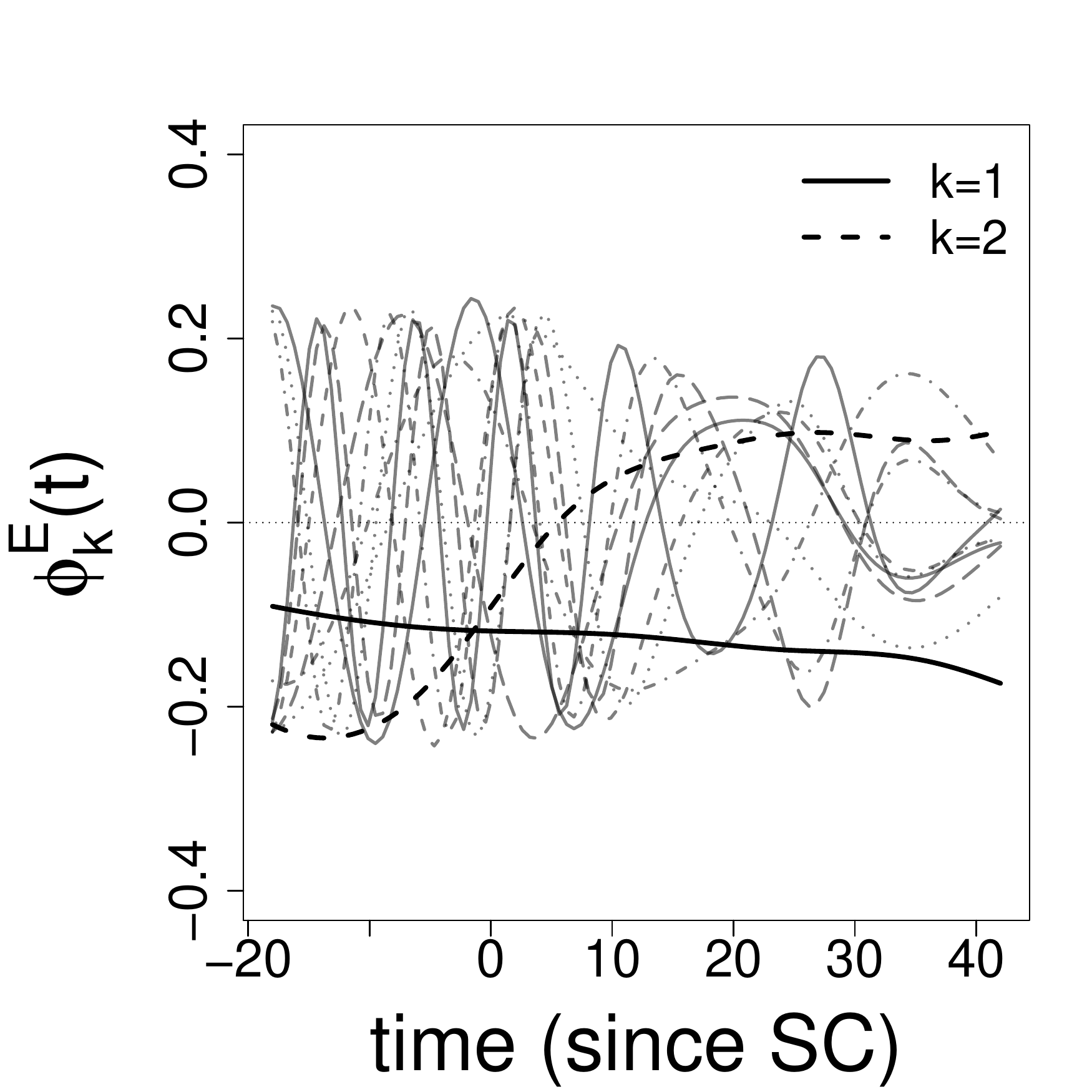}\\
\end{tabular}
\caption{Results for the curve-specific fRI for TRI-CONSTR, TRI-CONSTR-W, WHOLE, and TRI. Top row: estimated covariance surfaces. Middle row: contours of the estimated covariance surfaces. Bottom row: estimated corresponding eigenfunctions.}
\label{fig: cd4 covariances mine}
\end{center}
\end{figure}

\begin{figure}[h!]
\begin{center}
\begin{tabular}{cc}
 FACE & FACE-STEP-1\\ [-2ex]
\includegraphics[width=0.23\textwidth, page = 5]{cov_estimated_persp_image_26_Aug_26_Aug_26_Aug_all_covs_cov_paper.pdf} &
\includegraphics[width=0.23\textwidth, page = 6]{cov_estimated_persp_image_26_Aug_26_Aug_26_Aug_all_covs_cov_paper.pdf}\\[-4ex]
\includegraphics[width=0.23\textwidth, page = 11]{cov_estimated_persp_image_26_Aug_26_Aug_26_Aug_all_covs_cov_paper.pdf} &
\includegraphics[width=0.23\textwidth, page = 12]{cov_estimated_persp_image_26_Aug_26_Aug_26_Aug_all_covs_cov_paper.pdf}\\ [-2ex]
\includegraphics[width=0.23\textwidth]{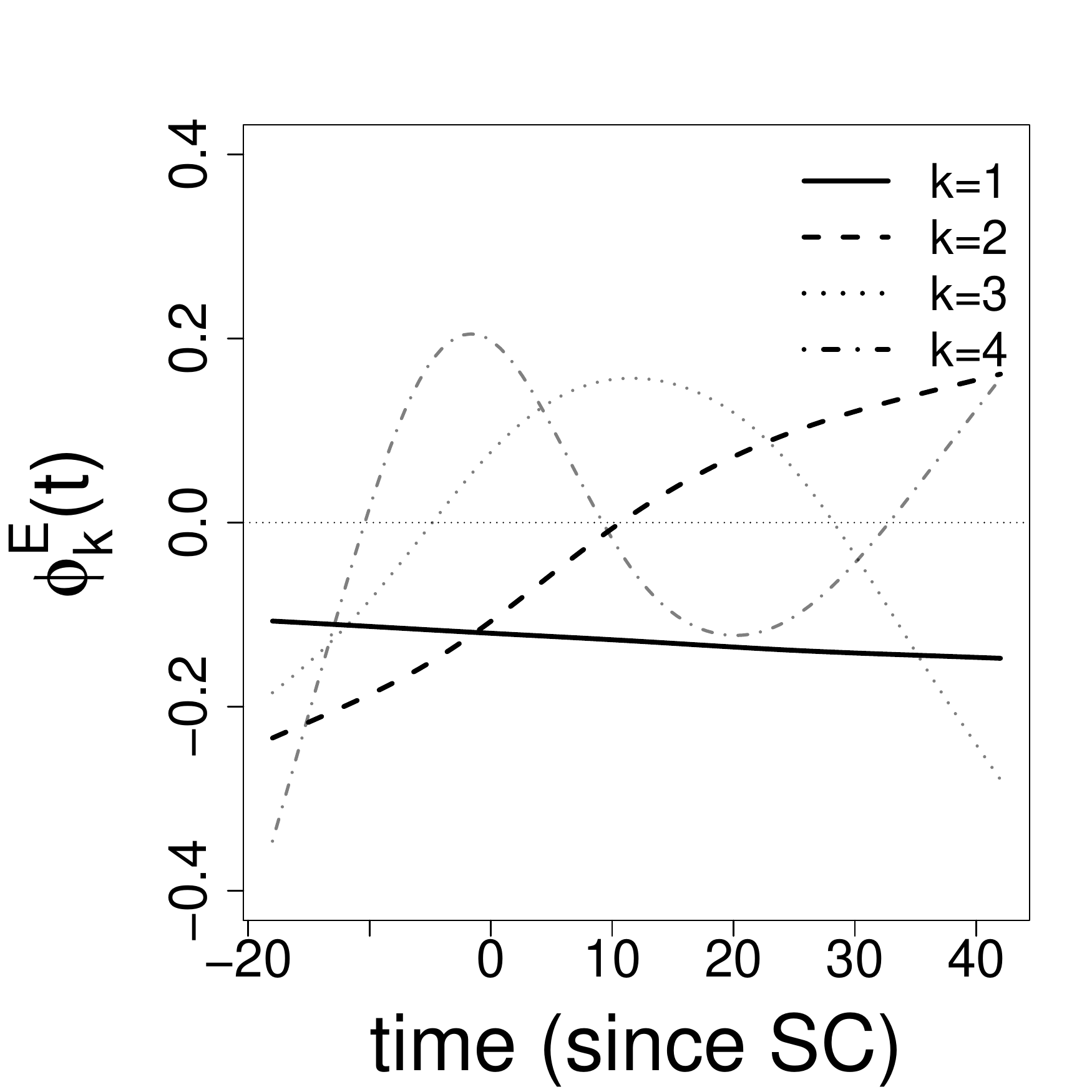} &
\includegraphics[width=0.23\textwidth]{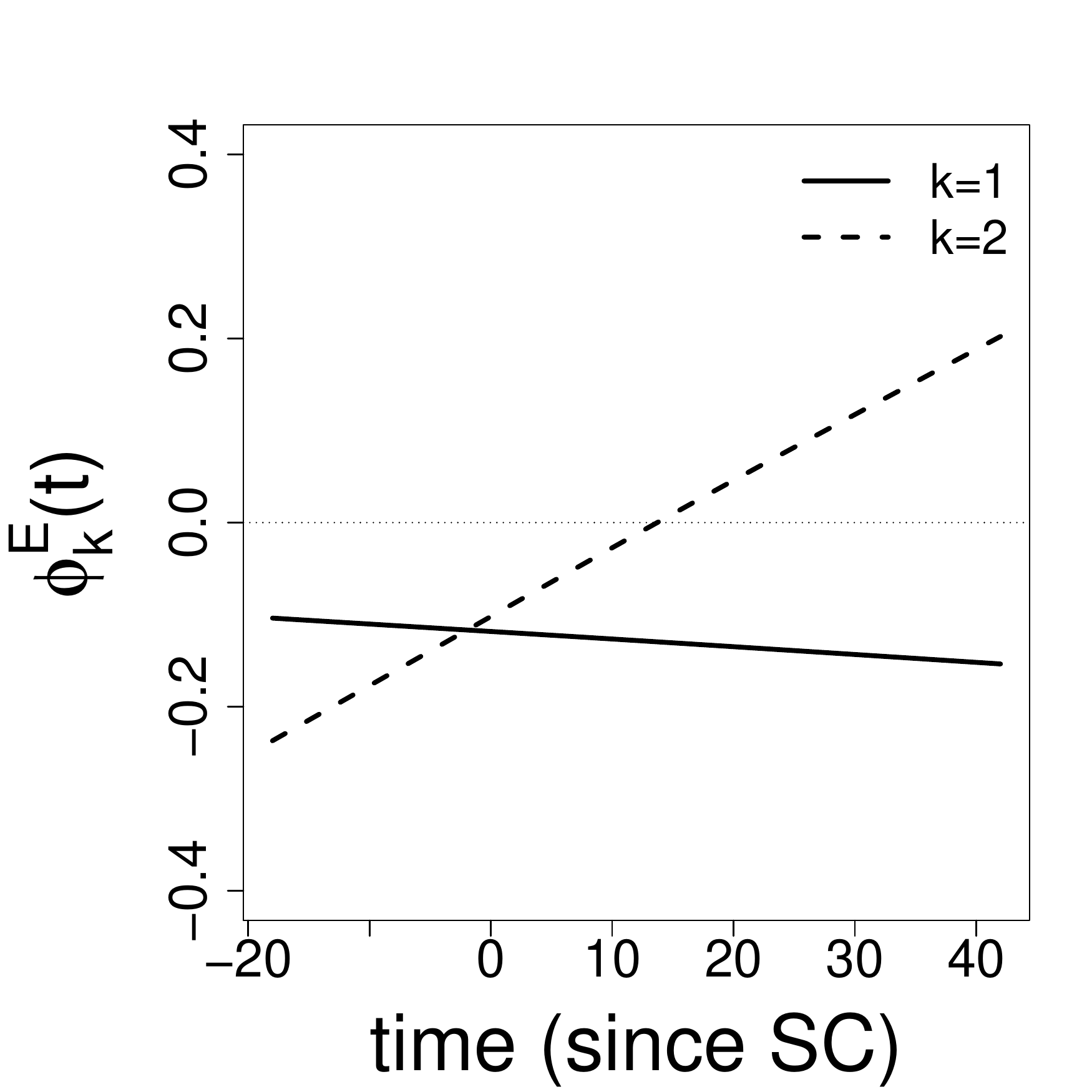} 
\end{tabular}
\caption{Results for the curve-specific fRI for FACE and FACE-STEP-1. Top row: estimated covariance surfaces. Middle row: contours of the estimated covariance surfaces. Bottom row: estimated corresponding eigenfunctions.}
\label{fig: cd4 covariances FACE}
\end{center}
\end{figure}

Figure \ref{fig: cd4 covariances mine} and Figure \ref{fig: cd4 covariances FACE} show the estimated covariance surfaces, reconstructed after truncation from the estimated eigenvalues and eigenfunctions, for our approach, WHOLE, and TRI and for FACE and FACE-STEP-1, respectively. In the bottom of the two figures, we also depict the truncated estimated eigenfunctions. Table \ref{table: cd4 variance decomposition} additionally gives the truncated estimated eigenvalues   and the estimated error variance. As in the application to the phonetics data in Section \ref{sec: phonetics}, we obtain the same number of eigenfunctions (two) for our approach and WHOLE  and a higher number (eleven) of wigglier eigenfunctions for TRI. FACE-STEP-1 also yields two eigenfunctions and FACE yields four eigenfunctions. We can see small differences in the resulting covariance surfaces of our approaches TRI-CONSTR and TRI-CONSTR-W. As expected, the latter is slightly more similar to that of WHOLE. The estimated surface of FACE-STEP-1 is also similar but a little smoother, whereas the the estimated surface of FACE is less smooth. As expected, the estimated surface of TRI shows a clear difference to the others on the diagonal, where it is much wigglier. The interpretation of the first and second eigenfunction is similar for all compared methods. The first eigenfunction is almost a vertical shift and thus describes the level of the (square root) CD4 cell counts. HIV infected individuals with negative basis weights for the first component tend to have a higher number of CD4 cells during the whole time interval [-18,42] than individuals with positive basis weights. The second eigenfunction gives insight in how fast the disease progresses. Individuals with negative basis weights for the second eigenfunction tend to have a faster decrease in CD4 cells than individuals with positive basis weights.


\begin{table}[h!]
\begin{center}
\caption{Truncated estimated eigenvalues and estimated error variance for all compared methods.}
\label{table: cd4 variance decomposition}
\begin{small}
 \setlength{\tabcolsep}{1.1mm}
\begin{tabular}{l|cccccccccccc}
& $\hat{\nu}_1$ &$\hat{\nu}_2$& $\hat{\nu}_3$ & $\hat{\nu}_4$ & $\hat{\nu}_5$ & $\hat{\nu}_6 $&$\hat{\nu}_7$ & $\hat{\nu}_8$  &  $\hat{\nu}_9$ & $\hat{\nu}_{10}$ &  $\hat{\nu}_{11}$  &$\hat{\sigma}^2$\\ \hline
TRI-CONSTR & 1170.37 &184.73 &&&&&&&&&& 15.54 \\
TRI-CONSTR-W &  1173.96 & 178.71&&&&&&&&&& 15.63  \\
WHOLE &  1174.96 & 178.03 &&&&&&&&&&  15.57\\
TRI &1191.80 &205.31 &61.88&26.95& 17.21&9.47&6.57&4.45& 3.55& 2.75& 2.30 &  12.13\\
FACE &  1162.51  &280.53   &22.28 &  12.87 & &&&&&&& 13.70\\
FACE-STEP-1 & 1161.84&  191.53&&&& &&&&&& 15.45
\end{tabular}
\end{small}
\end{center}
\end{table}

\subsection{Phonetics data}
In the following, we show additional results for our application to the phonetics data (cf.~Section \ref{sec: phonetics}), including the estimated auto-covariances for the smooth error $E_i(t)$, the estimated eigenvalues for both random processes, and the estimated error variance.

\begin{figure}[h!]
\begin{center}
\begin{tabular}{cccc}
TRI-CONSTR & TRI-CONSTR-W & WHOLE & TRI\\[-2ex]
\includegraphics[width=0.23\textwidth, page=1]{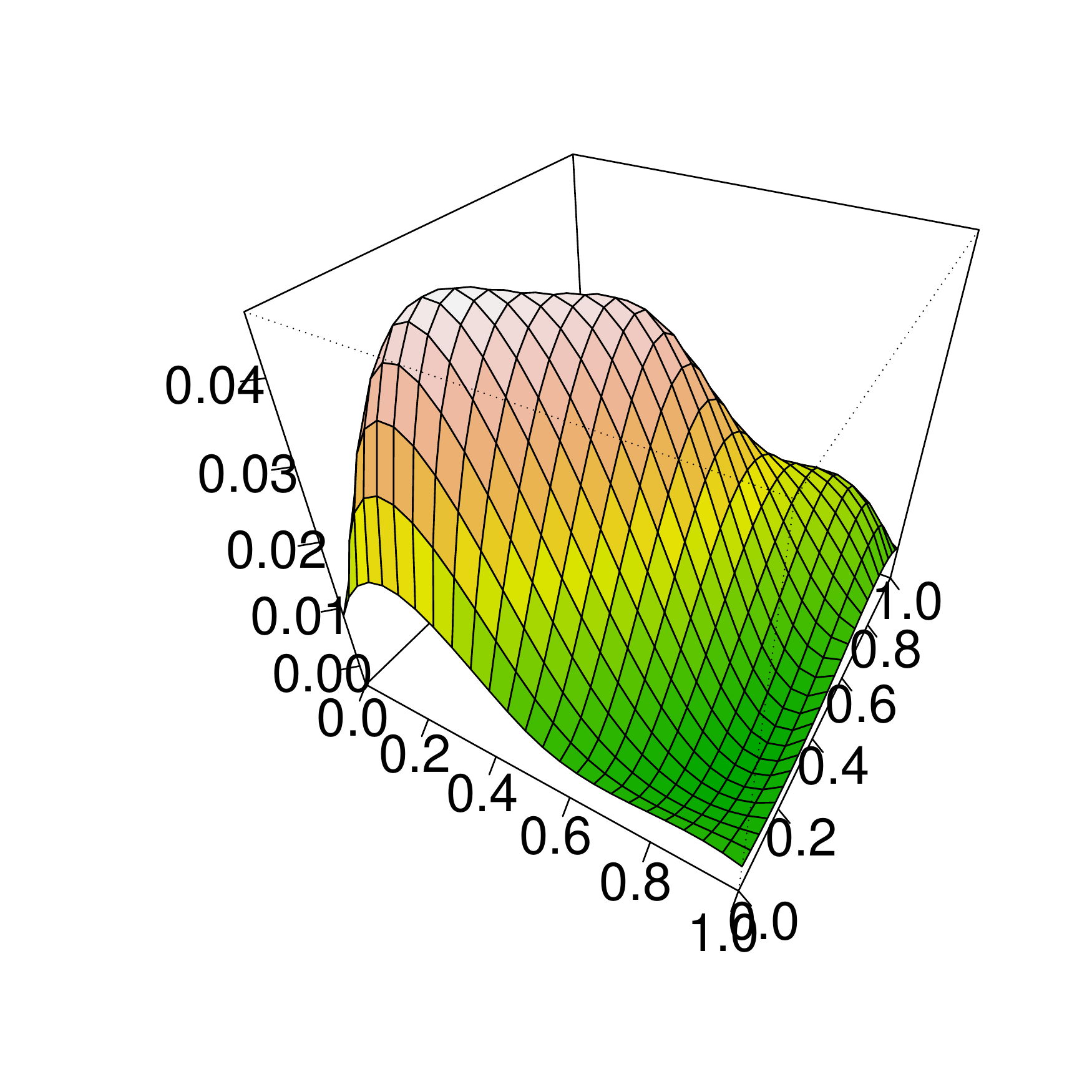} &
\includegraphics[width=0.23\textwidth, page=2]{covsE_estimated_persp_image_s_sh_acoustic_25_May_all_covs_cov_paper.pdf}&
\includegraphics[width=0.23\textwidth, page=3]{covsE_estimated_persp_image_s_sh_acoustic_25_May_all_covs_cov_paper.pdf} &
\includegraphics[width=0.23\textwidth, page=4]{covsE_estimated_persp_image_s_sh_acoustic_25_May_all_covs_cov_paper.pdf}\\[-4ex]
\includegraphics[width=0.23\textwidth, page=5]{covsE_estimated_persp_image_s_sh_acoustic_25_May_all_covs_cov_paper.pdf} &
\includegraphics[width=0.23\textwidth, page=6]{covsE_estimated_persp_image_s_sh_acoustic_25_May_all_covs_cov_paper.pdf}&
\includegraphics[width=0.23\textwidth, page=7]{covsE_estimated_persp_image_s_sh_acoustic_25_May_all_covs_cov_paper.pdf}&
\includegraphics[width=0.23\textwidth, page=8]{covsE_estimated_persp_image_s_sh_acoustic_25_May_all_covs_cov_paper.pdf}
\\[-2ex]
\includegraphics[width=0.23\textwidth, page=2]{FPCs_estimated_s_sh_acoustic_25_May_simultan_tri_constr_cov_paper.pdf}&
\includegraphics[width=0.23\textwidth, page=2]{FPCs_estimated_s_sh_acoustic_25_May_simultan_tri_constr_weights_cov_paper.pdf}&
\includegraphics[width=0.23\textwidth, page=2]{FPCs_estimated_s_sh_acoustic_25_May_simultan_whole_cov_paper.pdf}&
\includegraphics[width=0.23\textwidth, page=2]{FPCs_estimated_s_sh_acoustic_25_May_simultan_tri_cov_paper.pdf}
\end{tabular}
\caption{Results for the smooth error curve $E_i(t)$ using the four smoothing methods. Top row: estimated covariance surfaces. Middle row: contours of the estimated covariance surfaces. Bottom row: estimated corresponding eigenfunctions $\phi^E_k(t)$.}
\label{fig: phonetics covariances E}
\end{center}
\end{figure}

Figure \ref{fig: phonetics covariances E} depicts the estimated surfaces and contours of the auto-covariance of the smooth error $E_i(t)$, reconstructed after truncation from the estimated eigenvalues and eigenfunctions which are shown in the bottom of the figure. As for the auto-covariance of the fRI for speakers, we can see from Figure \ref{fig: phonetics covariances E} that the two estimates based on our symmetric smoother (TRI-CONSTR, TRI-CONSTR-W) are very similar to each other and to the one obtained by using all cross products (WHOLE). Again, we obtain wigglier estimates for TRI -- especially on the diagonal of the estimated surface. This corresponds to the fact that for TRI, the error variance is estimated to be zero. The first three eigenfunctions are very similar for all four compared methods. For TRI, however, nine more (high-frequency) eigenfunctions are chosen, yielding a wigglier surface estimate.

Table \ref{table: phonetics variance decomposition} gives the complete variance decomposition for our model. The upper table shows the truncated estimated eigenvalues of $\hat{K}^B$ for the four compared methods. The lower table shows the truncated estimated eigenvalues of the smooth error as well as the estimated error variance. For better display, all values are multiplied with $10^{3}$. It shows that the first two [three] estimated eigenvalues for $K^B(t,t^\prime)$ $[K^E(t,t^\prime)]$ are very similar for all smoothing methods and that TRI-CONSTR-W and WHOLE are most similar. The estimated error variance is slightly higher for TRI-CONSTR than for the other approaches.

\begin{table}[h!]
\begin{center}
\caption{Truncated estimated eigenvalues of $K^B(t,t^\prime)$, $\hat{\nu}^B_k \cdot 10^{3}$, and of $K^E(t,t^\prime)$, $\hat{\nu}^E_k \cdot 10^{3}$, and estimated error variance $\hat{\sigma}^2 \cdot 10^{3}$ for all compared methods.}
\label{table: phonetics variance decomposition}
\begin{small}
 \setlength{\tabcolsep}{1.1mm}
\begin{tabular}{l|cccc}
& $\hat{\nu}^B_1$ & $\hat{\nu}^B_2$ & $\hat{\nu}^B_3$ & $\hat{\nu}^B_4$\\ \hline
TRI-CONSTR & 5.84 & 3.24\\
TRI-CONSTR-W &  5.84 & 3.23  \\
WHOLE &  5.84 & 3.23 & \\
TRI & 5.85& 3.27& 0.42 &0.24
\end{tabular}

\vspace{0.5cm}
\begin{tabular}{l|ccccccccccccc}
& $\hat{\nu}^E_1$ & $\hat{\nu}^E_2$ & $\hat{\nu}^E_3$ & $\hat{\nu}^E_4$ & $\hat{\nu}^E_5$& $\hat{\nu}^E_6$ & $\hat{\nu}^E_7$ & $\hat{\nu}^E_8$ & $\hat{\nu}^E_9$ & $\hat{\nu}^E_{10}$ & $\hat{\nu}^E_{11}$ & $\hat{\nu}^E_{12}$  &$\hat{\sigma}^2$\\ \hline
TRI-CONSTR & 19.52 & 7.56 & 2.74 & &&&&&&&&&4.15\\
TRI-CONSTR-W & 19.53 & 7.59 & 2.73 &&&&&&&&&& 3.97\\
WHOLE &   19.53 & 7.59 & 2.73 &&&&&&&&&& 3.94\\
TRI & 19.67 & 7.77 & 2.95 & 1.37 & 0.93 & 0.63 & 0.50 & 0.36 & 0.30 & 0.23 & 0.20 & 0.16 &0.00
\end{tabular}
\end{small}
\end{center}
\end{table}

\clearpage

\section{Supplementary simulation details and results}\label{appendix: supplementary simulation details and results}

\subsection{Generation Details}
\begin{figure}[h!]
\begin{center}
\begin{tabular}{ccc}
B & C & E \\[-2ex]
\includegraphics[width=0.23\textwidth, page=1]{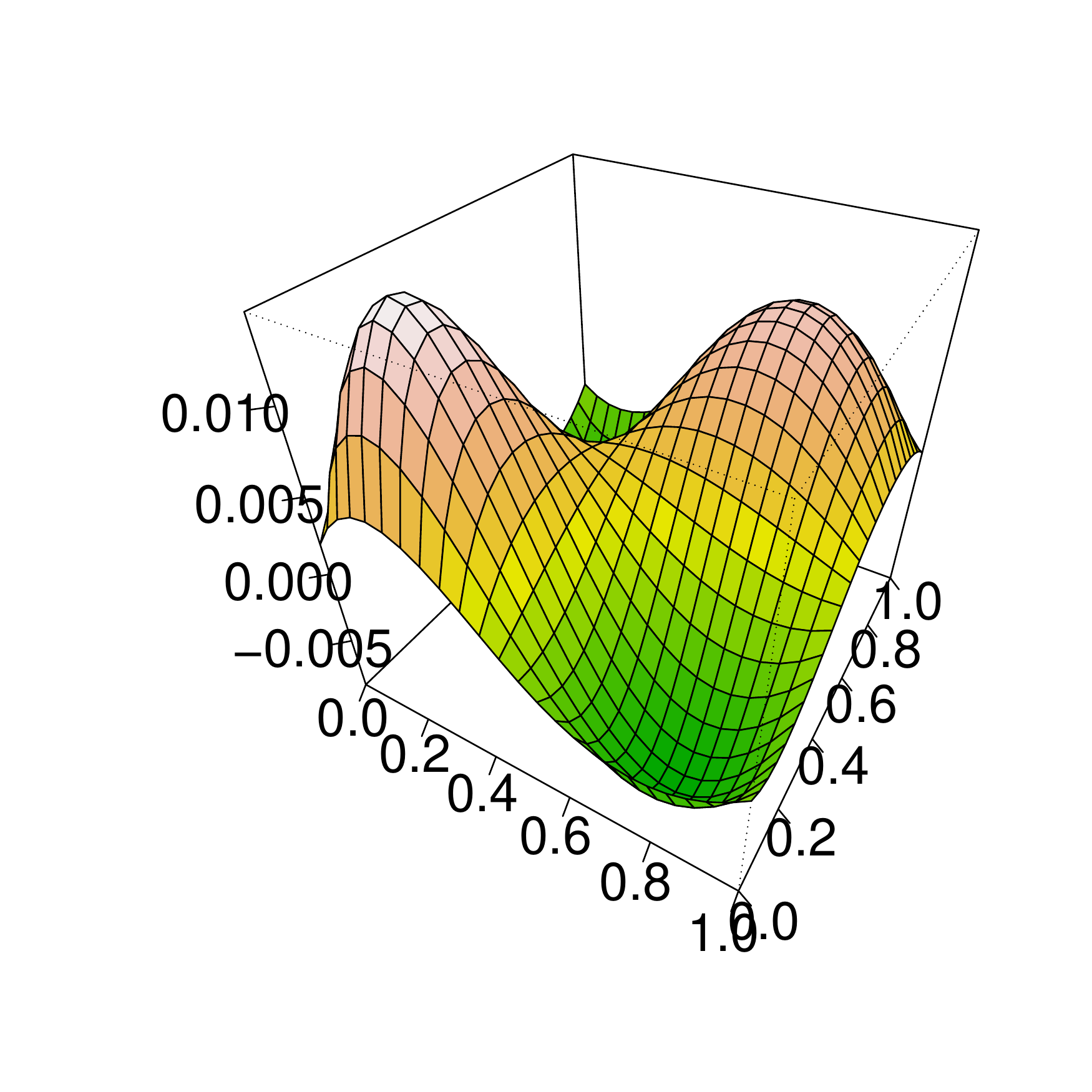} &
\includegraphics[width=0.23\textwidth, page=3]{plot_true_covs_crossed_as_data_26_Dec.pdf}&
\includegraphics[width=0.23\textwidth, page=5]{plot_true_covs_crossed_as_data_26_Dec.pdf}\\[-4ex]
\includegraphics[width=0.23\textwidth, page=2]{plot_true_covs_crossed_as_data_26_Dec.pdf}&
\includegraphics[width=0.23\textwidth, page=4]{plot_true_covs_crossed_as_data_26_Dec.pdf}&
\includegraphics[width=0.23\textwidth, page=6]{plot_true_covs_crossed_as_data_26_Dec.pdf}\\[-2ex]
\includegraphics[width=0.23\textwidth, page=1]{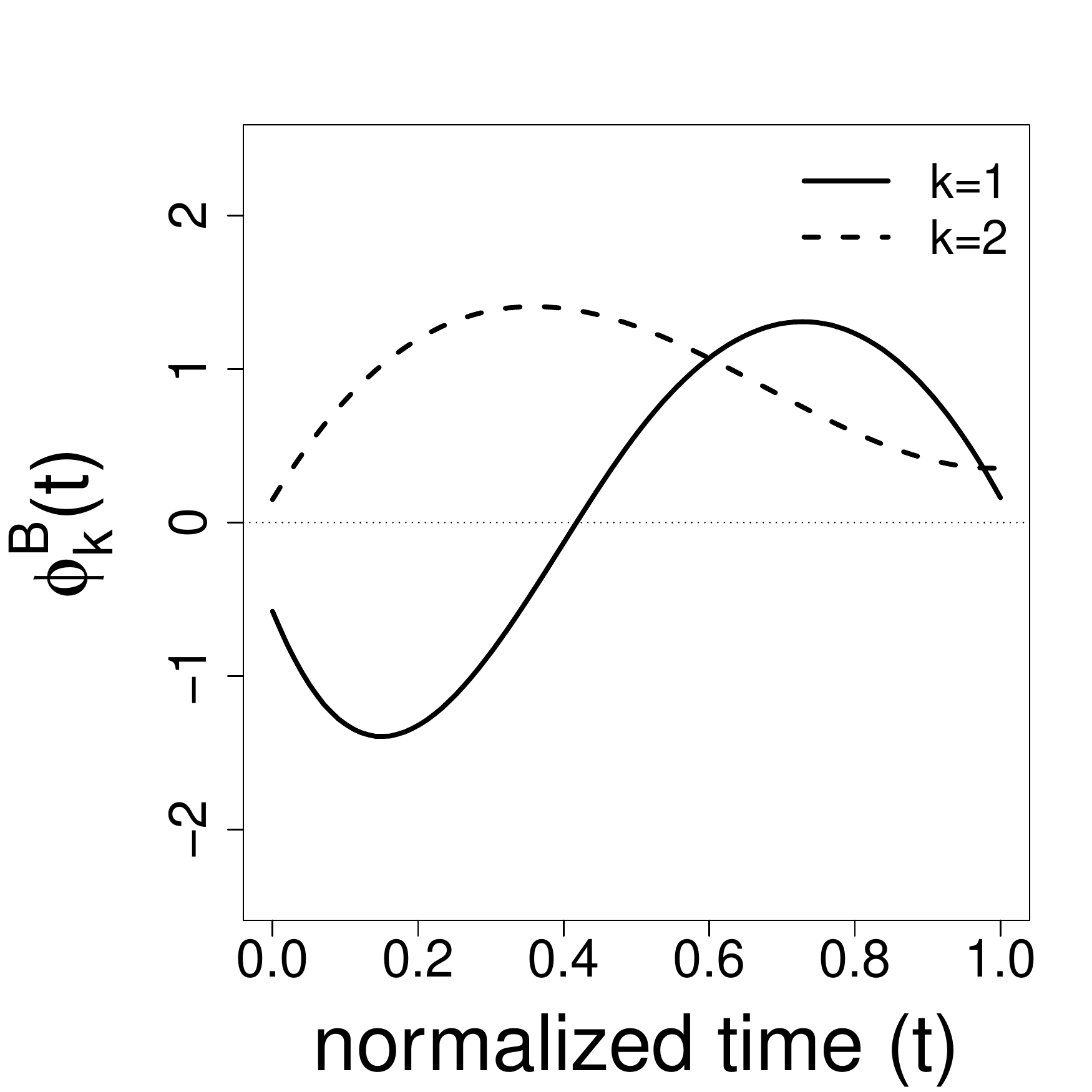}&
\includegraphics[width=0.23\textwidth, page=2]{plot_true_eigenfunctions_crossed_as_data_26_Dec.pdf}&
\includegraphics[width=0.23\textwidth, page=3]{plot_true_eigenfunctions_crossed_as_data_26_Dec.pdf}
\end{tabular}
\caption{Auto-covariances $K^B(t,t^\prime)$, $K^C(t,t^\prime)$, and $K^E(t,t^\prime)$ and their eigenfunctions used for the data generation for the scenario with crossed fRIs. Top row: covariance surfaces. Middle row: contours of the covariance surfaces. Bottom row: corresponding eigenfunctions $\phi^B_k(t)$, $\phi^C_k(t)$, and $\phi^E_k(t)$.}
\label{fig: true auto-covariances crossed}
\end{center}
\end{figure}

\begin{figure}[h!]
\begin{center}
\includegraphics[width=0.23\textwidth]{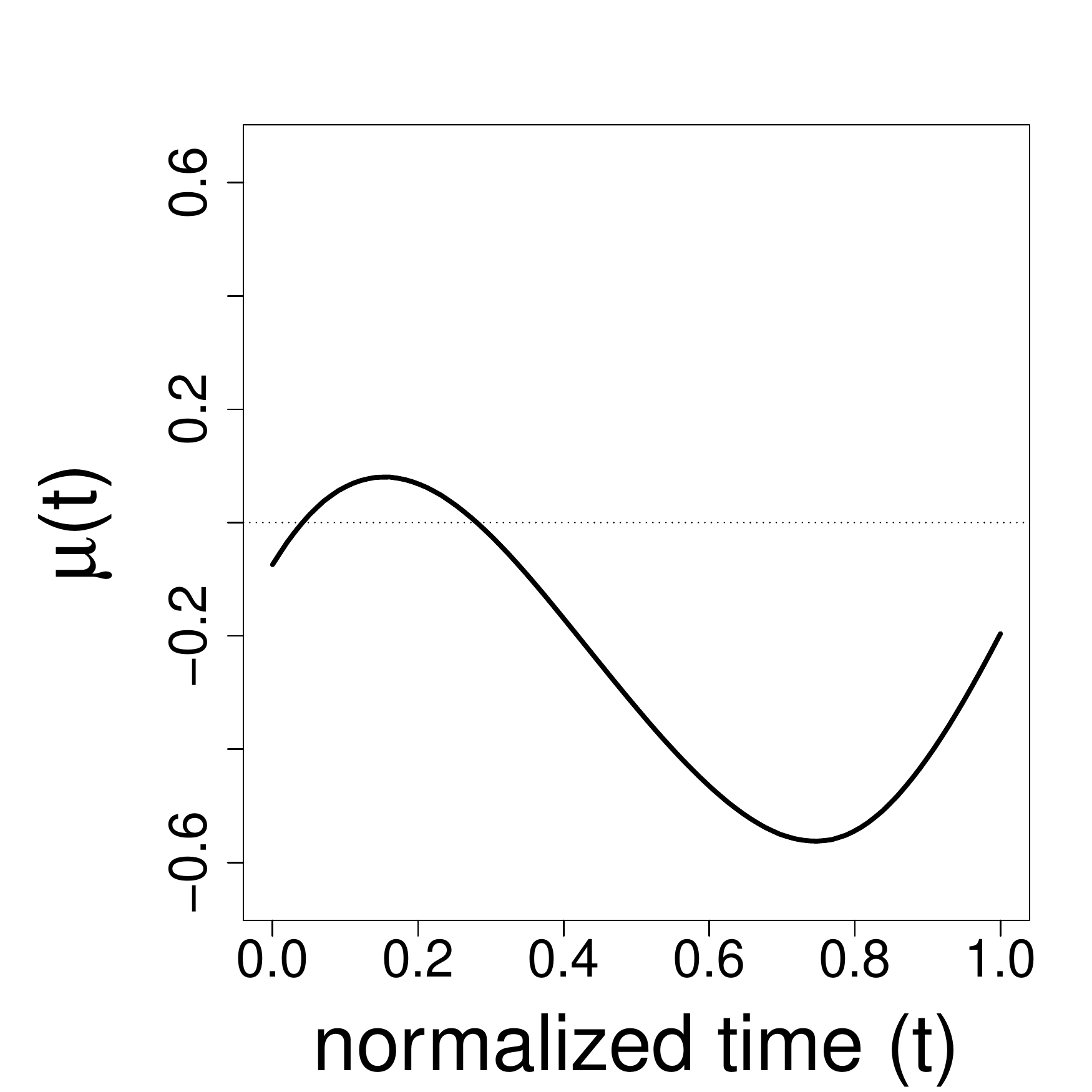}
\caption{Mean function used for the data generation for the scenario with crossed fRIs.}
\label{fig: true mean crossed}
\end{center}
\end{figure}

For the scenario with crossed fRIs, we use two, one, and three eigenfunctions, estimated from the phonetics data, for the generation of the auto-covariances of processes $B_i(t)$, $C_i(t)$, and $E_i(t)$, respectively. The resulting auto-covariance surfaces are shown in Figure \ref{fig: true auto-covariances crossed}. The eigenfunctions used for data generation are shown in the bottom of the Figure. The corresponding eigenvalues used for data generation are shown in Table \ref{tab: true eigenvalues crossed}, where also the error variance is given. The values are multiplied with $10^{3}$ for better display. The underlying mean function is depicted in Figure \ref{fig: true mean crossed}.

\begin{table}[h!]
\begin{center}
\caption{Eigenvalues $\nu_k^X \cdot 10^{3}$, $X\in \lbrace B,C,E\rbrace$, and error variance $\sigma^2 \cdot 10^{3}$ used for data generation for the scenario with crossed fRIs.}
\label{tab: true eigenvalues crossed}
\begin{tabular}{ccccccc}
$\nu_1^B$ & $\nu_2^B$ & $\nu_1^C$ & $\nu_1^E$& $\nu_2^E$& $\nu_3^E$ & $\sigma^2$\\ \hline
5.86 & 2.71 & 8.89 & 19.05 & 7.53 & 2.66 & 5.62
\end{tabular}
\end{center}
\end{table}

\subsection{Measures of goodness of fit}
We use root relative mean squared errors (rrMSEs) as measures of goodness of fit for all model components \citep[cp.][]{Cederbaum.2016}.

 For vector-valued estimates $\hat{\mtheta}$ of $\mtheta= \left(\theta_1,\ldots,\theta_L\right)^\top$, we define the rrMSE as 
\bea  \label{eq: rrMSE}
\rrMSE(\mtheta,\hat{\mtheta}) = \sqrt{\frac{\dfrac{1}{L}\sum_{l=1}^L \left(\theta_{l} - \hat{\theta}_{l} \right)^2}{\dfrac{1}{L}\sum_{l=1}^L {\theta_{l}}^2}}.
\eea 
We use that for the random basis weights $\xi^{X}_{lk}$, \eqref{eq: rrMSE} is approximately $\sqrt{\nicefrac{\nicefrac{1}{L^X}\sum_{l=1}^{L^X} \left(\xi^X_{lk} - \hat{\xi}^X_{lk} \right)^2}{\nu_k^X}}$, $X \in \lbrace B, C, E\rbrace$. 

The form of the rrMSE for scalar estimates results as special case of \ref{eq: rrMSE} with $L=1$. 

For all functions $\theta(t)$, we approximate the integrals by sums and obtain 
\bea \label{eq: rrISE}
\rrMSE\left[\theta(\cdot),\hat{\theta}(\cdot)\right] =  \sqrt{\frac{\dfrac{1}{D}\sum_{d=1}^D\left[\theta(t_d)-\hat{\theta}(t_d)\right]^2}{\dfrac{1}{D}\sum_{d=1}^D \theta(t_d)^2}}.
\eea
As the eigenfunctions are only unique up to sign, we also compute the rrMSEs of the estimated eigenfunctions mirrored around the x-axis and choose the smaller rrMSE. For the random processes, we additionally average over the respective levels. For centered processes, we use that the denominator simplifies to the average variance.

For bivariate functions, such as the auto-covariances, we define
\bea  \label{eq: rrISE bivariate}
\rrMSE\left[\theta(\cdot,\cdot),\hat{\theta}(\cdot,\cdot)\right] =  \sqrt{\frac{\dfrac{1}{D^2}\sum_{t_d,t_{d^\prime} =1}^D\left(\theta(t_d,t_{d^\pr})-\hat{\theta}(t_d,t_{d^\pr})\right)^2}{\dfrac{1}{D^2}\sum_{t_d,t_{d^\pr} =1}^D \theta(t_d,t_{d^\pr})^2}}.
\eea

\subsection{Results for the scenario with independent curves}\label{sec: results indep} 
\begin{figure}[p]
\begin{center}
\textbf{Setting 2}\\
\includegraphics[width= 1\textwidth]{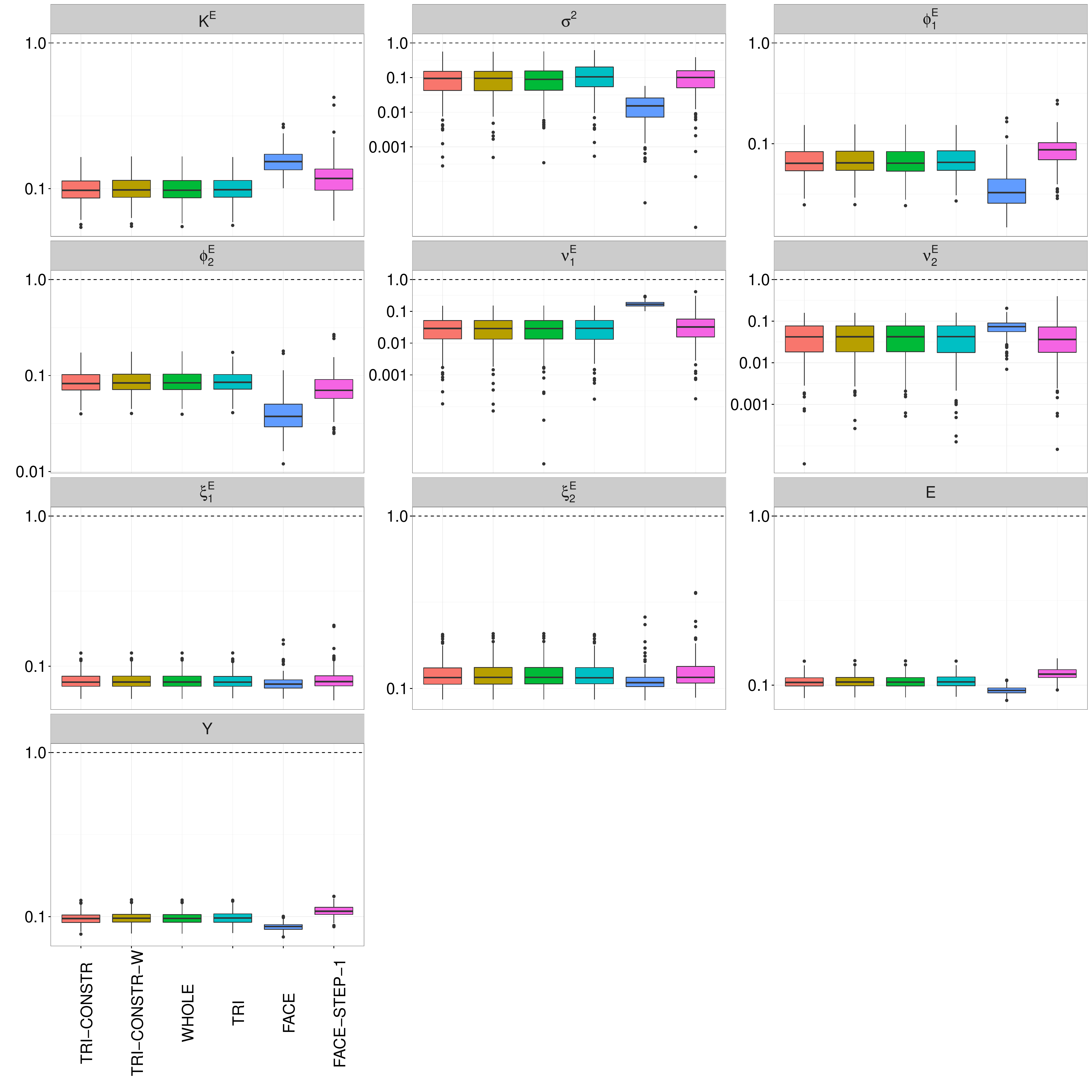}
\caption{Boxplots of the rrMSEs (log10 scale at y-axis) for Setting 2 of the scenario with independent curves.  Top row: rrMSEs for auto-covariance $K^E(t,t^\prime)$, error variance $\sigma^2$, and the first eigenfunction $\phi^E_1(t)$. Second row: rrMSEs for the second eigenfunction $\phi^E_2(t)$ and eigenvalues $\nu^E_1$, $\nu^E_2$. Third row: rrMSEs for the random basis weights $\xi^E_1$, $\xi^E_2$ and process $E_i(t)$. Bottom row: rrMSEs for curves $Y_i(t)$.}
\label{fig: boxplots rrMSEs indep curves 22_Mar}
\end{center}
\end{figure}

\begin{figure}[p]
\begin{center}
\textbf{Setting 3}\\
\includegraphics[width= 1\textwidth]{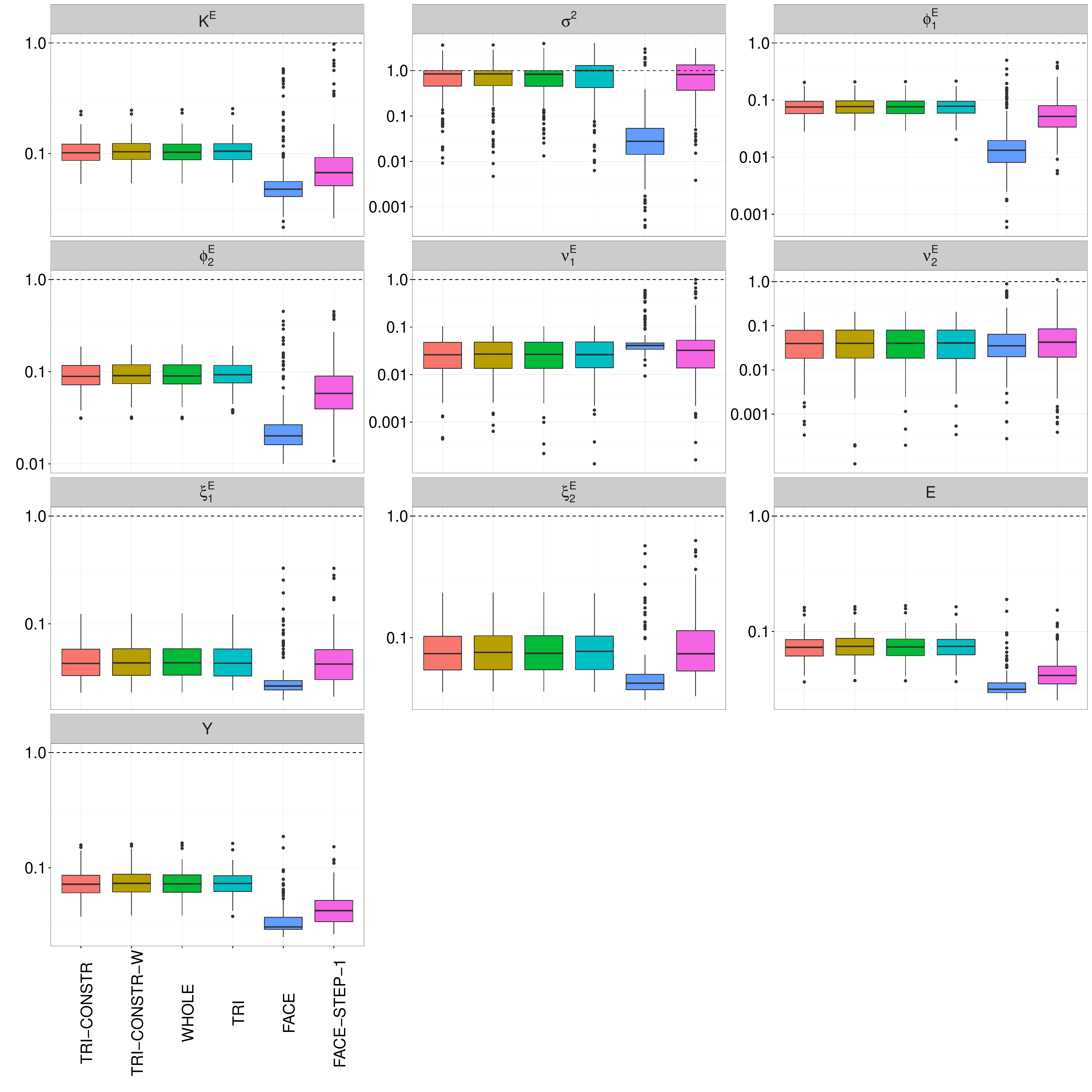}
\caption{Boxplots of the rrMSEs (log10 scale at y-axis) for Setting 3 of the scenario with independent curves. Top row: rrMSEs for auto-covariance $K^E(t,t^\prime)$, error variance $\sigma^2$, and the first eigenfunction $\phi^E_1(t)$. Second row: rrMSEs for the second eigenfunction $\phi^E_2(t)$ and eigenvalues $\nu^E_1$, $\nu^E_2$. Third row: rrMSEs for the random basis weights $\xi^E_1$, $\xi^E_2$ and process $E_i(t)$. Bottom row: rrMSEs for curves $Y_i(t)$.}
\label{fig: boxplots rrMSEs indep curves 23_Mar}
\end{center}
\end{figure}

\begin{figure}[p]
\begin{center}
\textbf{Setting 4}\\
\includegraphics[width= 1\textwidth]{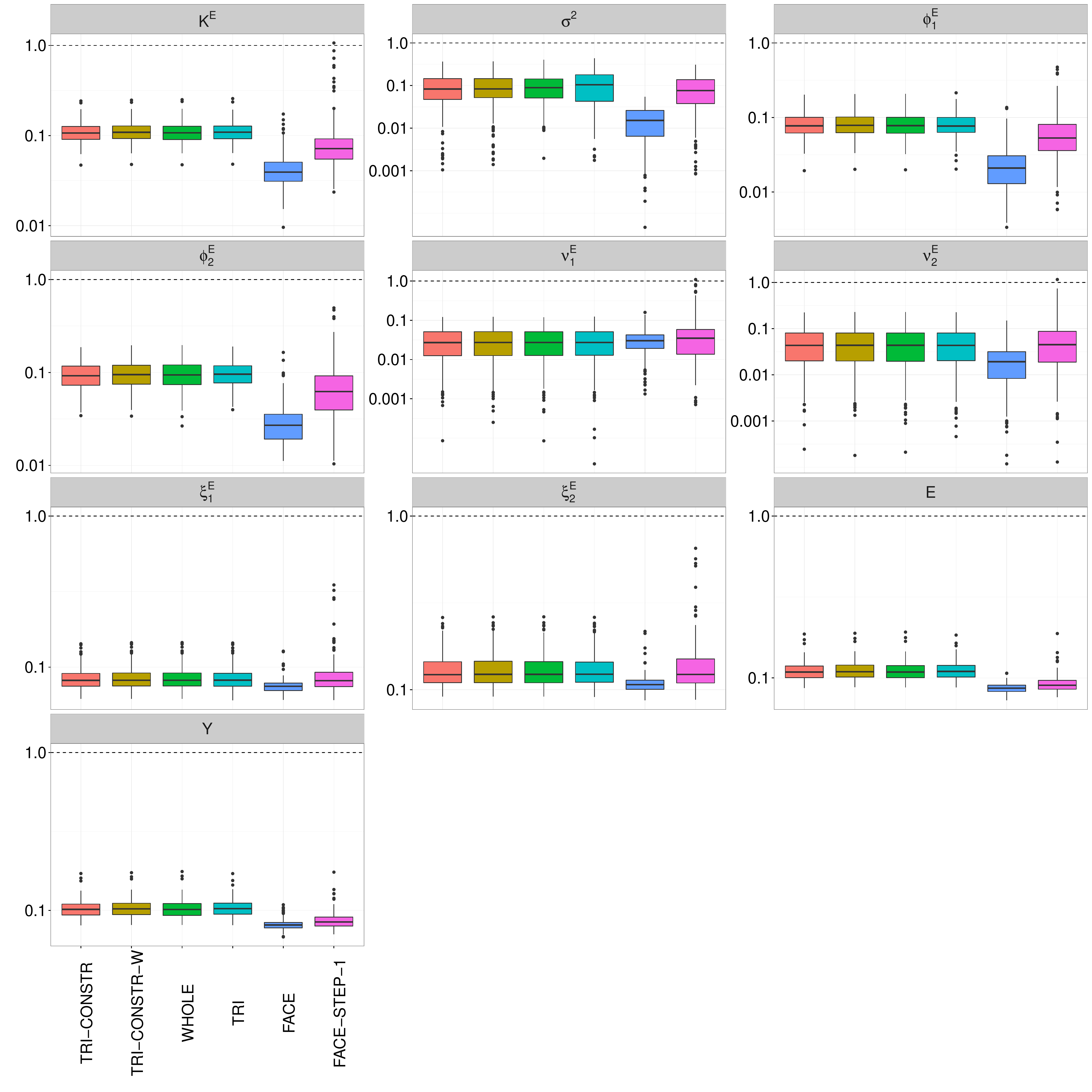}
\caption{Boxplots of the rrMSEs (log10 scale at y-axis) for Setting 4 of the scenario with independent curves. Top row: rrMSEs for auto-covariance $K^E(t,t^\prime)$, error variance $\sigma^2$, and the first eigenfunction $\phi^E_1(t)$. Second row: rrMSEs for the second eigenfunction $\phi^E_2(t)$ and eigenvalues $\nu^E_1$, $\nu^E_2$. Third row: rrMSEs for the random basis weights $\xi^E_1$, $\xi^E_2$ and process $E_i(t)$. Bottom row: rrMSEs for curves $Y_i(t)$.}
\label{fig: boxplots rrMSEs indep curves sigmasq05_23_Mar}
\end{center}
\end{figure}

\begin{figure}[p]
\begin{center}
\textbf{Setting 5}\\
\includegraphics[width= 1\textwidth]{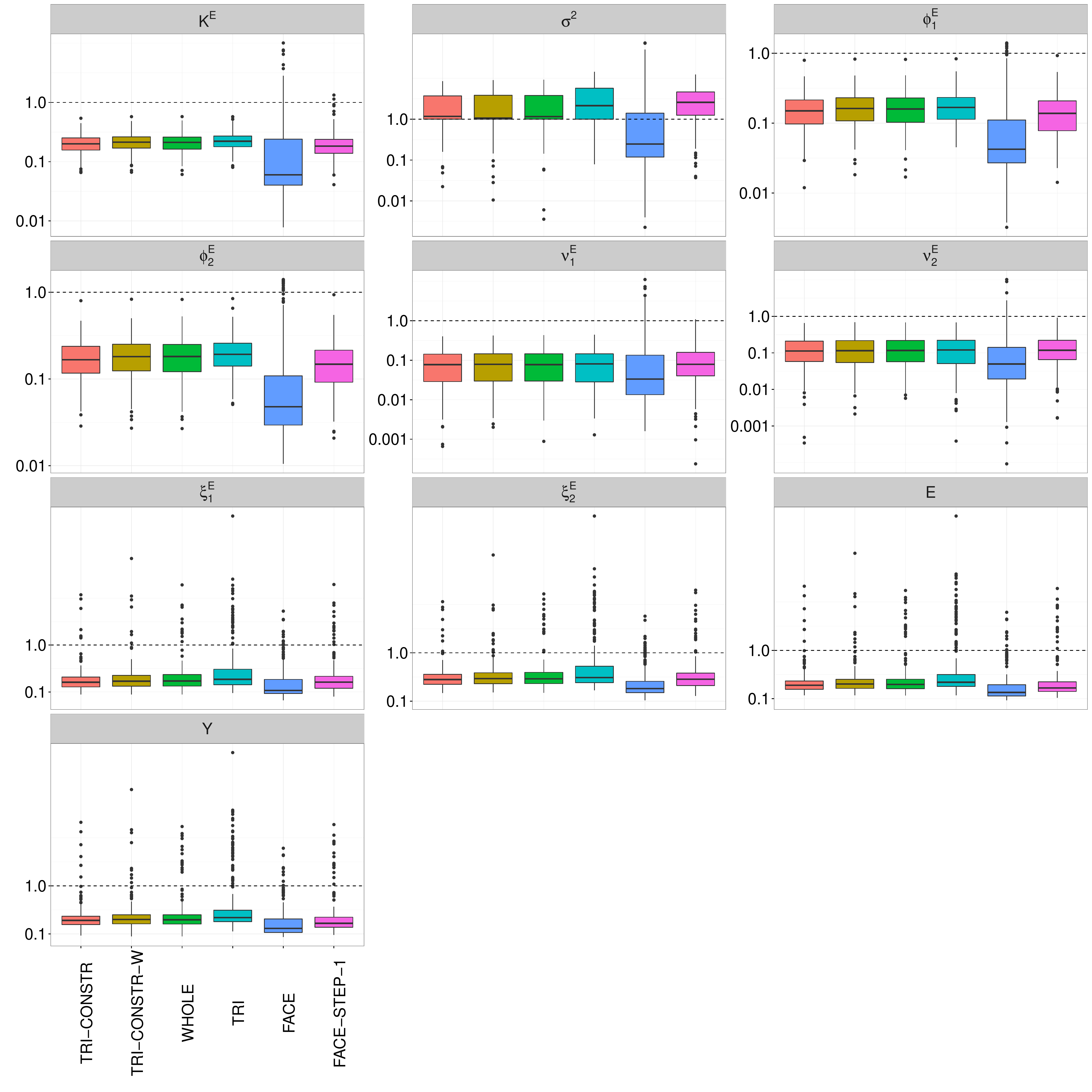}
\caption{Boxplots of the rrMSEs (log10 scale at y-axis) for Setting 5 of the scenario with independent curves. Top row: rrMSEs for auto-covariance $K^E(t,t^\prime)$, error variance $\sigma^2$, and the first eigenfunction $\phi^E_1(t)$. Second row: rrMSEs for the second eigenfunction $\phi^E_2(t)$ and eigenvalues $\nu^E_1$, $\nu^E_2$. Third row: rrMSEs for the random basis weights $\xi^E_1$, $\xi^E_2$ and process $E_i(t)$. Bottom row: rrMSEs for curves $Y_i(t)$.}
\label{fig: boxplots rrMSEs indep curves sparse_23_Mar}
\end{center}
\end{figure}

\begin{figure}[p]
\begin{center}
\textbf{Setting 6}\\
\includegraphics[width= 1\textwidth]{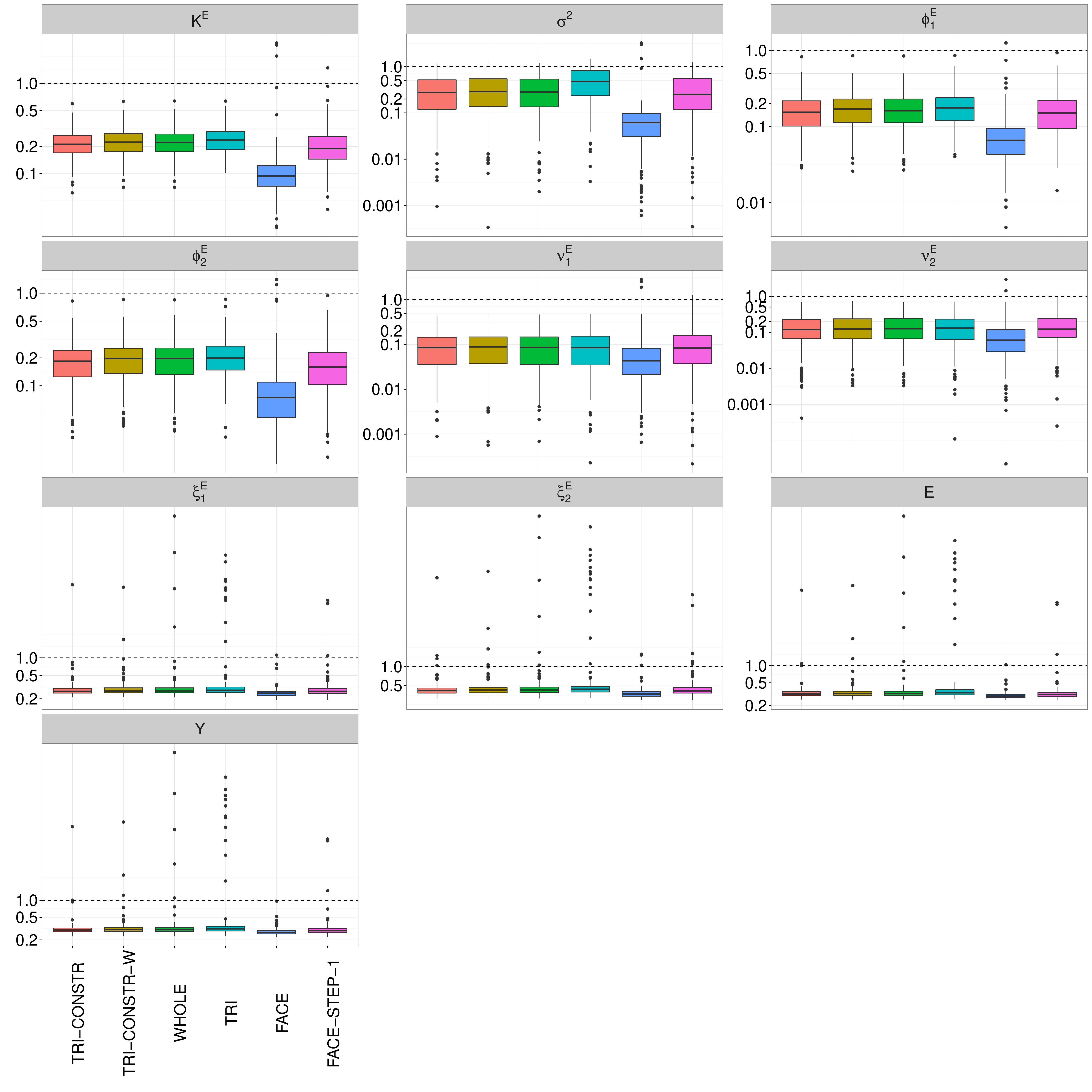}
\caption{Boxplots of the rrMSEs (log10 scale at y-axis) for Setting 6 of the scenario with independent curves. Top row: rrMSEs for auto-covariance $K^E(t,t^\prime)$, error variance $\sigma^2$, and the first eigenfunction $\phi^E_1(t)$. Second row: rrMSEs for the second eigenfunction $\phi^E_2(t)$ and eigenvalues $\nu^E_1$, $\nu^E_2$. Third row: rrMSEs for the random basis weights $\xi^E_1$, $\xi^E_2$ and process $E_i(t)$. Bottom row: rrMSEs for curves $Y_i(t)$.}
\label{fig: boxplots rrMSEs indep curves sparse_sigmasq05_23_Mar}
\end{center}
\end{figure}

\begin{figure}[p]
\begin{center}
\textbf{Setting 7}\\
\includegraphics[width= 1\textwidth]{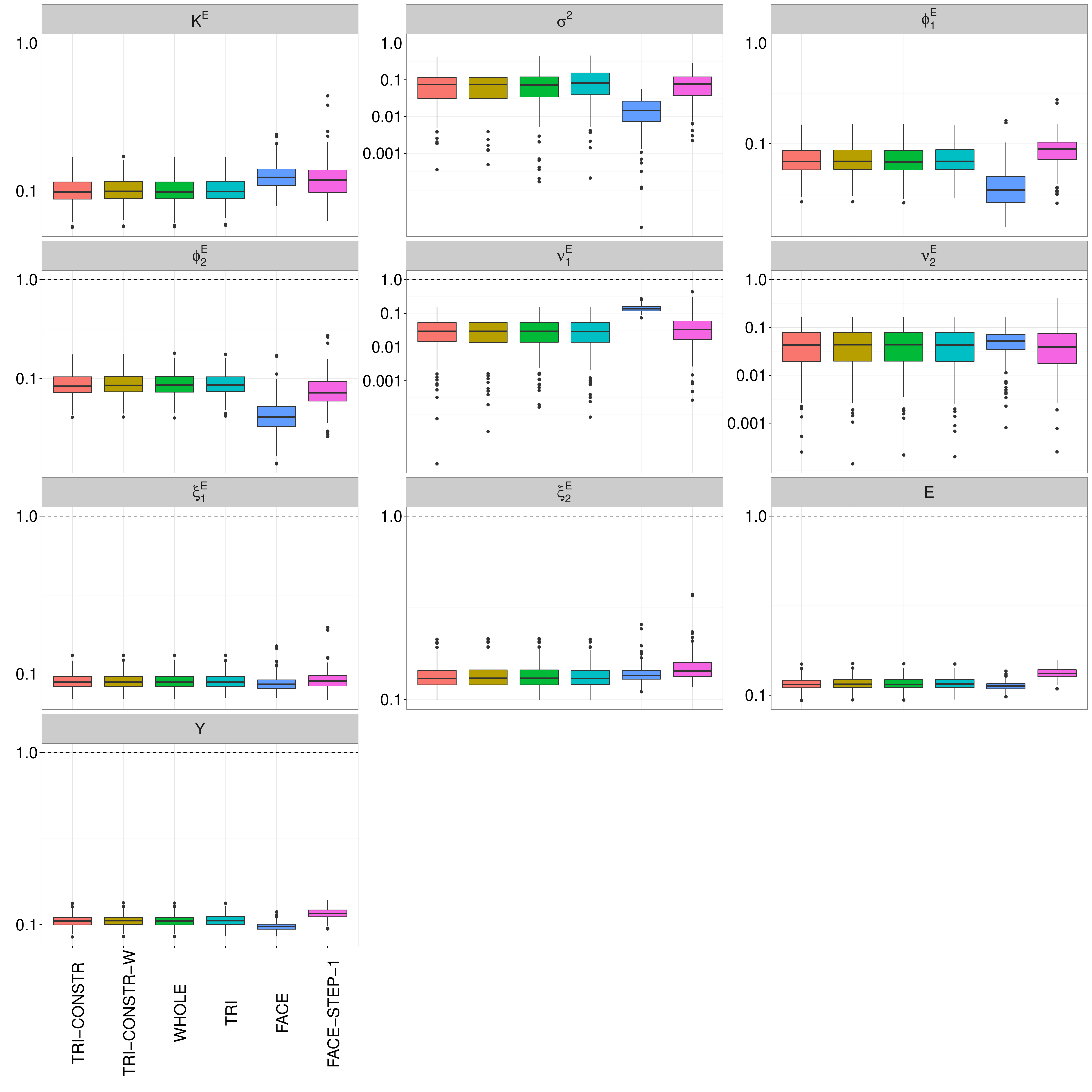}
\caption{Boxplots of the rrMSEs (log10 scale at y-axis) for Setting 7 of the scenario with independent curves. Top row: rrMSEs for auto-covariance $K^E(t,t^\prime)$, error variance $\sigma^2$, and the first eigenfunction $\phi^E_1(t)$. Second row: rrMSEs for the second eigenfunction $\phi^E_2(t)$ and eigenvalues $\nu^E_1$, $\nu^E_2$. Third row: rrMSEs for the random basis weights $\xi^E_1$, $\xi^E_2$ and process $E_i(t)$. Bottom row: rrMSEs for curves $Y_i(t)$.}
\label{fig: boxplots rrMSEs indep curves complexphi_30_Mar}
\end{center}
\end{figure}

\begin{figure}[p]
\begin{center}
\textbf{Setting 8}\\
\includegraphics[width= 1\textwidth]{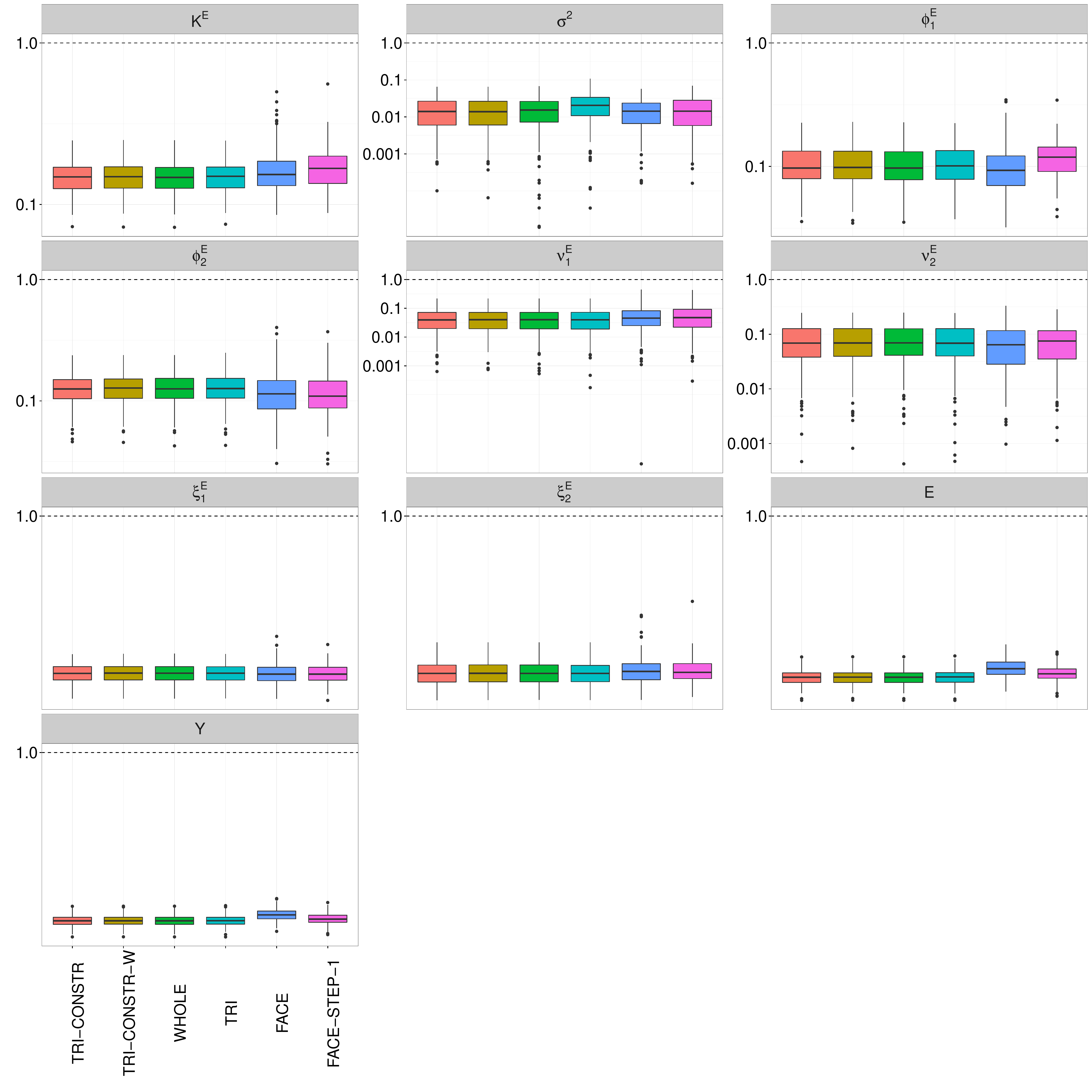}
\caption{Boxplots of the rrMSEs (log10 scale at y-axis) for Setting 8 of the scenario with independent curves. Top row: rrMSEs for auto-covariance $K^E(t,t^\prime)$, error variance $\sigma^2$, and the first eigenfunction $\phi^E_1(t)$. Second row: rrMSEs for the second eigenfunction $\phi^E_2(t)$ and eigenvalues $\nu^E_1$, $\nu^E_2$. Third row: rrMSEs for the random basis weights $\xi^E_1$, $\xi^E_2$ and process $E_i(t)$. Bottom row: rrMSEs for curves $Y_i(t)$.}
\label{fig: boxplots rrMSEs indep curves complexphi_sigmasq05_30_Mar}
\end{center}
\end{figure}

\begin{figure}[p]
\begin{center}
\textbf{Setting 9}\\
\includegraphics[width= 1\textwidth]{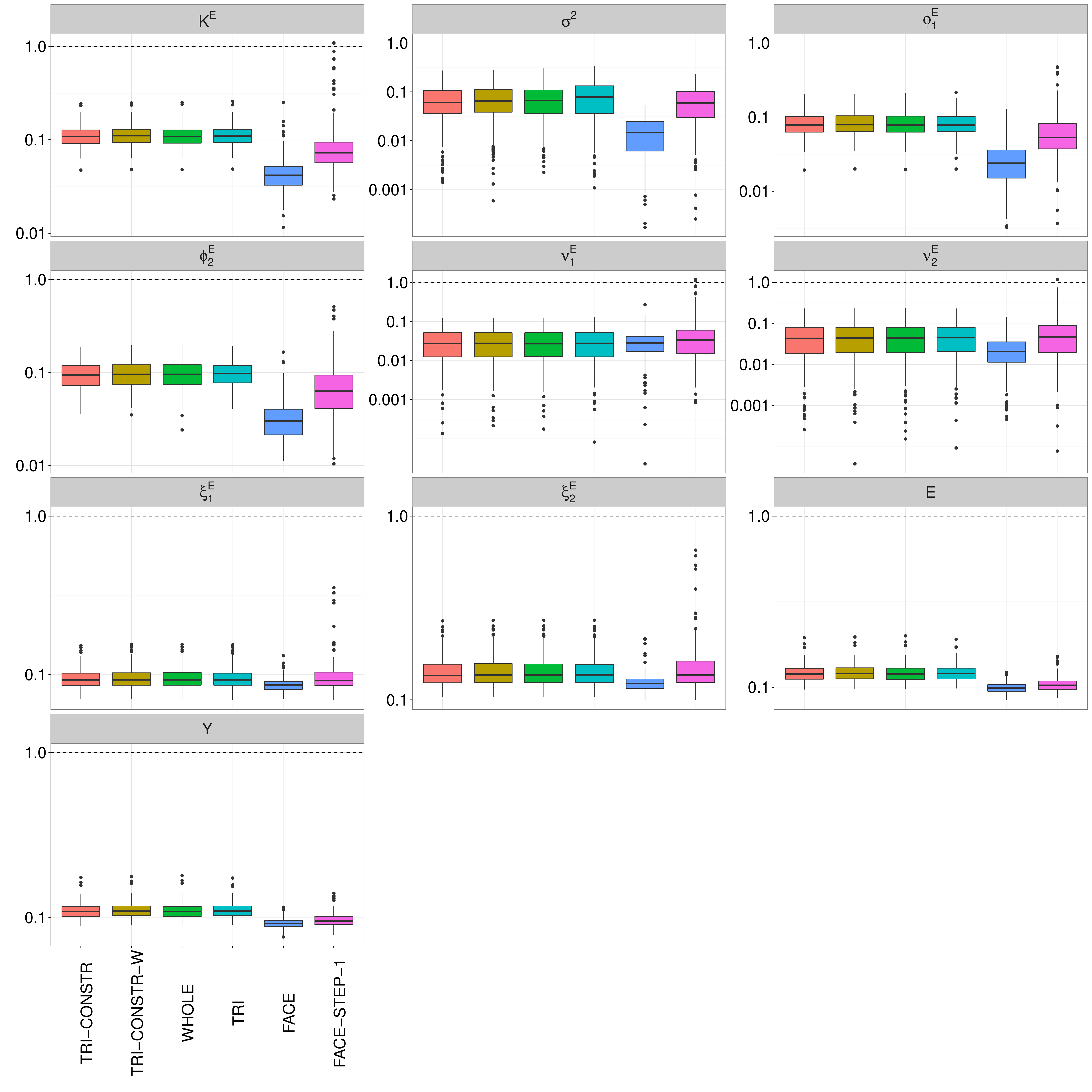}
\caption{Boxplots of the rrMSEs (log10 scale at y-axis) for Setting 9 of the scenario with independent curves. Top row: rrMSEs for auto-covariance $K^E(t,t^\prime)$, error variance $\sigma^2$, and the first eigenfunction $\phi^E_1(t)$. Second row: rrMSEs for the second eigenfunction $\phi^E_2(t)$ and eigenvalues $\nu^E_1$, $\nu^E_2$. Third row: rrMSEs for the random basis weights $\xi^E_1$, $\xi^E_2$ and process $E_i(t)$. Bottom row: rrMSEs for curves $Y_i(t)$.}
\label{fig: boxplots rrMSEs indep curves 30_Mar}
\end{center}
\end{figure}

\begin{figure}[p]
\begin{center}
\textbf{Setting 10}\\
\includegraphics[width= 1\textwidth]{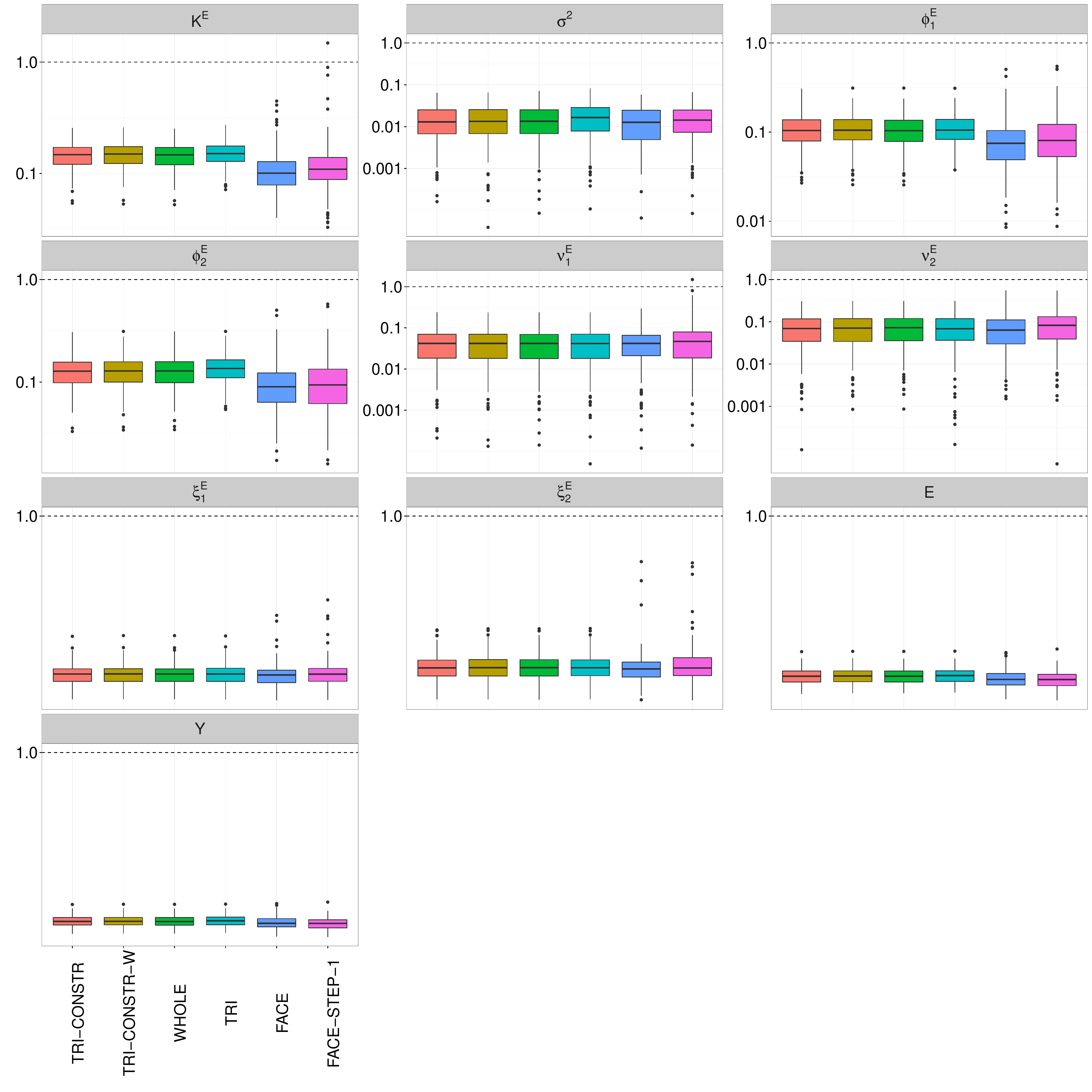}
\caption{Boxplots of the rrMSEs (log10 scale at y-axis) for Setting 10 of the scenario with independent curves. Top row: rrMSEs for auto-covariance $K^E(t,t^\prime)$, error variance $\sigma^2$, and the first eigenfunction $\phi^E_1(t)$. Second row: rrMSEs for the second eigenfunction $\phi^E_2(t)$ and eigenvalues $\nu^E_1$, $\nu^E_2$. Third row: rrMSEs for the random basis weights $\xi^E_1$, $\xi^E_2$ and process $E_i(t)$. Bottom row: rrMSEs for curves $Y_i(t)$.}
\label{fig: boxplots rrMSEs indep curves sigmasq05_30_Mar}
\end{center}
\end{figure}

\begin{figure}[p]
\begin{center}
\textbf{Setting 11}\\
\includegraphics[width= 1\textwidth]{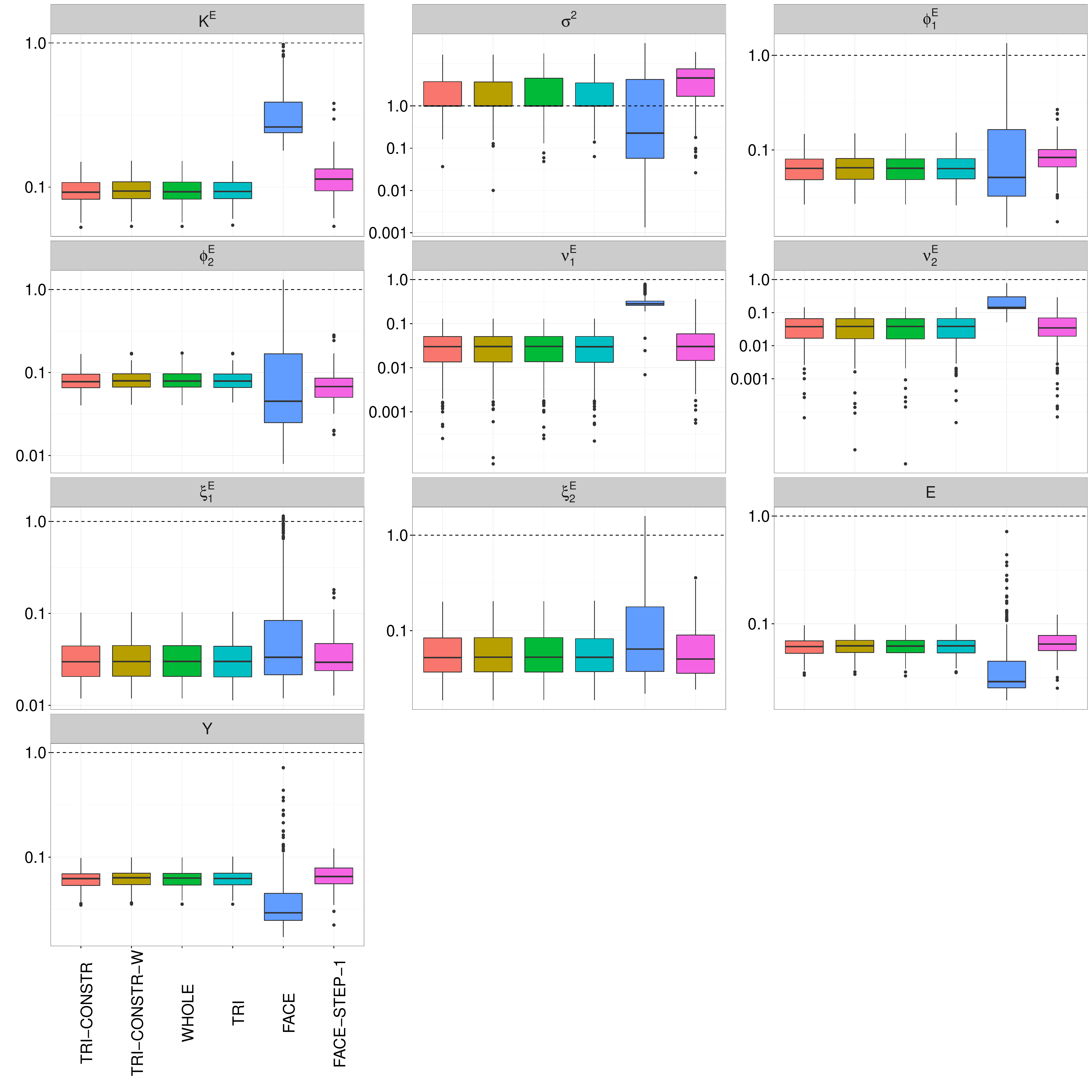}
\caption{Boxplots of the rrMSEs (log10 scale at y-axis) for Setting 11 of the scenario with independent curves. Top row: rrMSEs for auto-covariance $K^E(t,t^\prime)$, error variance $\sigma^2$, and the first eigenfunction $\phi^E_1(t)$. Second row: rrMSEs for the second eigenfunction $\phi^E_2(t)$ and eigenvalues $\nu^E_1$, $\nu^E_2$. Third row: rrMSEs for the random basis weights $\xi^E_1$, $\xi^E_2$ and process $E_i(t)$. Bottom row: rrMSEs for curves $Y_i(t)$.}
\label{fig: boxplots rrMSEs indep curves 31_Mar}
\end{center}
\end{figure}

In the following, we show the complete results for the remaining ten settings (Setting 2 -- Setting 11) for the scenario with independent curves. Table \ref{tab: settings indep scenario} lists the different settings we consider for this scenario. In Figure \ref{fig: boxplots rrMSEs indep curves 22_Mar} to Figure \ref{fig: boxplots rrMSEs indep curves 31_Mar}, we depict boxplots of the rrMSEs based on the 200 simulation runs for all model components.
\begin{table}[h!]
\begin{center}
\caption{Specification of the eleven considered settings for the scenario with independent curves. The simple eigenfunctions are given as $\lbrace \phi_1(t) = 1, \phi_2(t) = \sqrt{3}\left(2t-1\right) \rbrace$, the complex eigenfunctions are given as $\lbrace \phi_1(t) =\sin(2\pi t), \phi_2(t) = \cos(2\pi t)\rbrace$. The results for Setting 1 are shown in Section \ref{sec: simulation results}.}
\label{tab: settings indep scenario}
\begin{tabular}{l|llll}
Setting & grid & eigenfunctions & eigenvalues & error variance\\ \hline
Setting 1 &dense & complex & $\nu_1=2$, $\nu_2=1$& $\sigma^2=0.05$\\ \hdashline 
Setting 2 & dense & complex & $\nu_1=2$, $\nu_2=1$,&$\sigma^2=0.5$\\
Setting 3 & dense & simple & $\nu_1=2$, $\nu_2=1$ & $\sigma^2=0.05$\\
Setting 4 & dense & simple & $\nu_1=2$, $\nu_2=1$ & $\sigma^2=0.5$\\
Setting 5 & sparse & simple & $\nu_1=2$, $\nu_2=1$ & $\sigma^2=0.05$\\
Setting 6 & sparse & simple & $\nu_1=2$, $\nu_2=1$ & $\sigma^2=0.5$\\
Setting 7 & dense & complex  & $\nu_1=0.15$, $\nu_2=0.075$ & $\sigma^2=0.05$\\
Setting 8 & dense & complex & $\nu_1=0.15$, $\nu_2=0.075$ & $\sigma^2=0.5$\\
Setting 9 & dense & simple & $\nu_1=0.15$, $\nu_2=0.075$ &$\sigma^2=0.05$\\
Setting 10 & dense & simple & $\nu_1=0.15$, $\nu_2=0.075$ &$\sigma^2=0.5$\\
Setting 11 & dense & complex & $\nu_1=2$, $\nu_2=1$ & $\sigma^2=0.01$
\end{tabular}
\end{center}
\end{table}

\newpage

\clearpage
\subsection{Results for the scenario with crossed fRIs} \label{sec: results crossed}
\begin{figure}[p]
\begin{center}
\includegraphics[width=1\textwidth]{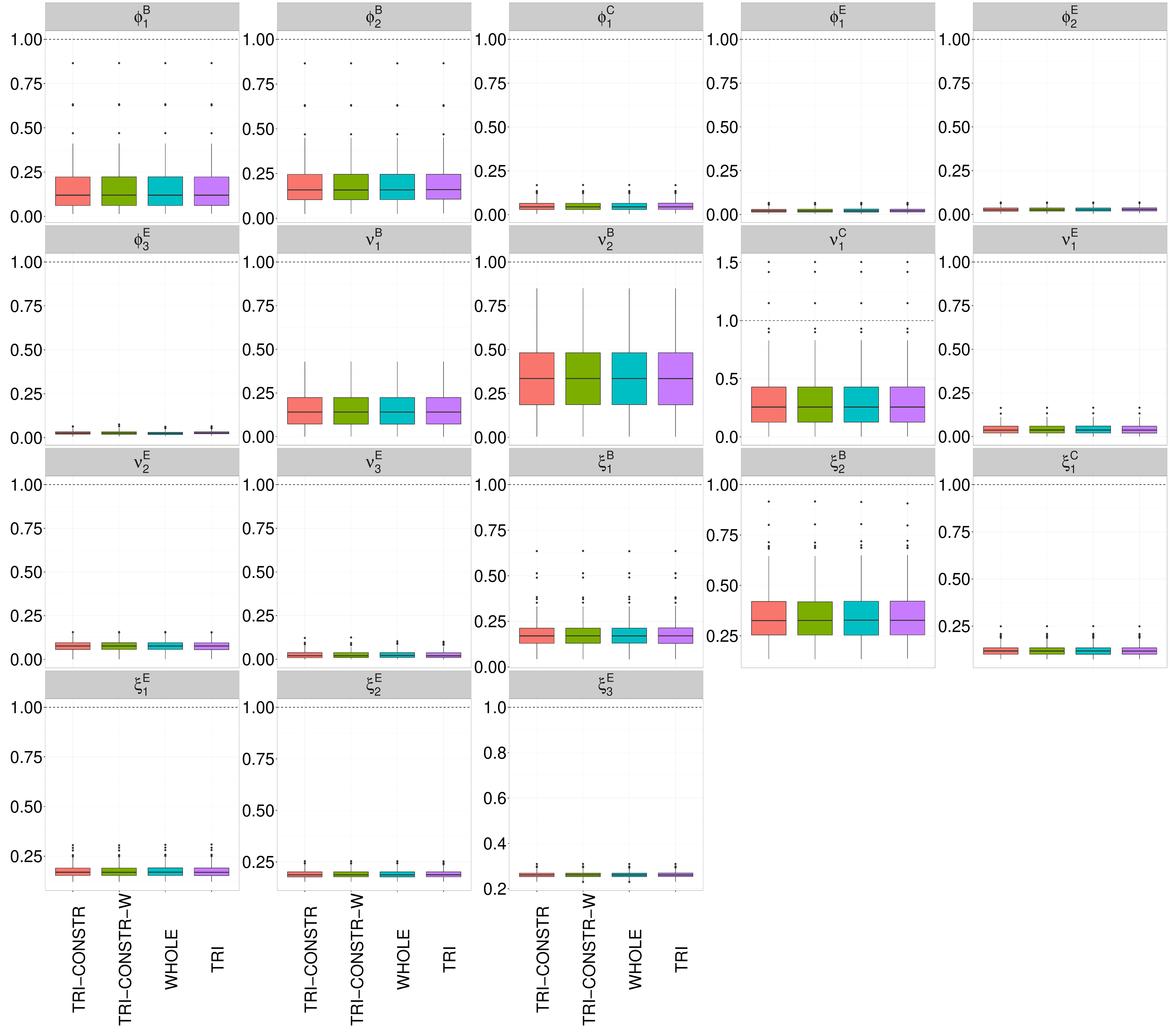}
\caption{Boxplots of the rrMSEs for the crossed fRIs setting. Shown are boxplots for all remaining model components, which are not shown in Section \ref{sec: simulation results}: The rrMSEs for the eigenfunctions and eigenvalues of the three auto-covariances $K^B(t,t^\prime)$, $K^C(t,t^\prime)$, and $K^E(t,t^\prime)$, as well as the rrMSEs for the corresponding random basis weights.}
\label{fig: boxplot rrMSEs crossed remaining curves 20 May}
\end{center}
\end{figure}

In the following, we show the remaining results for the scenario with crossed fRIs. Figure \ref{fig: boxplot rrMSEs crossed remaining curves 20 May}, shows boxplots of the rrMSEs for the estimated eigenfunctions and eigenvalues, as well as for the random basis weights for the three random processes $B_i(t)$, $C_i(t)$, and $E_i(t)$. All boxplots are based on 200 simulation runs.

\end{document}